\documentclass[ALICE,manyauthors]{cernphprep}

\usepackage[comma,square,numbers,sort&compress]{natbib}
\usepackage{amsmath,amssymb,amsfonts}

\usepackage{xspace}
\usepackage[utf8]{inputenc}
\usepackage[normalem]{ulem}
\usepackage[T1]{fontenc}
\usepackage{units}
\usepackage{bm}

\usepackage{multirow}
\usepackage{dcolumn}
\usepackage{tabularx}
\usepackage{booktabs}
\newcolumntype{C}{>{\centering\arraybackslash}X}

\usepackage{graphics}
\usepackage{graphicx}
\usepackage{subfigure}
\usepackage{grffile}
\usepackage{epsfig}
\usepackage{rotating}
\graphicspath{{figures/}}

\usepackage[dvipsnames]{xcolor}
\usepackage{hyperref}

\usepackage{listings}
\usepackage{lineno}
\usepackage{orcidlink}

\DeclareUnicodeCharacter{00A0}{}

\newcommand{\pPb}   {\mbox{p--Pb}\xspace}

\newcommand{\snn}          {\ensuremath{\sqrt{s_{\mathrm{NN}}}}\xspace}
\newcommand{\pt}           {\ensuremath{p_{\rm T}}\xspace}

\newcommand{\ycms}         {\ensuremath{y_{\rm CMS}}\xspace}

\newcommand{\nineH}         {$\sqrt{s}~=~0.9$~Te\kern-.1emV\xspace}
\newcommand{\seven}         {$\sqrt{s}~=~7$~Te\kern-.1emV\xspace}
\newcommand{\twoH}          {$\sqrt{s}~=~0.2$~Te\kern-.1emV\xspace}
\newcommand{\twosevensix}   {$\sqrt{s}~=~2.76$~Te\kern-.1emV\xspace}
\newcommand{\five}          {$\sqrt{s}~=~5.02$~Te\kern-.1emV\xspace}
\newcommand{\twosevensixnn} {$\sqrt{s_{\mathrm{NN}}}~=~2.76$~Te\kern-.1emV\xspace}
\newcommand{\fivenn}        {$\sqrt{s_{\mathrm{NN}}}~=~5.02$~Te\kern-.1emV\xspace}

\newcommand{\GeVc}          {Ge\kern-.1emV/$c$\xspace}
\newcommand{\MeVc}          {Me\kern-.1emV/$c$\xspace}
\newcommand{\TeV}           {Te\kern-.1emV\xspace}
\newcommand{\GeV}           {Ge\kern-.1emV\xspace}
\newcommand{\MeV}           {Me\kern-.1emV\xspace}
\newcommand{\GeVmass}       {Ge\kern-.2emV/$c^2$\xspace}
\newcommand{\MeVmass}       {Me\kern-.2emV/$c^2$\xspace}

\newcommand{\sNN}{\ensuremath{\snn}}

\newcommand{\pT}{\ensuremath{\pt}}

\newcommand{\thirteen} {$\sqrt{s}~=~13$~Te\kern-.1emV\xspace}

\newcommand{\cent}   [2] {$#1$--$#2\%$}

\DeclareUnicodeCharacter{00A0}{}

\begin{document}

\begin{titlepage}
\PHyear{2022}
\PHnumber{209}
\PHdate{11 October}

\title{Measurements of azimuthal anisotropies at forward and backward rapidity with muons in high-multiplicity \pPb collisions at $\mathbf{\sNN = 8.16}$~\TeV}
\ShortTitle{Muon azimuthal anisotropies in high-multiplicity p--Pb collisions}

\Collaboration{ALICE Collaboration\thanks{See Appendix~\ref{app:collab} for the list of collaboration members}}
\ShortAuthor{ALICE Collaboration}

\begin{abstract}

The study of the azimuthal anisotropy of inclusive muons produced in p--Pb collisions at $\sNN = 8.16$~\TeV, using the ALICE detector at the LHC is reported.
The measurement of the second-order Fourier coefficient of the particle azimuthal distribution, $v_2$, is performed as a function of transverse momentum $\pT$ in the \cent{0}{20} high-multiplicity interval at both forward ($2.03 < \ycms < 3.53$) and backward ($-4.46 < \ycms < -2.96$) rapidities over a wide $\pT$ range, $0.5 < \pT < 10$~\GeVc, in which a dominant contribution of muons from heavy-flavour hadron decays is expected at $\pT > 2$~\GeVc.
The $v_2$ coefficient of inclusive muons is extracted using two different techniques, namely two-particle cumulants, used for the first time for heavy-flavour measurements, and forward--central two-particle correlations.
Both techniques give compatible results.
A positive $v_2$ is measured at both forward and backward rapidities with a significance larger than $4.7\sigma$ and $7.6\sigma$, respectively, in the interval $2 < \pT < 6$~\GeVc.
Comparisons with previous measurements in \pPb collisions at \fivenn, and with AMPT and CGC-based theoretical calculations are discussed.
The findings impose new constraints on the theoretical interpretations of the origin of the collective behaviour in small collision systems.
\end{abstract}
\end{titlepage}

\setcounter{page}{2}


\section{Introduction}\label{sec1}

The study of high-energy heavy-ion collisions aims at investigating the properties of a state of strongly-interacting matter characterised by high energy density and temperature, called the quark--gluon plasma (QGP)~\cite{Muller:2006ee,Bazavov:2011nk}.
One of the key observables to address the transport properties of the QGP is the azimuthal anisotropy of produced particles~\cite{Ollitrault:1992bk}. 
In non-central collisions, the initial spatial anisotropy of the overlap region is converted into an anisotropy in momentum space via multiple interactions.
The magnitude of the azimuthal anisotropies is usually quantified via a Fourier decomposition of the particle azimuthal distribution given by
\begin{equation}
 {{\rm d}^2N \over {{\rm d}p_{\rm T}{\rm d}\varphi }} = {1\over 2\pi}
{{\rm d}N \over {\rm d} p_{\rm T}}
\bigg ( 1 + 2 \sum_{{n} = 1}^{\infty} v_{n} (p_{\rm T}) {\rm cos}
\lbrack {n} ( \varphi - \Psi_{n} ) \rbrack \bigg),
\label{eq:phi}
\end{equation}
where $\varphi$ and $p_{\rm T}$ are the particle azimuthal angle and transverse momentum, respectively.
The Fourier coefficients $v_{n}$ characterise the anisotropy of produced particles~\cite{Voloshin:1994mz} and $\Psi_{n}$ is the azimuthal angle of the symmetry plane for the $n^{\rm th}$ harmonic.
The largest contribution to the asymmetry of non-central collisions is provided by the second Fourier coefficient $v_2$ referred to as elliptic flow, and is expressed as $v_2 = \langle \rm cos \lbrack 2 (\varphi - \Psi_2) \rbrack\rangle$~\cite{Voloshin:1994mz,Ollitrault:1992bk}, where the brackets denote the average over all particles and all selected events.

Due to their large masses, heavy quarks (charm and beauty) are mostly produced via hard partonic scattering processes in the very early stage of the collisions before the formation of the QGP, and therefore they probe the properties and dynamics of the QGP through its full evolution~\cite{Averbeck:2013oga}.
During their propagation through the medium, they lose energy via elastic and inelastic processes.
Heavy-flavour hadrons and their decay products are thus sensitive probes of the QGP medium.
The open heavy-flavour $v_2$ coefficient is expected to provide information on the collective expansion of the medium and thermalisation of heavy quarks at low $\pT$~\cite{Ollitrault:1992bk,Kolb:2003dz}, and is sensitive to the path-length dependence of the in-medium energy loss at high $\pT$~\cite{Gyulassy:2000gk,Shuryak:2001me}.
These contributions are predicted to give positive $v_2$ values.
Moreover, the elliptic flow is also expected to be sensitive to the hadronisation processes at low and intermediate $\pT$~\cite{Greco:2003vf,Molnar:2004ph}. Extensive measurements of the elliptic flow of heavy flavours have been carried out in Pb--Pb collisions at a centre-of-mass energy $\sNN =$~2.76 and 5.02 TeV at the LHC, and in Au--Au collisions at $\sNN =$~200 GeV at the RHIC~\cite{STAR:2017kkh} where a positive elliptic flow coefficient was observed, see Refs.~\cite{ALICE:2021kfc,CMS:2020bnz,Aad:2020grf,Acharya:2020qvg,ALICE:2020iug,Aaboud:2018bdg,Acharya:2018bxo,Sirunyan:2017plt,ALICE:2017pbx,Adam:2016ssk,CMS:2016mah,Adam:2015pga,Abelev:2014ipa} for results on open heavy flavours. The nuclear modification factor $R_{\rm AA}$, defined as the ratio of the particle yields in nucleus--nucleus (AA) collisions at the same centre-of-mass energy to that in binary-scaled pp collisions, is also an important observable to quantify in-medium effects. In addition to the non-zero $v_2$ coefficient, the $R_{\rm AA}$ measurements in heavy-ion collisions at LHC energies show a strong suppression of the open heavy-flavour yields at intermediate and high $\pT$~\cite{ALICE:2021rxa,ALICE:2021kfc,ATLAS:2021xtw,ALICE:2020sjb,Acharya:2019mom,Aaboud:2018bdg,Acharya:2018upq,Acharya:2018hre,Sirunyan:2018ktu,CMS:2018eso,ATLAS:2018hqe,Sirunyan:2017xss,Sirunyan:2017oug,Sirunyan:2017isk,Adam:2016khe,CMS:2016mah}. These results reflect significant energy losses of heavy quarks due to strong interactions with the medium constituents.

The study of small collision systems such as p--Pb collisions at the LHC was initially proposed to address cold nuclear matter (CNM) effects relevant for the interpretation of the measurements in heavy-ion collisions, such as the nuclear modification of the parton distribution functions~\cite{Eskola:2009uj}, $k_{\rm T}$ broadening~\cite{Kopeliovich:2002yh}, and energy loss in cold nuclear matter~\cite{Kang:2014hha}. Surprisingly, long-range structures in two-particle correlations (``ridges'') associated with a positive $v_2$ were first observed for light-flavour hadrons in high-multiplicity p--Pb collisions at $\sNN = 5.02 $~TeV in the midrapidity region by the ALICE, ATLAS, and CMS collaborations~\cite{CMS:2012qk,Abelev:2012ola,Aad:2012gla,ABELEV:2013wsa,Aaboud:2016yar,Adam:2015sza} and at forward rapidity by the LHCb collaboration~\cite{Aaij:2015qcq}, using the two-particle azimuthal correlation method. In the heavy-flavour sector, the ALICE and CMS collaborations revealed also hints of collectivity with the measurement of a positive $v_2$ of inclusive muons~\cite{Adam:2015bka}, inclusive and prompt J/$\psi$~\cite{Acharya:2017tfn,Sirunyan:2018kiz}, prompt and non-prompt D mesons~\cite{ALICE:2016clc,Acharya:2019icl,Sirunyan:2018toe,Sirunyan:2020obi}, and electrons from heavy-flavour hadron decays~\cite{Acharya:2018dxy}, using two-particle correlations. A positive $v_2$ of muons from heavy-flavour hadron decays was measured by ATLAS in high-multiplicity pp collisions at $\sNN$ = 13 TeV~\cite{ATLAS:2019xqc} as well. Long-range correlations were also observed at lower beam energies, in d--Au~\cite{Adare:2013piz,Adare:2014keg,Adamczyk:2015xjc} and $^3$He--Au collisions~\cite{Adare:2015ctn}, by the PHENIX and STAR collaborations at RHIC. Finally, the PHENIX collaboration studied the pseudorapidity ($\eta$) dependence of the charged-particle $v_2$ at high multiplicity in various asymmetric collision systems~\cite{PHENIX:2018hho} as well. The $v_2$ signal was found to increase from the smallest to the largest system, and was also more pronounced in the backward rapidity region ($-3 < \eta < -1$) than at forward rapidity ($1 < \eta < 3$). Several strategies were developed to subtract the correlations not related to collectivity but rather due to jet correlations and resonance decays, referred to as nonflow effects. These nonflow contributions were usually suppressed by requiring a pseudorapidity separation between particles forming the pair and by subtracting correlations measured in low-multiplicity collisions~\cite{Abelev:2012ola,Aad:2014lta}.
Later on, a standard template fit procedure~\cite{Aad:2019ajj} was implemented to isolate the long-range correlations, which was further corrected to consider the multiplicity dependence of the Fourier coefficients $v_n$~\cite{ATLAS:2018ngv}. It was also observed that multi-particle
cumulants~\cite{Khachatryan:2015waa,Aaboud:2017acw,Aaboud:2017blb,Sirunyan:2019pbr,Aaboud:2016yar,Acharya:2019vdf,Abelev:2014mda} strongly suppress nonflow correlations.

On the other hand, the simultaneous observation of a positive $v_2$ and particle yields compatible with those in binary-scaled pp collisions, i.e. a nuclear modification factor $R_{\rm pPb} \sim 1$, for high-$\pT$ charged particles measured in high-multiplicity p--Pb collisions at $\sNN =$~5.02 TeV, is also puzzling~\cite{Adam:2014qja,Aad:2016zif,Aad:2019ajj}. As reported in Ref.~\cite{Aad:2019ajj}, a jet quenching calculation with two different initial geometries~\cite{Zhang:2013oca} slightly underestimates the measured $v_2$ and fails in describing the $R_{\rm pPb}$ data which are in favour of no significant energy loss effects. Several other possible scenarios (see a review in Ref.~\cite{Dusling:2015gta}) relying either on final-state effects involving a hydrodynamic evolution of the produced particles~\cite{Bozek:2011if,Bozek:2012gr,Nagle:2018nvi} or initial-state effects such as gluon saturation in the framework of the colour glass condensate (CGC)~\cite{Dusling:2012cg,Dusling:2013oia,Dusling:2015gta}, are investigated. Colour-charge exchanges in the final state~\cite{Dusling:2012cg} or the anisotropic escape of partons from the surface of the interaction region~\cite{He:2015hfa} might also contribute to the flow-like effects evidenced in small collision systems. A MultiPhase Transport (AMPT) model~\cite{Lin:2004en,Li:2018leh,Lin:2021mdn} addresses non-equilibrium dynamics and provides a microscopic evolution of parton interactions, hence including also the parton escape mechanism. The latter scenario is further investigated using the string-melting version of the AMPT model~\cite{Lin:2021mdn}, employed here for the comparison with the data (see Section~\ref{sec:res}). In this version, the initial strings are converted into partons and the interactions are described via a parton cascade model~\cite{Zhang:1997ej}. The partons are then combined into hadrons via a spatial quark coalescence model and the rescatterings between hadrons are described by a relativistic transport (ART) model~\cite{Li:1995pra}.

In order to shed more light on the origin of the azimuthal anisotropies in small systems, this letter presents new results concerning the $v_2$ coefficient of inclusive muons in high-multiplicity p--Pb collisions at $\sNN = 8.16$~TeV with the ALICE detector at the LHC. These measurements are performed at forward ($2.03 < \ycms < 3.53$) and backward ($-4.46 < \ycms < -2.96$) rapidities and cover the transverse momentum interval $0.5 < \pT < 10$~GeV/$c$. They are obtained in a significantly extended $\pT$ range dominated by decays of heavy-flavour hadrons at
$\pT > 2$~GeV/$c$. The total uncertainties are reduced by a factor up to about 2.1 and 1.3 at forward and backward rapidities, respectively, compared to the previous ALICE muon results in p--Pb collisions at $\sNN =$~5.02 TeV in the interval $0.5 < \pT < 4$~GeV/$c$~\cite{Adam:2015bka}. Various analysis methods which exhibit different sensitivity
to nonflow effects, are implemented to measure the $\pT$-differential $v_2$ of inclusive muons. The two-particle correlation method and the two-particle cumulants with generic framework~\cite{Bilandzic:2013kga}, the latter applied for the first time for heavy-flavour $v_2$ measurements with ALICE, are used.
In order to reduce nonflow contributions, a novel procedure based on the subtraction of correlations measured in low-multiplicity events is developed.

The letter is organised as follows. Section~\ref{sec:det} presents the ALICE apparatus with an emphasis on the detectors used in the analysis and the data taking conditions. Section~\ref{sec:ana} contains a description of the flow methods and a presentation of the analysis details. Section~\ref{sec:res} presents the results, namely the $\pT$-differential muon $v_2$ at forward and backward rapidities, measured using several multiplicity estimators with the two-particle correlation and two-particle cumulant methods. Comparisons with published measurements performed at $\sNN =$~5.02 TeV in various kinematic regions are reported. Detailed comparisons with model calculations based on the colour glass condensate and AMPT predictions are also discussed. A summary and concluding remarks are given in Section~\ref{sec:concl}.

\section{Experimental apparatus and data samples}\label{sec:det}

A detailed description of the ALICE apparatus and its performance can be found in Refs.~\cite{Aamodt:2008zz,Abelev:2014ffa}. The main subdetectors used in the analysis are presented in the following. The analysis uses muons reconstructed in the muon spectrometer which covers the pseudorapidity interval $-4.0 < \eta < -2.5$. The muon spectrometer consists of a 10 nuclear interaction length ($\lambda_{\rm I}$) absorber in front of five tracking stations, each composed of two planes of cathode pad chambers, with the central one inside a dipole magnet of 3~Tm integrated field. The tracking system is completed with two trigger stations, each equipped with two planes of resistive plate chambers, behind a 1.2 m thick iron wall (7.2~$\lambda_{\rm I}$). The latter stops secondary hadrons escaping from the front absorber as well as low momentum muons from light-hadron decays. A conical absorber
protects the muon spectrometer throughout its full length against secondary particles produced by the interaction with
the beam pipe of primary particles at large $\eta$.
Among the central barrel detectors, the two innermost layers of the Inner Tracking System composing the Silicon Pixel Detector (SPD) cover the pseudorapidity intervals $\vert \eta \vert < 2$ and
$\vert \eta \vert < 1.4$. The SPD is employed for the determination of the position of the primary interaction vertex. The SPD tracklets, track segments joining hits in the two SPD layers, are also used in the flow analysis, either as associated particles in the two-particle correlation method or to determine the reference flow with two-particle cumulants (Section~\ref{sec:ana}).
The V0 detector composed of two scintillator arrays, covers the pseudorapidity intervals $ 2.8 < \eta < 5.1$ (V0A) and $-3.7 < \eta < -1.7$ (V0C). It provides the minimum bias (MB) trigger defined by the coincidence of signals in the two sets of scintillators. The V0 is also used for the luminosity determination, and an independent measurement is obtained with the two T0 arrays of Cherenkov detectors located in the regions $4.6 < \eta < 4.9$ and $-3.3 < \eta < -3.0$. The two sets of Zero Degree Calorimeters (ZDC), each including a neutron calorimeter (ZN) and a proton calorimeter (ZP), are located on both sides of the interaction point at $z = \pm 112.5$~m, along the beam line. The timing information delivered by the V0 and ZDC is exploited offline to reject the beam-induced background. These two detectors are also used to estimate the event activity.

The results reported in this letter are obtained with the data samples recorded by ALICE during the 2016 p--Pb run at $\sNN =$~8.16 TeV, with two different beam configurations obtained by reversing the rotation direction of the proton and lead beams. Due to the asymmetry of the beam energy per nucleon, the nucleon$-$nucleon centre of mass is shifted in rapidity with respect to the laboratory frame by $\Delta y =$~0.465 in the direction of the proton beam. Therefore, inclusive muons are measured in the forward rapidity interval $2.03 < \ycms < 3.53$ with the proton beam travelling in the direction of the muon spectrometer (p-going direction, p--Pb configuration) and in the backward rapidity interval $-4.46 < \ycms < -2.96$ (Pb-going direction, Pb--p configuration). The analysis is based on muon-triggered events, requiring the MB condition and at least a track registered in the muon trigger stations with a $\pT$ above a programmable threshold value. Data were collected with two programmable $\pT$ thresholds set to about 0.5 GeV/$c$ and 4.2 GeV/$c$, referred to as MSL and MSH, respectively. These two thresholds are not sharp, and correspond to a $\sim$50\% efficiency for
muons~\cite{Bossu:2012jt}. Based on statistical considerations, the measurement of the $v_2$ of inclusive muons is performed by combining MSL- and MSH-triggered events which are used up to $\pT =$~2~GeV/$c$ and above this value, respectively. It has been checked that in the overlapping region, both samples give same inclusive muon $v_2$ results within uncertainties. After applying an event selection which uses the information of the V0 and SPD together with an algorithm to tag events with multiple vertices, the pile-up contribution is found negligible for the data samples considered in this analysis. Moreover, only events with a primary vertex along the beam axis $z_{\rm vtx}$ within $\pm$7~cm  are considered in order to reduce non-uniform acceptance effects. After the event selection, the integrated luminosity of the p--Pb and Pb--p data samples corresponds to about 0.22 (5.8)~nb$^{-1}$ and 0.22 (8.2)~nb$^{-1}$ for MSL-(MSH-)triggered events, respectively.
The p--Pb and Pb--p data samples are further classified according to their activity using multiplicity-based estimators~\cite{Acharya:2018egz} such as the total charge deposited in the two V0 arrays (V0M), the number of clusters in the outer layer of the SPD (CL1),
and the energy deposited by spectator neutrons in the ZN located in the direction of the Pb beam. The ZN estimator minimises biases on the binary scaling of hard processes~\cite{Acharya:2018egz}. The multiplicity classes are defined as percentile intervals of the p--Pb and Pb--p hadronic cross section. In this analysis, the 0--20\% and 60--90\% multiplicity classes are studied. The evolution of the charged-particle pseudorapidity density at midrapidity in different multiplicity classes and with several multiplicity estimators is reported in Ref.~\cite{ALICE:2018wma}.

The same selection criteria as in previous analyses~\cite{pubPbPb,Acharya:2019mky} are applied to the muon candidates. Tracks in the muon spectrometer are reconstructed within the pseudorapidity region $-4 < \eta < -2.5$ to reject tracks at the edge of the muon spectrometer acceptance. The track polar angle at the end of the front absorber $\theta_{\rm abs}$ has to satisfy the condition $170^\circ < \theta_{\rm abs} < 178^\circ$ in order to remove tracks crossing the high-density region of the absorber that undergo significant scattering. Tracks reconstructed in the muon tracking chambers are identified as muons by requiring their matching with corresponding track segments in the muon trigger chambers. Finally, a selection on the minimum distance of the track to the primary vertex in the transverse plane (DCA) weighted by its momentum ($p$) is applied to reject fake tracks and remaining beam-induced background tracks. The SPD tracklets are selected in the fiducial region $\vert \eta \vert < 1$.
Moreover, a condition on the pseudorapidity dependence on the longitudinal position of the primary vertex $z_{\rm vtx}$ is applied for the removal of edge effects in the region where the SPD acceptance is small. As in previous publications ~\cite{Adam:2015bka,Acharya:2017tfn}, a condition on the difference between the azimuthal angles of clusters in the two SPD layers with respect to the primary vertex, $\Delta \varphi_{\rm SPD} < 5$~mrad, is applied in order to select in average reference tracks with larger $\pT$, which exhibit a larger flow, and to reduce the contribution from fake and secondary tracklets.

\section{Analysis methods and associated systematic uncertainties}\label{sec:ana}

\subsection{Two-particle correlations}\label{sec:2correl}

The method using two-particle correlations to extract the azimuthal anisotropy is extensively discussed
in Refs.~\cite{CMS:2012qk,Abelev:2012ola,Aad:2012gla,ABELEV:2013wsa,Aaij:2015qcq,Adam:2015bka,Aaboud:2016yar,Acharya:2017tfn,Acharya:2018dxy}. The two-particle correlation between pairs of trigger particles, inclusive muons at forward or backward rapidities, and associated particles, SPD tracklets at midrapidity, is measured as a function of their azimuthal angle difference ($\Delta \varphi$) and pseudorapidity difference ($\Delta \eta$). The correlation is expressed in terms of $Y$, the associated yield per trigger particle defined as
\begin{linenomath}
\begin{equation}
Y = \frac{1}{N_{\rm trig}} \frac{{\rm d}^2N_{\rm assoc}}{{\rm d}\Delta \eta {\rm d}\Delta \varphi} =  \frac{S (\Delta \eta, \Delta \varphi)}{B (\Delta \eta, \Delta \varphi)},
\label{eq:2pc}
\end{equation}
\end{linenomath}
where $N_{\rm trig}$ is the total number of trigger particles, i.e. the number of inclusive muons in a given multiplicity class, $z_{\rm vtx}$ interval, and $\pT$ interval. The signal distribution $S(\Delta \eta, \Delta \varphi)$ given by $\displaystyle{\frac{1}{N_{\rm trig}} \frac{{\rm d}^2 N_{\rm same}}{{\rm d}\Delta \eta {\rm d} \Delta\varphi}}$, corresponds to the  associated yield per trigger particle for particle pairs from the same event. The background distribution $ \displaystyle B(\Delta \eta, \Delta \varphi) = \alpha \frac{{\rm d}^2 N_{\rm mix}}{{\rm d}\Delta \eta {\rm d} \Delta\varphi}$ is obtained by correlating trigger particles in an event with associated particles from other events in the same multiplicity class and $z_{\rm vtx}$ interval. The parameter $\alpha$ is introduced to normalise the background distribution to unity in the region of maximum pair acceptance. Both the signal and background distributions are determined considering the same multiplicity class and same $z_{\rm vtx}$ interval of 1~cm width. The final associated yield per trigger particle is obtained from an average over the $z_{\rm vtx}$ intervals weighted by $N_{\rm trig}$.

\begin{figure}[!t]
\begin{center}
\hspace*{-1.6cm}\includegraphics[width=0.45\textwidth]{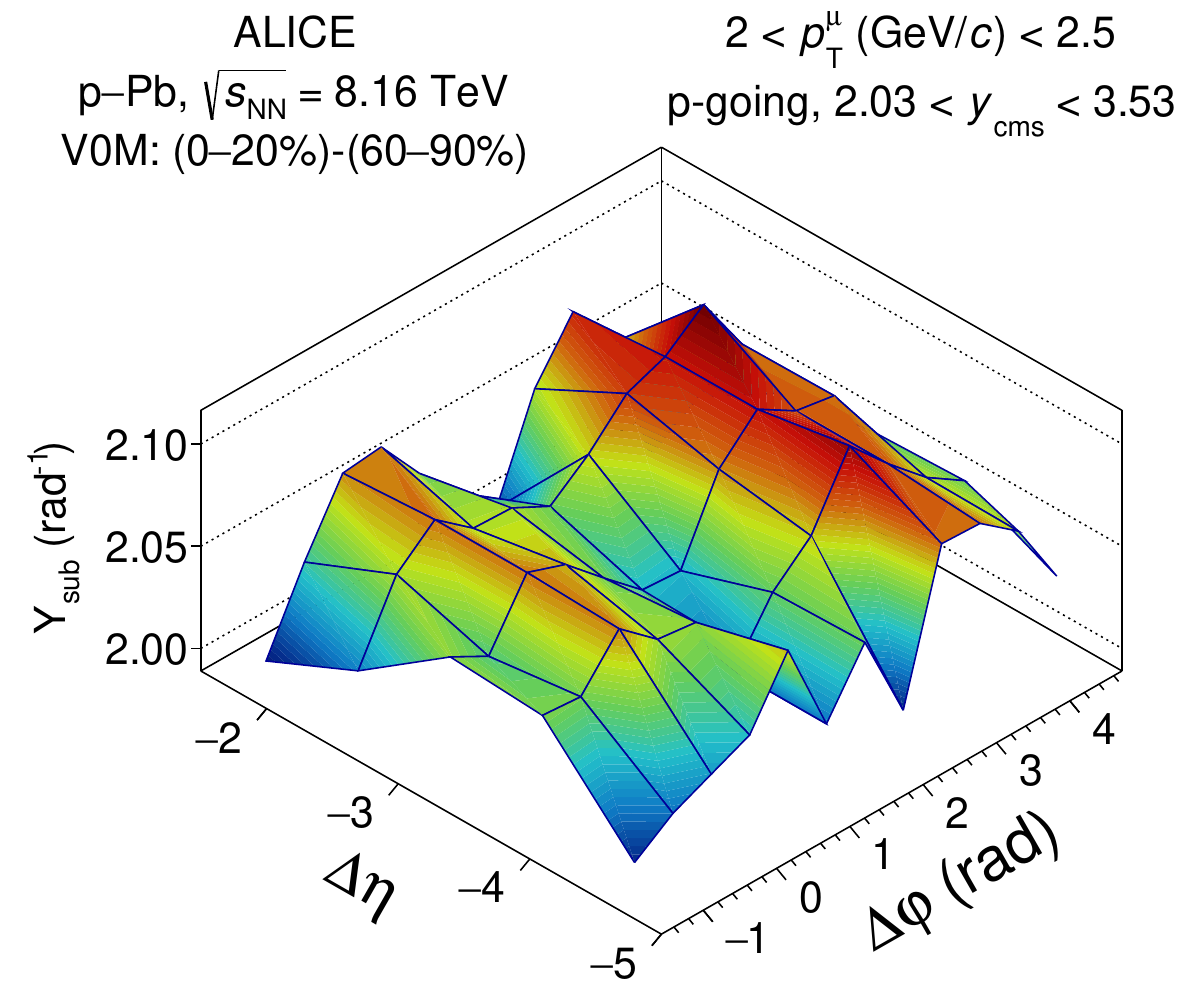}
\hspace*{0.8cm}\includegraphics[width=0.45\textwidth]{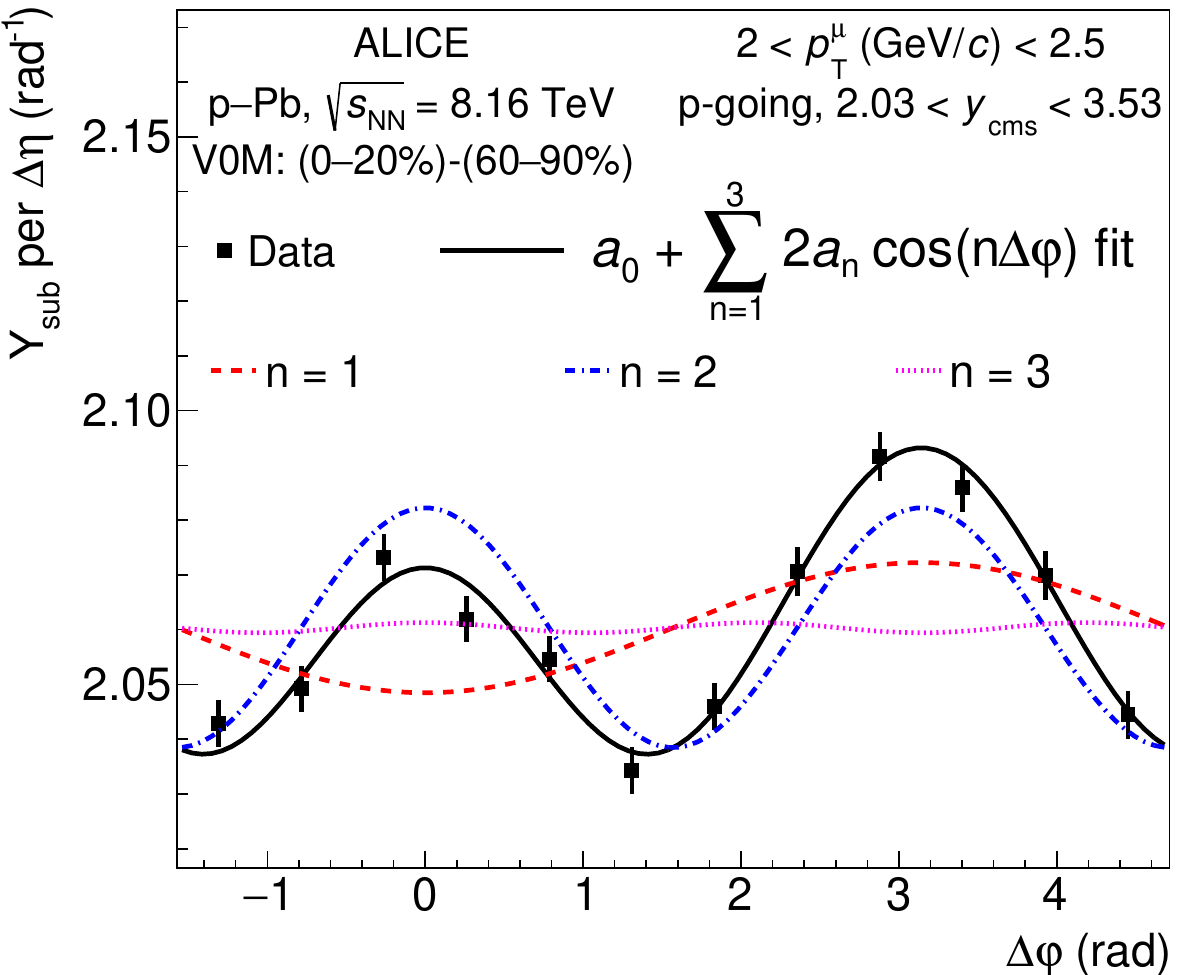}
\hspace*{-1.6cm}\includegraphics[width=0.45\textwidth]{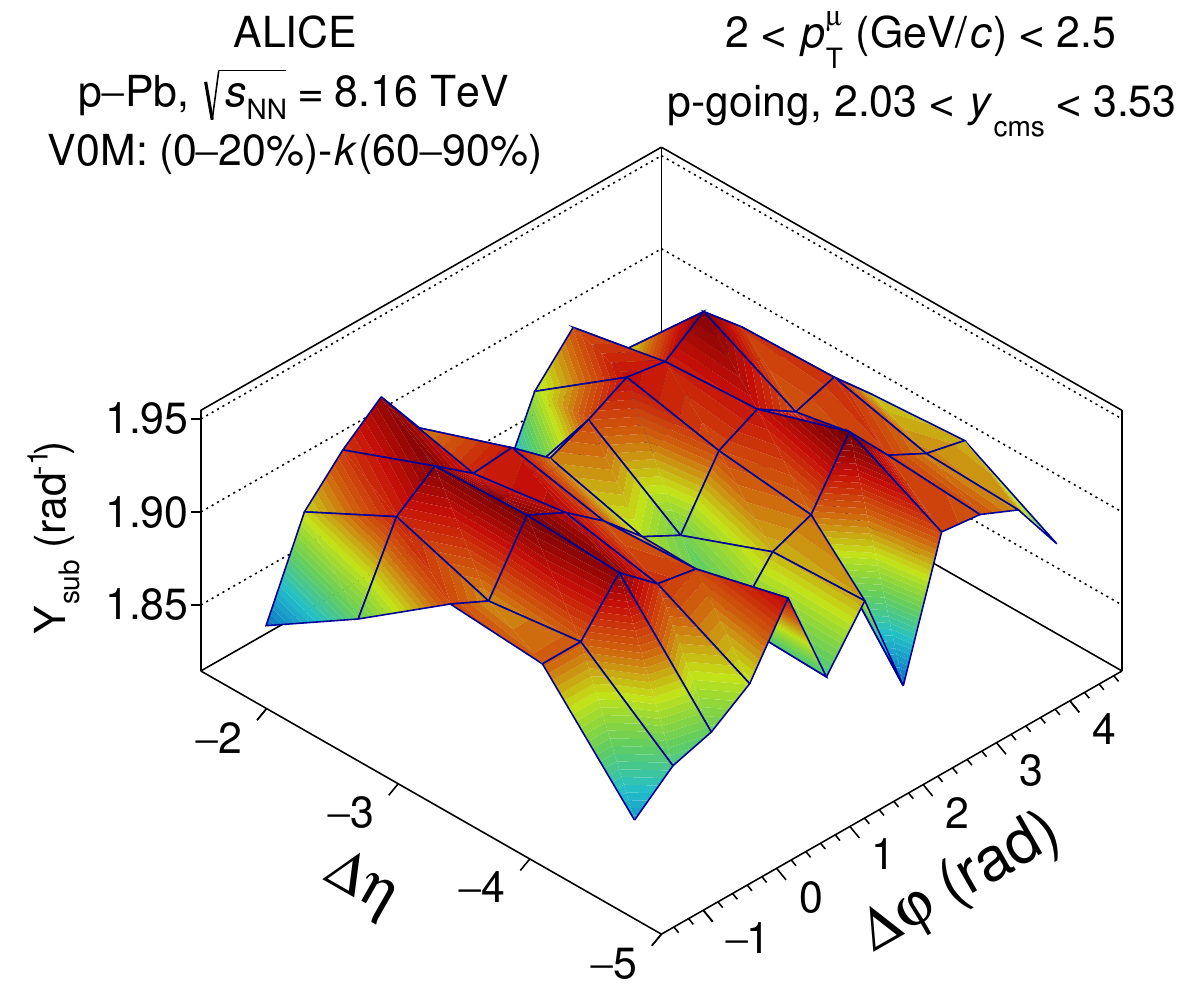}
\hspace*{0.8cm}\includegraphics[width=0.45\textwidth]{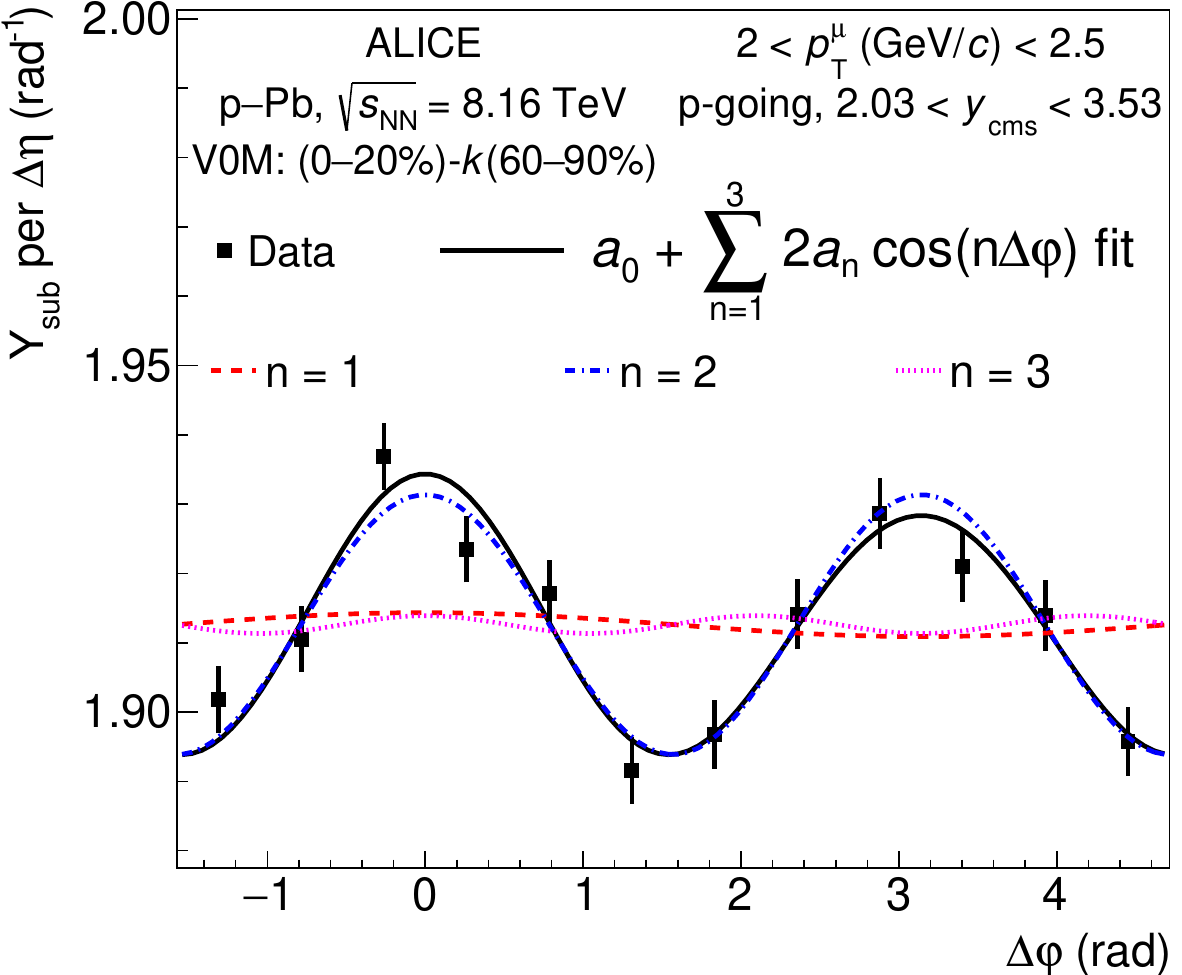}
\caption{Left: associated yield per trigger particle as a function of $\Delta \eta$ and $\Delta \varphi$ between inclusive muons with $ 2 < \pT < 2.5$~GeV/$c$ and SPD tracklets, $Y_{\rm sub}$, in high-multiplicity (0--20\%) p--Pb collisions at $\sNN =$~8.16 TeV after the subtraction of the unscaled (top) and scaled (bottom) muon--tracklet correlation in low-multiplicity (60--90\%) events. Right: fit of the corresponding correlation distributions in $-5 < \Delta\eta < -1.5$ projected onto $\Delta\varphi$ with a Fourier decomposition, see Eq.~(\ref{eq:fourier}). The three first harmonics are also presented. }
\label{Fig:mu-tracklet}
\end{center}
\end{figure}

The distribution of the associated yield per trigger particle (Eq.~(\ref{eq:2pc})) measured in high-multiplicity collisions is usually composed of correlations arising from collective and nonflow effects, the latter consisting of near-side ($\vert \Delta \varphi \vert < \pi/2$) and away-side ($ \pi/2 < \Delta \varphi < 3\pi/2$) jet structures. These nonflow effects can be reduced by subtracting the per-trigger yield distribution measured in low-multiplicity collisions~\cite{Abelev:2012ola}. A typical example of associated yield per trigger particle for muon--tracklet correlations as a function of $\Delta \eta$ and $\Delta \varphi$ after the subtraction of the per-trigger yield measured in low-multiplicity (60--90\%) collisions~\cite{Abelev:2012ola}, labelled as $Y_{\rm sub}$, is shown in Fig.~\ref{Fig:mu-tracklet} (top-left panel) for high-multiplicity (0--20\%) p--Pb collisions at $\sNN =$~8.16 TeV. The selected $\pT$ interval of trigger particles is $2 < \pT < 2.5$~GeV/$c$.
A double-ridge structure is observed with a near-side ridge and a away-side ridge centred at $\Delta \varphi = 0$ and $\Delta \varphi = \pi$, respectively. This double-ridge structure indicates the presence of collective effects in p--Pb collisions. In order to quantify these remaining correlations, the resulting two-dimensional subtracted distribution in $-5 < \Delta \eta < -1.5$ is projected onto $\Delta \varphi$ and is further fitted with a Fourier series up to the third order

\begin{linenomath}
\begin{equation}
\frac{1}{N_{\rm trig}}
\frac{{\rm d}N_{\rm assoc}}{{\rm d} \Delta \varphi} = a_0 + \sum_{ {\rm n } = 1}^{3} 2 a_{\rm n} {\rm  cos} ({\rm n} \Delta\varphi),
\label{eq:fourier}
\end{equation}
\end{linenomath}
where, as shown in Fig.~\ref{Fig:mu-tracklet} (top-right panel), an azimuthal anisotropy dominated by the second-order coefficient $a_2$ is observed.

After the subtraction of the muon--tracklet distribution measured in low-multiplicity collisions, the away-side jet yield in high-multiplicity collisions has strongly decreased. However, since the jet correlation is observed to be diminished in low-multiplicity collisions~\cite{Abelev:2014mva}, a residual jet contamination in the subtracted distribution, in particular in the away side, is still present. The subtraction of this potential remaining jet component in the away side relies on the scaling of the low-multiplicity event class as described hereafter (see also Refs.~\cite{Abelev:2014mva,Adam:2015bka,Aaboud:2016yar}). The per-trigger associated yield in the low-multiplicity event class is first fitted with a Gaussian function centered at $\Delta\varphi = \pi$. The per-trigger yield in the high-multiplicity event class is further adjusted with Eq.~(\ref{eq:fourier}) replacing the cos($\Delta \varphi$) term by a Gaussian function whose width is fixed to the value obtained from the fit in low-multiplicity events. The per-trigger yield distribution in low-multiplicity events is scaled by a factor $k$ defined as the ratio of the corresponding yields in the away side in high-multiplicity collisions to those in low-multiplicity collisions and the subtraction procedure is applied~\cite{Adam:2015bka,Acharya:2017tfn}. The $k$ factors are larger in the p-going than in the Pb-going direction. They reach maximum values up to 1.43 (p-going) and 1.23 (Pb-going) with the V0M estimator. The values increase up to 1.74 (1.43) and 1.32 (1.0) in the p-going (Pb-going) direction, with the CL1 and ZN multiplicity estimators. Figure~\ref{Fig:mu-tracklet} presents the resulting associated yield per trigger particle as a function of $\Delta \eta$ and $\Delta \varphi$ and the projected distribution onto $\Delta\varphi$ in bottom-left and bottom-right panels, respectively. One can notice that the amplitudes of the near-side and away-side ridges become comparable.

The subtraction of the scaled per-trigger yield in low-multiplicity collisions as well as the fit of the projected $\Delta\varphi$ distribution with Eq.~(\ref{eq:fourier}) are repeated for each $\pT$ trigger-particle interval. The values of the reduced $\chi^2$ are all smaller than 1.5, confirming that the distributions are well described by this Fourier series. The $V_{{\rm n} \Delta}^{{\rm \mu}-{\rm tracklet}}$ Fourier coefficients are further extracted from the fit parameters according to $V_{{\rm n} \Delta}^{{\rm \mu}-{\rm tracklet}} = a_{\rm n}/(a_0+ b)$, $b$ corresponding to the baseline of the scaled 60--90\% low-multiplicity class estimated from the integral in $\Delta \varphi$ of the correlation distribution around the minimum. By assuming that $V_{2 \Delta}^{{\rm \mu}-{\rm tracklet}}$ can be factorised as the product of the anisotropies of single muons and SPD tracklets~\cite{Aad:2014lta}, the single muon second-order coefficient ($v_2^{\rm \mu} \lbrace{\rm 2PC}\rbrace$) can be expressed as
\begin{linenomath}
\begin{equation}
v_2^{\rm \mu} \lbrace{\rm 2PC}\rbrace  = \frac{V_{2 \Delta}^{{\rm \mu}-{\rm tracklet}}}{\sqrt{V_{2 \Delta}^{{\rm tracklet}-{\rm tracklet}}}}.
\label{eq:v2PCsub}
\end{equation}
\end{linenomath}
In order to extract the $V_{2 \Delta}^{{\rm tracklet}-{\rm tracklet}}$, the analysis is repeated by correlating only SPD tracklets, as also done in Refs.~\cite{Abelev:2012ola,ABELEV:2013wsa},  and the region $\vert \Delta \eta \vert$ < 1.2 is excluded to reduce a bias due to a possible remaining jet peak on the near side at ($\Delta \eta \sim 0$, $\Delta \varphi \sim 0$) after the subtraction of the correlation distribution in low-multiplicity collisions.

The systematic uncertainties affecting the $v_2^{\rm \mu}\lbrace {\rm 2PC}\rbrace$ coefficient originate from the measurement of the muon--tracklet and tracklet--tracklet correlation distributions. They arise from the selection criteria of SPD tracklets, the procedure to remove the jet contamination in the near side and away side, the $v_2^{\rm \mu}\lbrace {\rm 2PC}\rbrace$ calculation including the fit stability and baseline determination, and the effect of the muon-track resolution.

The effect of the SPD acceptance is investigated by varying the range of the position of the reconstructed vertex along the beam axis, which is decreased down to $\vert z_{\rm vtx} \vert < 5$~cm.

The systematic effect related to the procedure implemented to subtract the jet contamination is investigated. A residual near-side jet peak in the tracklet--tracklet correlation is negligible since the $V_{2\Delta}^{\rm tracklet-tracklet}$ coefficient remains unchanged when varying the $\vert \Delta \eta \vert$ gap in the range 0.8--1.2 units.
The effect of a possible incomplete away-side jet subtraction is studied with the scaling procedure of the correlations in low-multiplicity collisions prior to its subtraction, as previously discussed. The influence of such an effect is also investigated fitting the subtracted correlation distributions, (0--20\%)~$-$~(60--90\%), with Eq.~(\ref{eq:fourier}) replacing the first order term by a Gaussian function with the width fixed to the one extracted from the fit of correlations in low-multiplicity collisions in the away side. The differences between the two procedures give an estimation of the corresponding systematic uncertainty.
A potential bias could result from long-range correlations remaining in the 60--90\% low-multiplicity class, which may lead to an oversubtraction. Such effect is estimated by changing the multiplicity interval of the low-multiplicity event class from the nominal 60--90\% to 70--90\%.

The stability of the $v_2^{\rm \mu}\lbrace {\rm 2PC}\rbrace$ results is also investigated by excluding the third-order coefficient in the fit of the muon--tracklet and track-tracklet correlations with Fourier series. The quality of the fits is getting worse, although no significant change in the extracted $v_2^{\rm \mu}\lbrace {\rm 2PC}\rbrace$ values is seen. A systematic uncertainty also arises from the procedure employed for the $\Delta \varphi$ projection. The baseline is a parameter which could influence the fits and is estimated following the same strategy as in Ref.~\cite{Adam:2015bka}. The latter is obtained by fitting the correlation distributions in low-multiplicity collisions using a Gaussian function in the away side and a constant for the baseline. Alternatively, the baseline can also be  calculated in high-multiplicity collisions from the integral or from a second-order polynomial fit around the minimum at $\Delta \varphi \sim \pi/2$. Finally, the $\Delta \varphi$ projection is obtained from a constant fit instead of a first-order polynomial fit along $\Delta \eta$ for each $\Delta \varphi$ interval.

The systematic effect due to the angular and momentum resolution of reconstructed muon tracks is evaluated by means of a dedicated Monte Carlo simulation based on the DPMJET event generator~\cite{Roesler:2000he}, which uses the GEANT4 transport code~\cite{GEANT4:2002zbu,Asai:2015xno} and the afterburner flow technique~\cite{Masera:2009zz}.

The various systematic uncertainties coming from each source are reported in Table~\ref{tab:SystUnc2PC} at forward
(p-going) and backward (Pb-going) rapidities for the V0M multiplicity estimator. They are added in quadrature to obtain the overall systematic uncertainty on $v_2^{\rm \mu}\lbrace {\rm 2PC}\rbrace$.
Comparable uncertainty values are estimated in the 0--20\% event class selected with CL1 and ZN multiplicity estimators.

\begin{table}
\caption{Summary of absolute systematic uncertainties affecting the $v_2^{\rm \mu}\lbrace {\rm 2PC} \rbrace$ coefficients measured in high-multiplicity (0--20\%) p--Pb collisions at $\sNN =$~8.16 TeV in p-going and Pb-going directions. The values are reported for the event class selected with the V0M multiplicity estimator. The systematic uncertainties vary within the indicated intervals depending on the inclusive muon $\pT$.}
\centering
\begin{tabular}{l|l|l}
Source & \multicolumn{2}{c}{
V0M} \\
& p-going ($\times 10^{3}$) & Pb-going ($\times 10^{3}$) \\ \hline
SPD acceptance & 0.2--8.0 & 0.9--6.5 \\
Residual jet & 0.6--6.7 & 0.3--7.1 \\
Remaining ridge in 60--90\% & 0.3--6.2 & 0.05--13.7   \\
$v_2^{\rm \mu} \lbrace {\rm 2PC} \rbrace$ calculation &  0.2--1.7 & 0.4--3.4 \\
Resolution effects & 0.03--0.7 & 0.2--0.8 \\
\hline
Total  & 1.5--10.7 & 1.4--17.1 \\
\end{tabular}
\label{tab:SystUnc2PC}
\end{table}

\subsection{Two-particle cumulants}\label{sec:2cum}

Two-particle cumulants, which exhibit different sensitivities to nonflow compared to the two-particle correlation method, are also employed for the study of the second-order $v_2$ coefficient of inclusive muons~\cite{Adam:2015pga}. For the first time in the heavy-flavour sector, the muon $v_2$ is measured with the two-particle cumulants using the framework presented in Ref.~\cite{Bilandzic:2013kga} where a method to correct for non-uniform acceptance and inefficiencies is provided. The SPD tracklets (reference particles, RP) are selected under the same selection criteria as with two-particle correlations. Apart from that, a more restrictive selection of the primary vertex position along the beam direction of $-5 < z_{\rm vtx} < 3$~cm is considered to account for dead zones in the SPD detector in the plane $z_{\rm vtx}-\varphi$, which cannot be corrected with particle weights. The standard selections discussed in Section~\ref{sec:2correl} are applied to identify muon tracks, the so-called particles of interest (POI). The two-particle second-order reference and $\pT$-differential cumulants, $c_2\lbrace 2 \rbrace$ and $d_2^{\rm \mu}\lbrace 2 \rbrace (\pT)$, are based on the construction of weighted $Q$ and $p$ vectors\footnote{Since no autocorrelations between RP and POI are present, the $q_{{\rm n,l}}$-vector constructed with only particles labeled both as RP and POI is not needed in the analysis.}~\cite{Bilandzic:2013kga}, defined on an event-by-event basis for a given multiplicity class and data sample (MSL- or MSH-triggered events) as
\begin{linenomath}
\begin{equation}
Q_{{\rm n,l}} = \sum_{ {\rm k = 1}}^{M} w_{\rm k}^{\rm l}(\eta, \varphi, z_{\rm vtx}) e^{{\rm i}({\rm n}\varphi_{\rm k})}\  {\rm and} \
p_{{\rm n,l}} = \sum_{{\rm k = 1}}^{m_{\rm p}} w_{\rm k}^{\rm l}(\pT,\eta, \varphi, z_{\rm vtx})e^{{\rm i}({\rm n}\varphi_{\rm k})},
\label{eq:cumbis}
\end{equation}
\end{linenomath}
where ${\rm n}$ is the ${\rm n}^{\rm th}$ harmonic number, ${\rm l}$ is an integer exponent of the ${\rm k^{\rm th}}$-particle (RP or POI) weight $w_{\rm k}^{\rm l}$, $M$ is the multiplicity of the SPD tracklets, $m_{\rm p}$ is the muon multiplicity, and $\varphi_{\rm k}$ is the SPD tracklet (muon) azimuthal angle needed for the computation of
$Q_{{\rm n,l}}$ ($p_{{\rm n,l}}$). After applying the weights for both SPD tracklets and inclusive muons, the non-uniformities in the azimuthal acceptance and inefficiencies are found negligible.

The calculation of the $\pT$-differential second-order coefficient of inclusive muons $v_2^{\rm \mu}\lbrace 2 \rbrace$ in a given multiplicity interval is performed as
\begin{linenomath}
\begin{equation}
 v_2^{\rm \mu}\lbrace 2 \rbrace(\pT) = \frac{d_2^{\rm \mu}\lbrace 2 \rbrace (\pT)}{\sqrt{c_2\lbrace 2 \rbrace}},
\label{eq:cum}
\end{equation}
\end{linenomath}
$c_2\lbrace 2 \rbrace$ being related to the reference second-order coefficient $V_2\lbrace 2 \rbrace$ as
\begin{linenomath}
\begin{equation}
V_2\lbrace 2 \rbrace =\sqrt{c_2\lbrace 2 \rbrace}.
\label{eq:cumref}
\end{equation}
\end{linenomath}

The measurement of the muon $\pT$-differential second-order coefficient with two-particle cumulants is influenced by nonflow effects which need to be isolated and subtracted. The $v_2^{\rm \mu}\lbrace {\rm 2}\rbrace$ coefficient can be extracted from the equation
  \begin{linenomath}
\begin{equation}
v_2^{\rm \mu}\lbrace {\rm 2} \rbrace (\pT) = \frac{\lbrack d_2^{\rm \mu}\lbrace {\rm 2} \rbrace (\pT) \rbrack_{\rm (0-20\%)}- f \cdot g \cdot \lbrack d_2^{\rm \mu} \lbrace {\rm 2} \rbrace (\pT) \rbrack_{\rm (60-90\%)}}{f_{\rm R
P} \cdot \sqrt{\lbrack c_2 \lbrace {2} \rbrace \rbrack_{(0-20\%)} - f \cdot \lbrack c_2 \lbrace {2} \rbrace \rbrack_{(60-90\%)}}} \cdot f_{\Delta \eta}.
\label{eq:cumsub2}
\end{equation}
\end{linenomath}

As done in the two-particle correlation method~\cite{Adam:2015bka}, the long-range jet correlations are estimated in the low-multiplicity event class 60--90\%. Then, the two-particle second-order reference and differential cumulants in these low-multiplicity collisions are scaled by a factor $f$ defined as the ratio of the mean SPD-tracklet multiplicity in low-multiplicity collisions to that in high-multiplicity collisions and subtracted as shown in Eq.~(\ref{eq:cumsub2}). Short-range jet correlations are usually suppressed applying a pseudorapidity gap between the correlated particles, i.e.\ for both muons and SPD tracklets. As a pseudorapidity gap between muons and SPD tracklets is naturally present, the procedure cannot be applied straightforwardly for SPD tracklets due to the dead zones in the SPD acceptance. Therefore, these nonflow effects are suppressed via the scaling of the muon $v_2^{\rm \mu}\lbrace {\rm 2} \rbrace$ by another factor, $f_{\Delta \eta}$, estimated by means of AMPT simulations~\cite{Lin:2004en,Xu:2011fi} and defined as the ratio of $v_2^{\rm \mu}\lbrace {\rm 2} \rbrace$ extracted with $\vert \Delta \eta \vert > 0.4$ to that obtained without applying a pseudorapidity gap. The optimised condition $\vert \Delta \eta \vert$ > 0.4 results from a compromise between statistical considerations and suppression of nonflow effects. The value of this scaling factor is about 1.2 and is found to be independent of $\pT$ and the multiplicity estimator.

As already discussed with the technique of two-particle correlations (see Section~\ref{sec:2correl}), remaining long-range jet correlations are not excluded even after the subtraction of correlations in low-multiplicity collisions. These remaining jet correlations can be suppressed by applying two additional correction factors. The factor $g$, applied to the $\pT$-differential cumulants measured in low-multiplicity collisions, is estimated from the muon--tracklet correlation distribution as discussed in Section~\ref{sec:2correl}.
The factor $f_{\rm RP}$, applied to $V_2 \lbrace 2 \rbrace$, is calculated from the tracklet--tracklet correlation function. This is the ratio of $V_2 \lbrace 2 \rbrace$ from tracklet--tracklet correlations extracted with the scaling of the remaining jet contribution to that obtained without any scaling procedure.

\begin{figure}[!t]
\begin{center}
\includegraphics[width=0.49\textwidth]{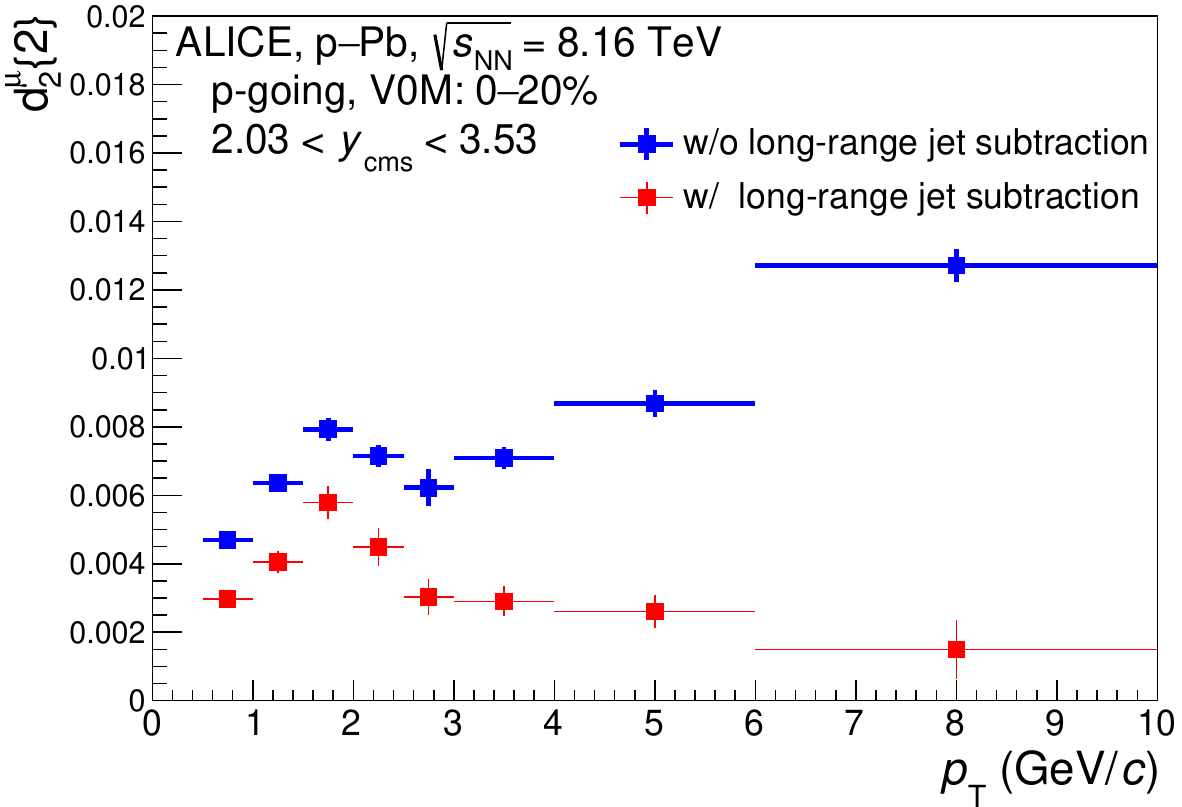}
\includegraphics[width=0.49\textwidth]{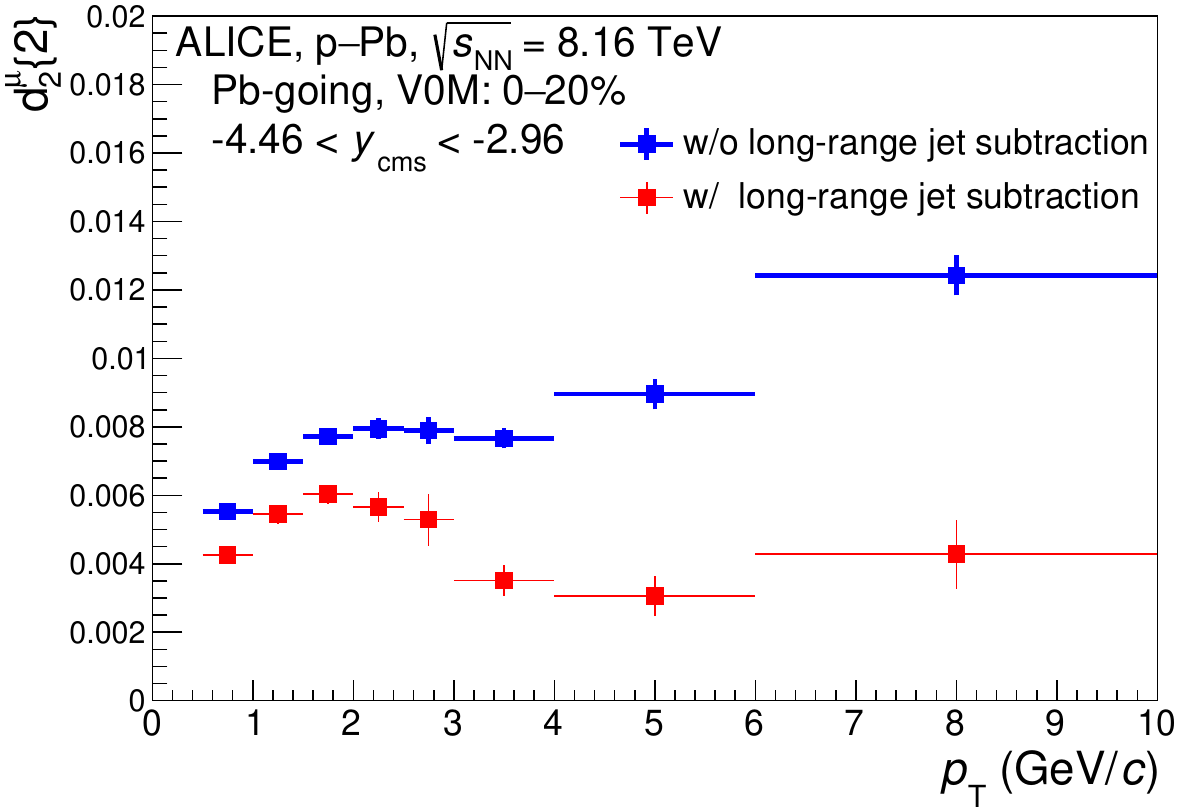}
\caption{Comparison of the two-particle second-order $\pT$-differential muon cumulant before and after the subtraction of the $\pT$-differential cumulant in low-multiplicity collisions and remaining long-range jet correlations at forward (left) and backward (right) rapidities in high-multiplicity p--Pb collisions at $\sNN =$~8.16 TeV. The comparison is presented with the V0M estimator. Only statistical uncertainties are shown.}
\label{Fig:cum-nonflow}
\end{center}
\end{figure}
Figure~\ref{Fig:cum-nonflow} presents the two-particle second-order $\pT$-differential muon cumulant in high-multiplicity (0--20\%) collisions without and with the subtraction of the scaled $\pT$-differential muon cumulant in low-multiplicity collisions and remaining long-range jet correlations at forward (left) and backward (right) rapidities, using V0M for the multiplicity estimation. Large deviations appear at high $\pT$, indicating the importance to subtract long-range jet correlations at both forward and backward rapidities.
These trends could result from the larger long-range jet contamination in the p-fragmentation region (forward rapidity, left panel) compared to the Pb-fragmentation region (backward rapidity, right panel) at high $\pT$, in particular. A similar behaviour is also seen with CL1 and ZN multiplicity estimators.

Several potential sources of systematic uncertainties affecting the measurement of the $v_2^{\rm \mu} \lbrace 2 \rbrace$ coefficient are considered.

 The non-uniform acceptance correction for SPD tracklets is expected to be independent of the analysed data sample. It is tested with the MB and muon-triggered samples. The difference of the results obtained with muon-triggered events with respect to the MB sample is taken as the corresponding systematic uncertainty.

The sensitivity of the results to the selection criteria of SPD tracklets is investigated by varying the $\eta$ interval of SPD tracklets from $\vert \eta \vert < 1$ to $\vert \eta \vert < 1.2$.

The systematic uncertainty related to the procedure implemented for the subtraction of short-range jet correlations includes two components. A first contribution to the systematic uncertainty comes from the variation of the pseudorapidity gap up to $\vert \Delta \eta \vert > 0.8$.
A second contribution is obtained conservatively from the comparison of the $v_2^{\rm \mu} \lbrace 2 \rbrace$ computed with the scaling factor $f_{\rm \Delta \eta}$ estimated from AMPT simulations and the measured one, using $\vert \Delta \eta \vert > 0$ ignoring that the effects due to non-uniformities in the azimuthal acceptance depend on $z_{\rm vtx}$.
In order to study the uncertainty from the away-side jet subtraction in the calculation of the $\pT$-differential cumulant, the subtracted $\pT$-differential cumulant, $\lbrack d_2^{\rm \mu}\lbrace {\rm 2} \rbrace (\pT) \rbrack_{\rm (0-20\%)}- f \cdot \lbrack d_2^{\rm \mu} \lbrace {\rm 2} \rbrace (\pT) \rbrack_{\rm (60-90\%)}$, i.e. the numerator of Eq.~(\ref{eq:cumsub2}) with $g$ set to unity, is scaled by a factor derived from the two-particle correlation method. This factor is estimated as the ratio of the $V_{2\Delta}^{{\rm \mu-}{\rm tracklet}}$ extracted fitting the subtracted correlation with Eq.~(\ref{eq:fourier}) to that obtained replacing the first order term with the Gaussian function discussed in Section~\ref{sec:2correl}.

The systematic effect coming from the possible influence of long-range correlations in low-multiplicity events is assessed by changing the low-multiplicity interval from 60--90\% to 70--90\%.

Finally, the systematic uncertainty related to the angular and momentum resolution of the muon spectrometer is evaluated following the same strategy as with two-particle correlations (see Section~\ref{sec:2correl}).

\begin{figure}[!bht]
\begin{center}
\includegraphics[width=0.49\textwidth]{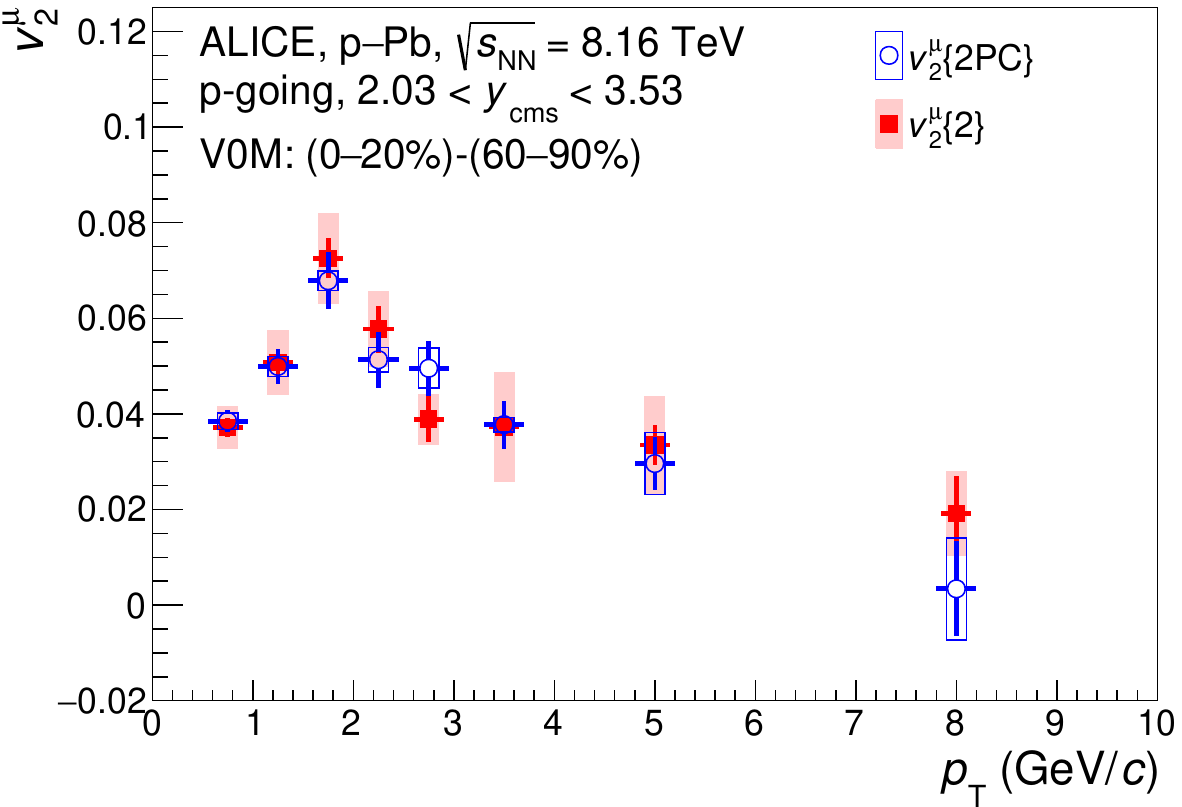}
\includegraphics[width=0.49\textwidth]{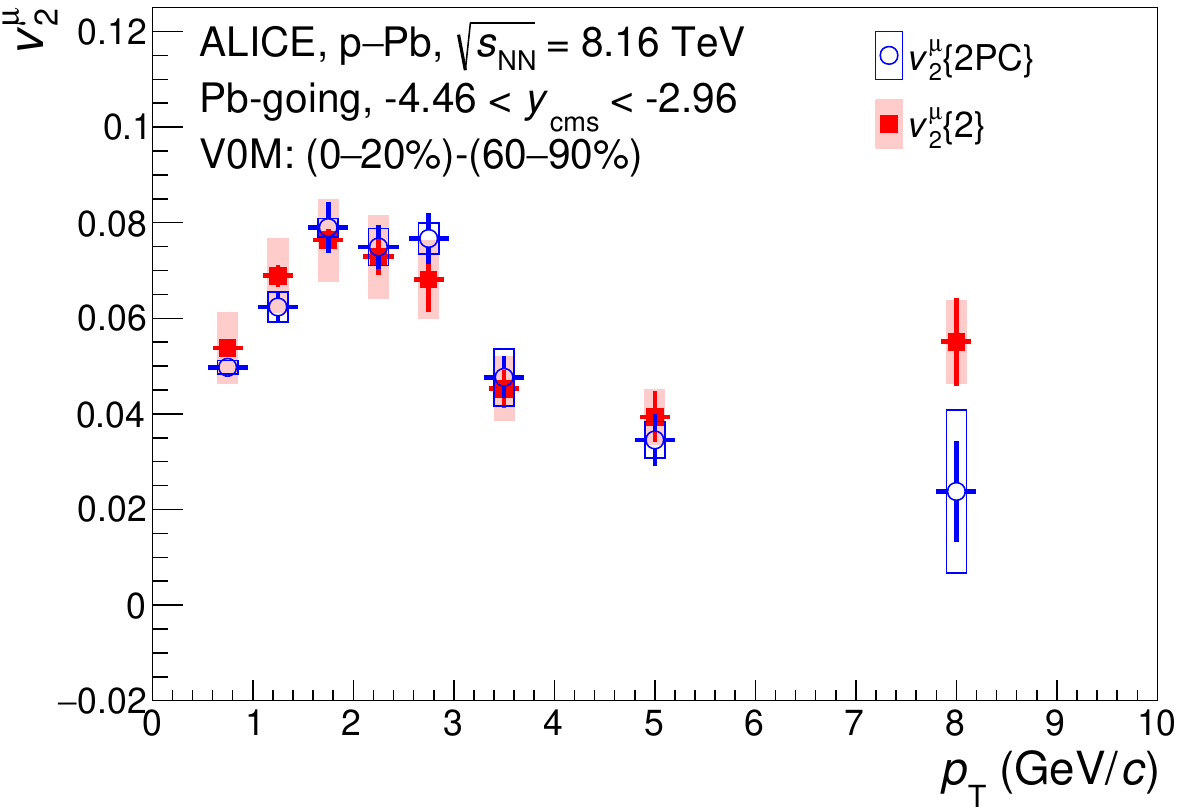}
\caption{Inclusive muon $v_2^{\rm \mu}$ as a function of $\pT$ at forward (left) and backward (right) rapidities in high-multiplicity p--Pb collisions at $\sNN =$~8.16 TeV. The event activity is estimated with the V0M estimator. Open and full symbols refer to the measurements with two-particle correlations and two-particle cumulants, respectively.}
\label{Fig:res1}
\end{center}
\end{figure}

\begin{table}[!htb]
\caption{Summary of absolute systematic uncertainties affecting the $v_2^{\rm \mu}\lbrace {\rm 2} \rbrace$ coefficients measured in high-multiplicity (0--20\%) p--Pb collisions at $\sNN =$~8.16 TeV in p-going and Pb-going directions. The values are reported for the event class selected with the V0M mutiplicity estimator.  The systematic uncertainties vary within the indicated intervals depending on the inclusive muon $\pT$.
}
\centering
\begin{tabular}{l|l|l}
Source & \multicolumn{2}{c}{V0M} \\
& p-going ($\times 10^{3}$) & Pb-going ($\times 10^{3}$) \\ \hline
Trigger bias & 0.06--4.3 & 0.04--1.2 \\
SPD acceptance & 1.3--4.3 & 1.6--4.4 \\
Short-range jet correlations: $\vert \Delta \eta \vert$ gap & 0.9--3.6 & 1.3--2.5\\
Short-range jet correlations: $\vert \Delta \eta \vert$ data vs. AMPT & 1.8--7.1 & 3.8--7.4 \\
Residual jet& 0.2--6.6 & 1.5--5.1 \\
Remaining ridge in 60--90\% & 0.4--9.5 & 0.1--3.1  \\
Resolution effects &  0.2--0.7 & 0.4--0.8  \\
\hline
Total  & 4.5--11.3 & 5.8--8.7  \\
\end{tabular}
\label{tab:SystUncCum}
\end{table}

The overall systematic uncertainty of $v_2^{\rm \mu} \lbrace 2 \rbrace$ is evaluated summing in quadrature the various systematic uncertainties coming from each source. These uncertainties are summarised in Table~\ref{tab:SystUncCum} at both forward (p-going) and backward (Pb-going) rapidities for the V0M multiplicity estimator. The systematic uncertainties of $v_2^{\rm \mu} \lbrace 2 \rbrace$ determined with CL1 or ZN mutiplicity estimators follow the same trends as a function of $\pT$ and their values are compatible.

\section{Results and model comparisons}\label{sec:res}

Figure~\ref{Fig:res1} presents the $\pT$-differential second-order coefficient $v_2^{\rm \mu}$ of inclusive muons after the subtraction of nonflow effects in the 0--20\% high-multiplicity class for p--Pb collisions at $\sNN =$~8.16 TeV. The results at forward and backward rapidities are depicted in the left and right panel, respectively. They are reported for the first time in a wide transverse momentum interval $0.5 < \pT < 10$~GeV/$c$. The event activity is determined with the V0M multiplicity estimator. Comparisons with the $v_2^{\rm \mu}$ coefficient of inclusive muons measured in high-multiplicity events selected using CL1 and ZN multiplicity estimators will be discussed later. The measurements are carried out with two-particle correlations (open symbols) and two-particle cumulants (full symbols), which are denoted $v_2^{\rm \mu} \lbrace {\rm 2PC} \rbrace$ and $v_2^{\rm \mu} \lbrace 2 \rbrace$, respectively. The statistical uncertainties (vertical bars) with the two-particle cumulants are estimated according to the Jackknife
method~\cite{Efron:98913}. The latter is based on the resampling technique, where 6 out of 12 sub-samples are used to extract the RMS of the distribution for each $\pT$ interval which is further divided by $\sqrt 2$ in order to obtain the statistical uncertainty. In the two-particle correlation method, the statistical uncertainty is obtained from the Fourier fit procedure. The systematic uncertainties are the empty and filled boxes.

The two methods give compatible results within uncertainties at both forward and backward rapidities. One observes first an increase of the $v_2^{\rm \mu}$ signal with increasing $\pT$ where it reaches a maximum value of about 0.07 (0.08) at $\pT \sim 2$~GeV/$c$ in the forward (backward) rapidity region, followed by a decrease as $\pT$ increases.
Although the $\pT$ dependence is similar in the two rapidity regions, there is a hint for a higher elliptic flow signal at backward rapidity than at forward rapidity as observed in Ref.~\cite{Adam:2015bka}. This is further investigated by the measurement of the $\pT$-differential ratio of the measured $v_2^{\rm \mu} \lbrace {\rm 2PC} \rbrace$ at backward rapidity to that at forward rapidity. This ratio exhibits a uniform behaviour within uncertainties and can be fitted with a constant function (not shown here). It amounts to $1.28 \pm 0.07$, the uncertainty being from the fit.
This asymmetry could be a consequence of decorrelation effects of the flow vectors in different rapidity regions~\cite{Khachatryan:2015oea,Aaboud:2017tql,ALICE:2017lyf,Acharya:2018pjd}.

In the $\pT$ interval of interest, muons originate mainly from charged-pion and kaon decays at low $\pT$ ($\pT < 2$~GeV/$c$), while muons from heavy-flavour hadron decays dominate over muons from light-hadron decays at higher $\pT$~\cite{Adam:2015bka}. The contributions of these muon sources are estimated by means of simulations using the DMPJET event generator~\cite{Roesler:2000he} and the GEANT4 transport package~\cite{GEANT4:2002zbu}. The relative contribution of muons from primary charged-pion and kaon decays amounts to about 67\% (68\%) at $0.5 < p_{\rm T} < 1$~GeV/$c$, and it decreases with increasing $\pT$ down to about 7.5\% (14.5\%) at $6 < p_{\rm T} < 10$~GeV/$c$ in the p-going (Pb-going) direction.  The fraction of muons originating from charm and beauty decays represents about 60\% (58\%) of the total muon yield at $\pT =$~2 GeV/$c$ and it reaches about 87\% (78\%) in the p-going (Pb-going) direction at $6 < p_{\rm T} < 10$~GeV/$c$. A similar fraction of muons from heavy-flavour hadron decays in $0.5 < p_{\rm T} < 4$~GeV/$c$ is reported for p--Pb collisions at $\sNN =$~5.02 TeV~\cite{Adam:2015bka}. Moreover, it is worth to mention that based on fixed-order plus next-to-leading logarithms (FONLL) calculations~\cite{Cacciari:1998it,Cacciari:2012ny} more than 60\% of muons from heavy-hadron decays originate from beauty quarks in the highest $\pT$ interval ($6 < \pT < 10$~GeV/$c$).
The measured $v_2^{\rm \mu}$ coefficient is positive with a significance which reaches values of 4.7$\sigma$--12$\sigma$ (7.6$\sigma$--11.9$\sigma$) at intermediate $\pT$ ($2 < \pT < 6$~GeV/$c$) in the forward (backward) rapidity region, depending on the analysis technique. These results might suggest the existence of a collective behaviour of heavy quarks in high-multiplicity (0--20\%) p--Pb collisions at $\sNN =$~8.16 TeV at forward and backward rapidities. In the highest $\pT$ interval accessible in this analysis
($6 < \pT < 10$~GeV/$c$), where muons from beauty-hadron decays take over charm as the dominant muon component, there is a hint for a positive $v_2^{\rm \mu}$ although not significant within uncertainties (significance of $0.2\sigma$ ($1.2\sigma$) at forward (backward) rapidity with two-particle correlations).

\begin{figure}[!t]
\begin{center}
\includegraphics[width=0.48\textwidth]{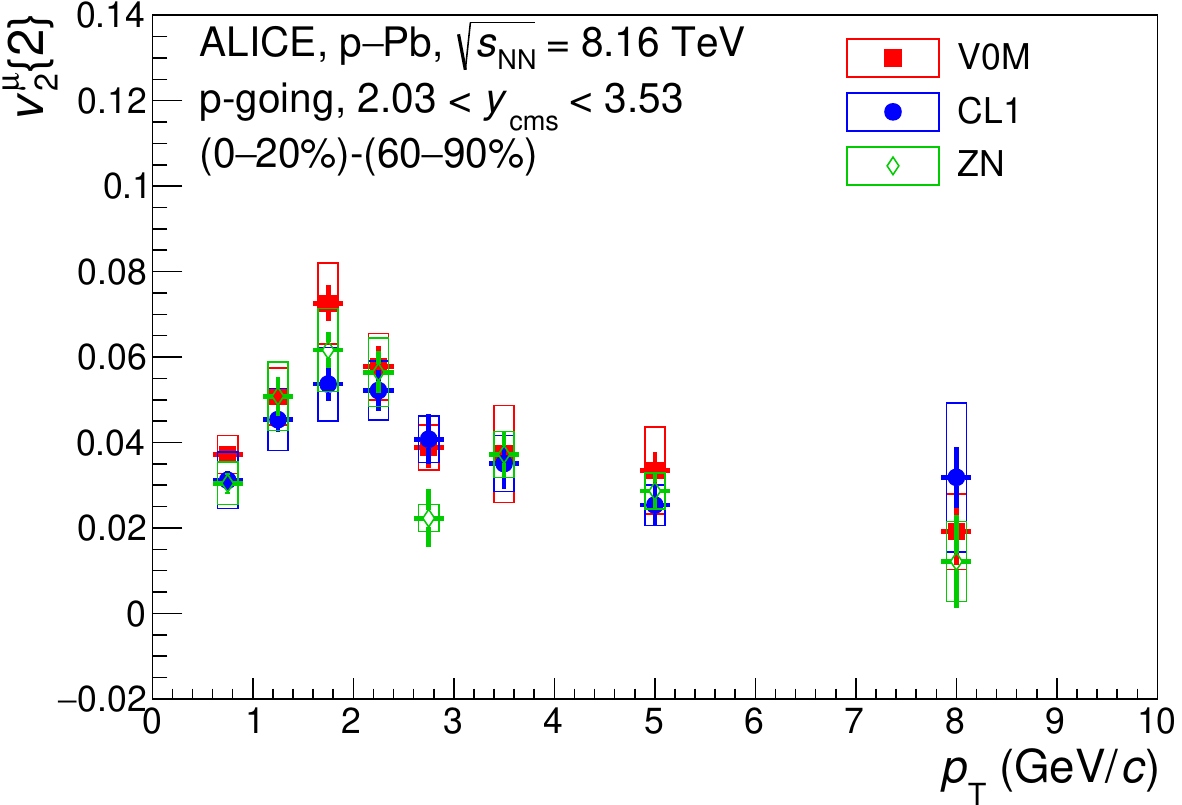}
\includegraphics[width=0.48\textwidth]{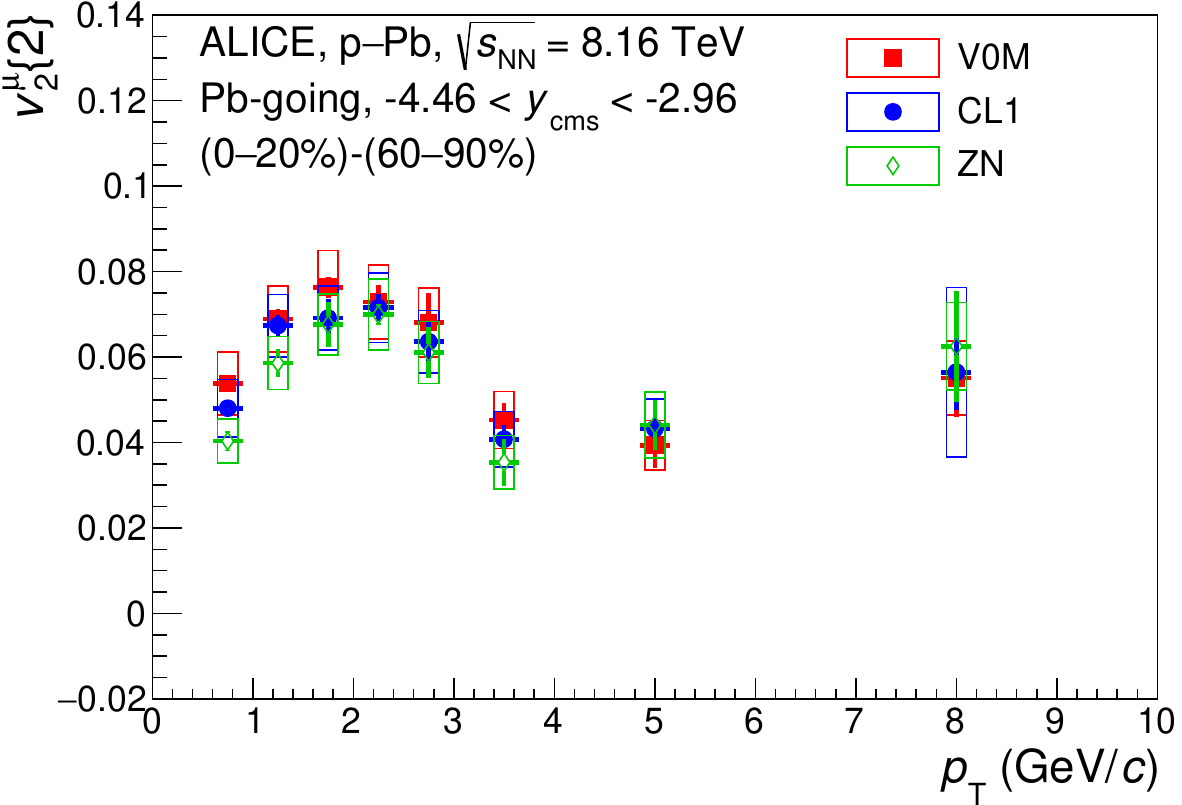}
\includegraphics[width=0.48\textwidth]{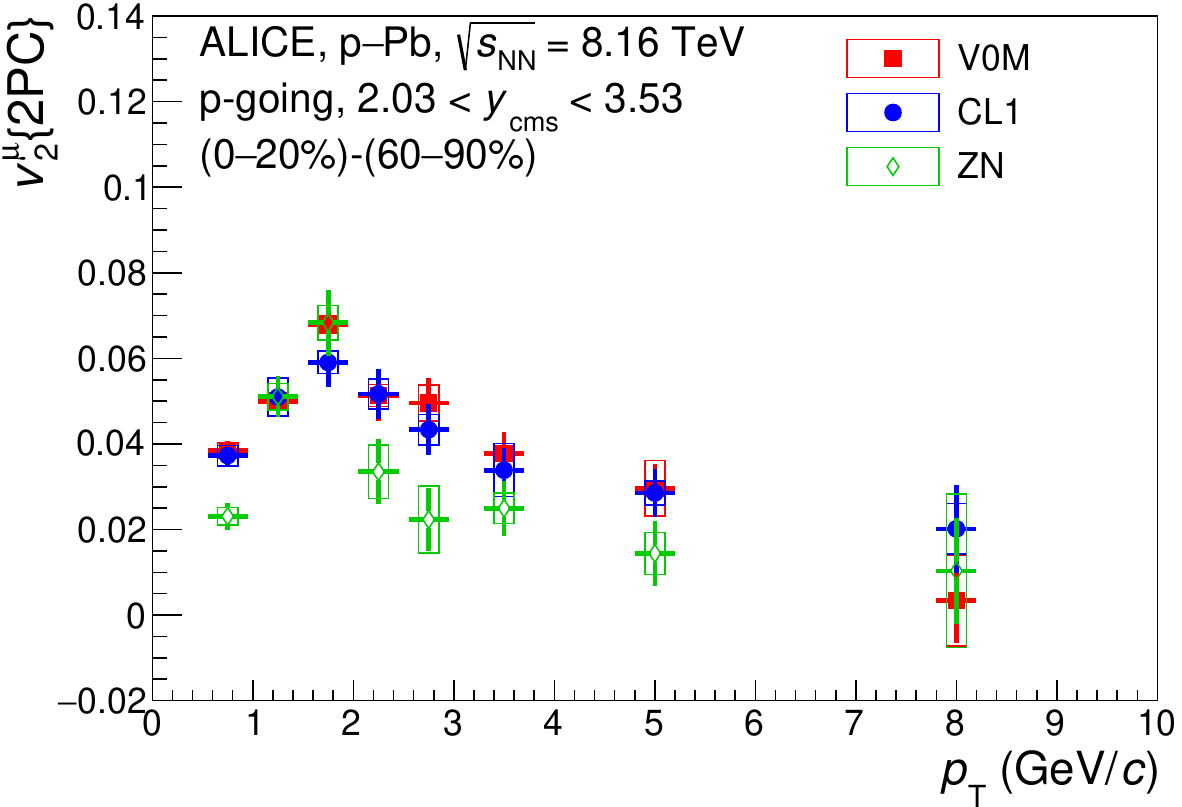}
\includegraphics[width=0.48\textwidth]{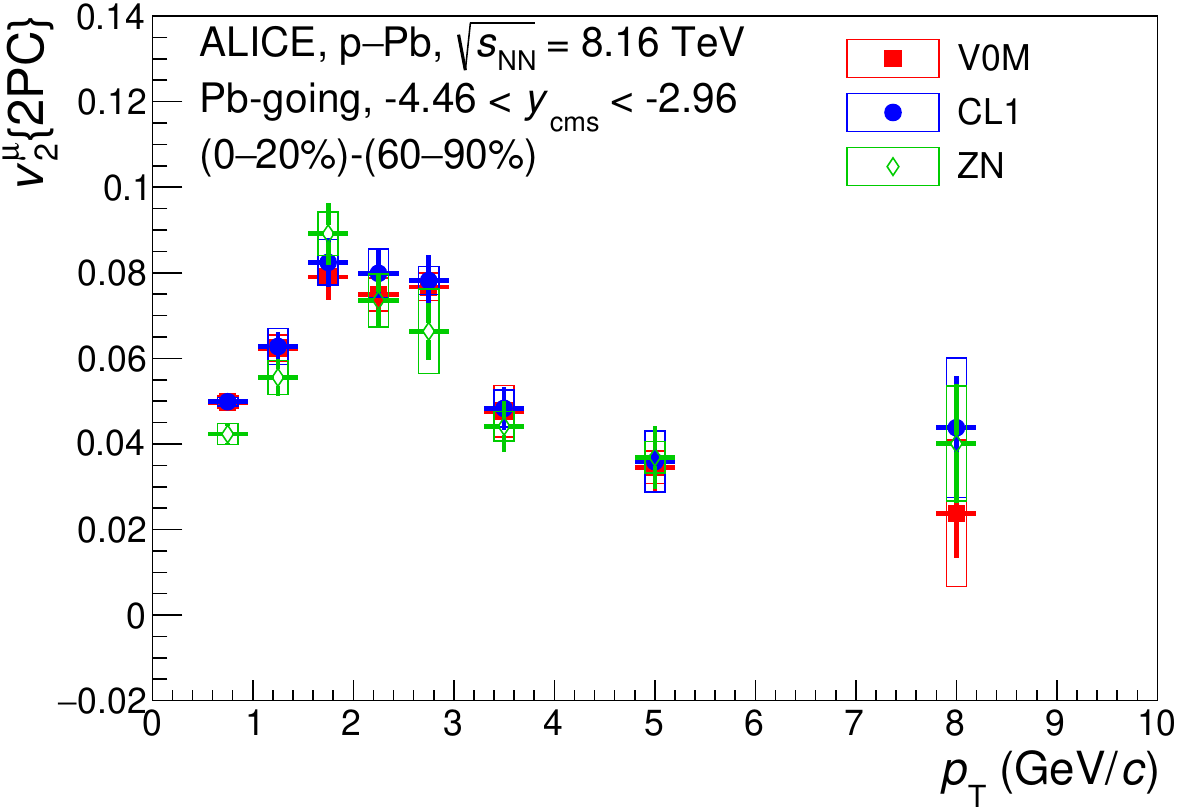}
\caption{Inclusive muon $v_2^{\rm \mu}$ as a function of $\pT$ at forward (left) and backward (right) rapidity in high-multiplicity p--Pb collisions at $\sNN =$~8.16 TeV, extracted with two-particle cumulants (top) and two-particle correlations (bottom). The results are obtained with three different estimators of the event activity: V0M, CL1 and ZN.}
\label{Fig:res2}
\end{center}
\end{figure}

The analysis is repeated using the number of clusters in the outer layer of the SPD (CL1) and the energy deposited in the neutron ZDC (ZN) for the event activity selection. Figure~\ref{Fig:res2} displays the measurements obtained with two-particle cumulants (top panels) and two-particle correlations (bottom panels) at forward (left panels) and backward (right panels) rapidities. It has been demonstrated that these multiplicity estimators select different event classes associated with different mean charged-particle multiplicity~\cite{Adam:2014qja}. The ZN is expected to be the least-biased estimator (see Section~\ref{sec:det} and Ref.~\cite{Acharya:2018egz}) but the correlation between the energy measured with the ZDC and the charged-particle multiplicity at midrapidity is known to be weak~\cite{Adam:2014qja}. On the other hand, autocorrelation effects are present when using CL1 for the event class selection (Section~\ref{sec:det}) since the reference flow is also calculated with the SPD tracklets. Even in this context, the computed $v_2^{\rm \mu}$ values are compatible within uncertainties, although there is a hint for smaller $v_2^{\rm \mu}$ values with the ZN multiplicity estimator than with V0M and CL1 multiplicity estimators. The ZN estimator selects on the average smaller charged-particle multiplicity density in the high-multiplicity class than V0M and CL1 estimators, while the opposite trend is observed for the low-multiplicity class~\cite{Acharya:2018egz,ALICE-PUBLIC-2018-011}. Consequently, a smaller $v_2^{\rm \mu}$ signal is expected after the subtraction of correlations from low-multiplicity events with the ZN estimator.

\begin{figure}[!t]
\begin{center}
\includegraphics[width=0.48\textwidth]{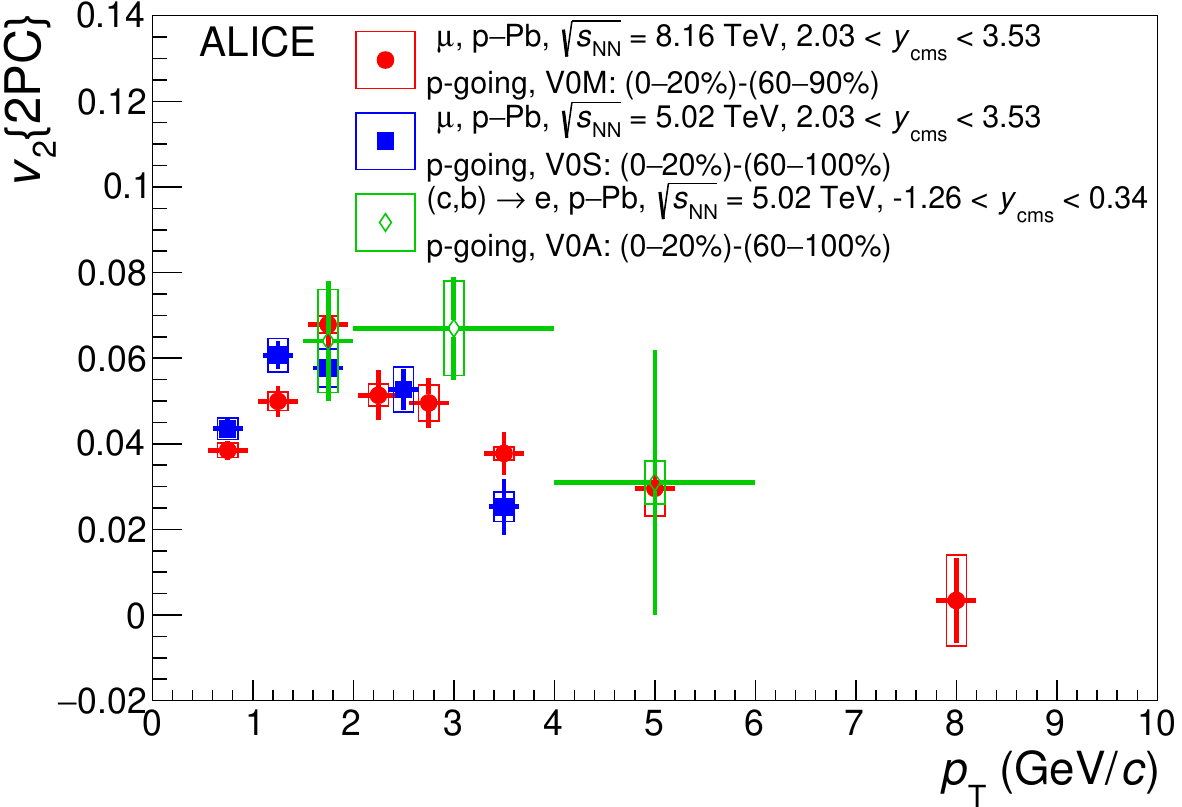}
\includegraphics[width=0.48\textwidth]{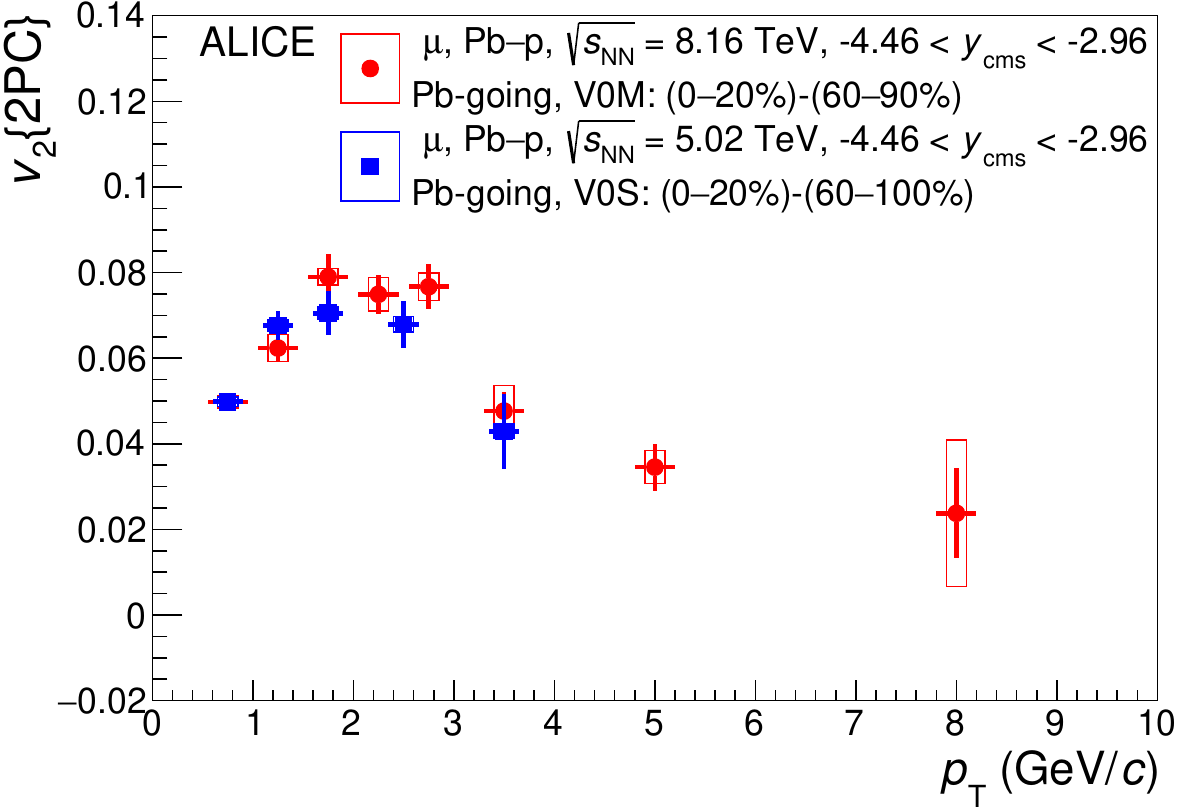}
\caption{Comparison of the $\pT$-differential $v_2\lbrace {\rm 2PC} \rbrace$ of inclusive muons at forward (left) and backward (right) rapidity in high-multiplicity p--Pb collisions at $\sNN =$~8.16 TeV, extracted with the two-particle correlation method with previous measurements performed in p--Pb collisions at $\sNN =$~5.02 TeV for inclusive muons~\cite{Adam:2015bka} and electrons from heavy-flavour hadron decays~\cite{Acharya:2018dxy}. }
\label{Fig:compdata}
\end{center}
\end{figure}
The present results are compared in Fig.~\ref{Fig:compdata} with previously published results of inclusive muons obtained by the ALICE collaboration in p--Pb collisions at $\sNN =$~5.02 TeV with two-particle correlations~\cite{Adam:2015bka}. The larger data sample collected at $\sNN =$~8.16 TeV allows us to perform the measurements in a wider $\pT$ interval, the $\pT$ reach being extended from $\pT =$~4 GeV/$c$ to $\pT =$~10 GeV/$c$. No significant $\sNN$ dependence is seen on the extracted $v_2$ values at $\sNN =$~5.02 TeV and 8.16 TeV in the interval $0.5 < \pT < 4$~GeV/$c$ which contains a significant fraction of muons from light-hadron decays for $\pT < 2$~GeV/$c$. The results are also in good agreement within uncertainties with those obtained by ALICE for electrons from heavy-flavour hadrons measured in p--Pb collisions at $\sNN =$~5.02 TeV at midrapidity ($-0.8 < \eta < 0.8$) and for $1.5 < \pT < 6$~GeV/$c$~\cite{Acharya:2018dxy}. It is interesting to point out that the magnitude of the $v_2$ of inclusive muons is comparable within uncertainties to the one of inclusive muons obtained at forward rapidity by ALICE in semicentral Pb--Pb collisions at $\sNN =$~2.76 TeV for $2 < \pT < 10$~GeV/$c$~\cite{Adam:2015pga}.
A positive $v_2$ also was reported by ALICE for $\rm J/\psi$ measured in $3 < p_{\rm T}^{\rm J/\psi} < 6$~GeV/$c$ and same rapidity intervals in p--Pb collisions at $\sNN =$~8.16 TeV, where the contribution from recombination of thermalised charm quarks in the medium is expected to be negligible and path-length dependent effects are smaller with respect to Pb--Pb collisions~\cite{Acharya:2017tfn}. The present inclusive muon $v_2$ results also complement those obtained in high-multiplicity p--Pb collisions at $\sNN =$~5.02 TeV for unidentified particles as well as  pions, kaons and protons, although in a different kinematic region (different $\pT$ and $y$ intervals)~\cite{ALICE:2013snk,ALICE:2022wpn}.

Further insights in the understanding of the observed azimuthal anisotropies in small collision systems can be gained by comparing the measurements with model predictions such as the AMPT and colour glass condensate (CGC) calculations. The v2-26t7b string-melting version of the AMPT model~\cite{Lin:2004en,Li:2018leh} which includes the improvements discussed in Ref.~\cite{Lin:2021mdn} is employed to compute the $v_2$ of heavy-flavour hadrons ($\rm D^0$ and $\rm B$ mesons) and primary charged pions and kaons by means of the forward--central two-particle corrections (Section~\ref{sec:2correl}). The event selection is performed by counting the charged particles in the acceptance of the V0 detector. The $v_2$ coefficient is then computed separately for muons from charm hadrons, beauty hadrons, charged pions and charged kaons by means of fast simulations which use the input $\pT$ and $v_2$ distributions of $\rm D^0$ and B mesons, and primary charged pions and kaons, as well as PYTHIA 6.4~\cite{Sjostrand:2014zea} for the decay kinematics. The D-meson species are assumed to have the same $v_2$ coefficient as $\rm D^0$ mesons\footnote{The elliptic flow of prompt $\rm D^0$, $\rm D^+$, $\rm D^{*+}$, and $\rm D^+_{\rm s}$ measured by ALICE in Pb--Pb  collisions at $\sNN =$~5.02 TeV is found to be compatible within uncertainties~\cite{ALICE:2017pbx}.}. The \pT-differential inclusive muon $v_2^{\rm \mu}$ is obtained from a weighted sum of the $v_2$ coefficient of muons from heavy-flavour hadron decays, $v_2^{\rm \mu \leftarrow b, c}$, and the $v_2$ of muons from charged pion and kaon decays, $v_2^{\rm \mu \leftarrow \pi, K}$, as $v_2^{\rm \mu} = (1 - f) \cdot  v_2^{\rm \mu \leftarrow b, c} + f \cdot v_2^{\rm \mu \leftarrow \pi, K}$, $f$ being the relative abundance of muons from the decay of charged pions and kaons estimated from Monte Carlo simulations with the DPMJET event generator~\cite{Roesler:2000he}. Similarly, the $v_2^{\rm \mu \leftarrow b, c}$ coefficient is computed as a weighted sum of the $v_2$ coefficient of muons of charm-hadron and beauty-hadron decays, $v_2^{\rm \mu \leftarrow c}$ and $v_2^{\rm \mu \leftarrow b}$, as $f^{\rm c} \cdot v_2^{\rm \mu \leftarrow c} + f^{\rm b} \cdot v_2^{\rm \mu \leftarrow b}$. The corresponding fractions of muons from charm-hadron and beauty-hadron decays with respect to the total yield of muons from heavy-flavour hadron decays, $f^{\rm c}$ and $f^{\rm b}$, are obtained by means of the fixed-order plus next-to-leading logarithms (FONLL) approach~\cite{Cacciari:1998it,Cacciari:2012ny}.

\begin{figure}[!tbh]
\begin{center}
\includegraphics[width=0.48\textwidth]{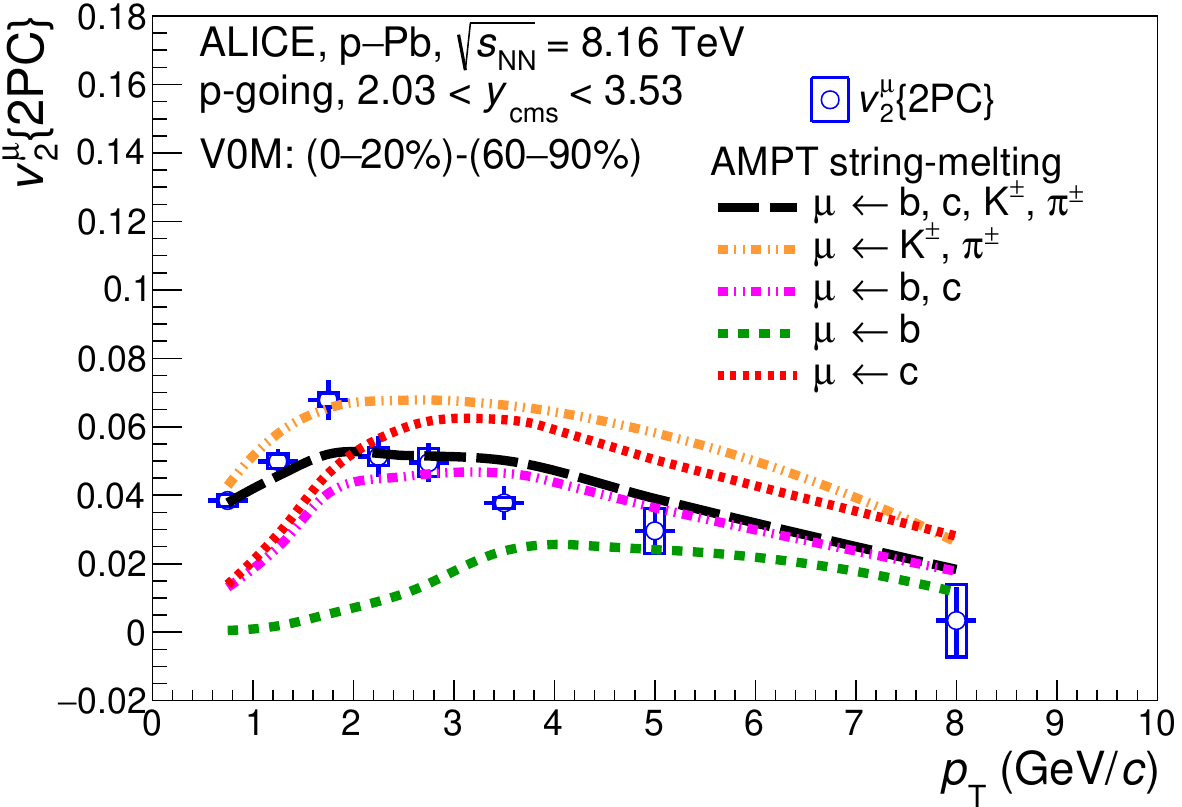}
\includegraphics[width=0.48\textwidth]{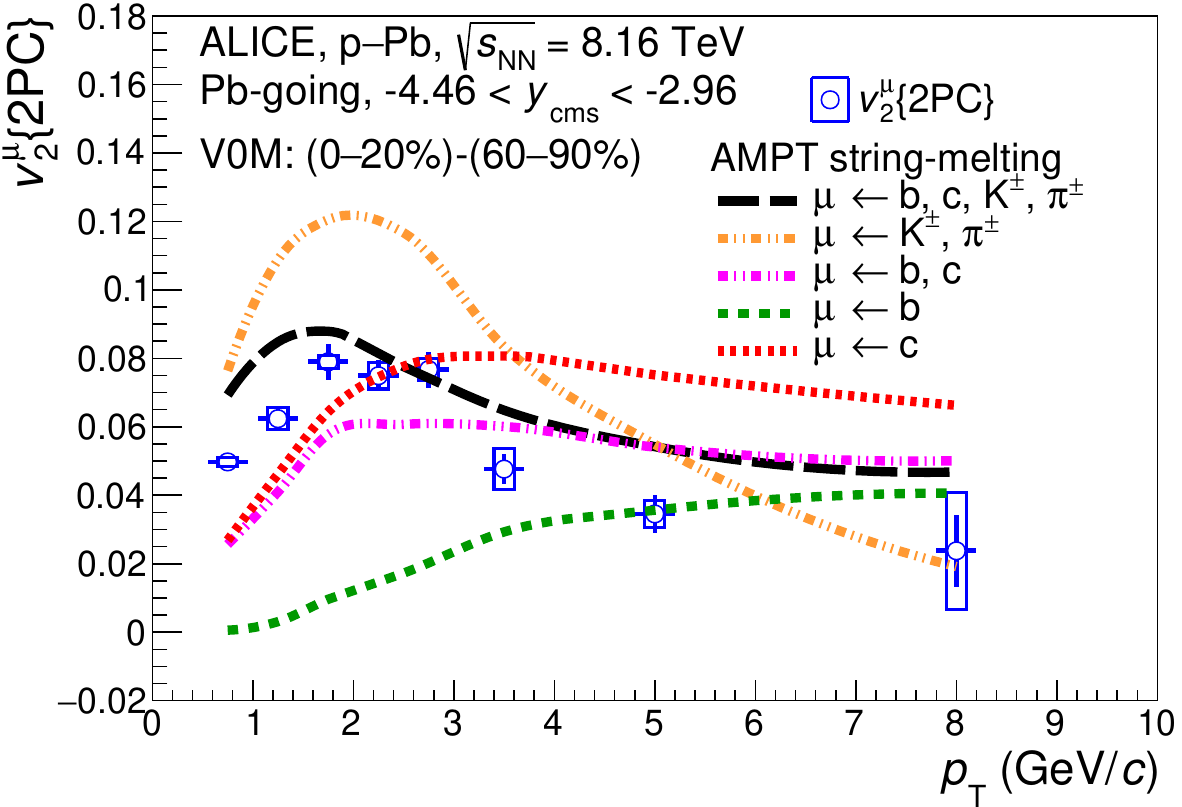}
\caption{Comparison of the $\pT$-differential $v_2^{\rm \mu} \lbrace {\rm 2PC} \rbrace$ of inclusive muons at forward (left) and backward (right) rapidities in high-multiplicity p--Pb collisions at $\sNN =$~8.16 TeV with AMPT calculations~\cite{Lin:2004en,Li:2018leh,Lin:2021mdn}. The predictions are shown for muons from charm-hadron decays and beauty-hadron decays, separately, muons from charged-pion and kaon decays, and for muons from the combination of the various sources. The contribution of muons from both charm- and beauty-hadron decays is also displayed. }
\label{Fig:ampt}
\end{center}
\end{figure}

Figure~\ref{Fig:ampt} presents a comparison of the \pT-differential inclusive muon $v_2^{\rm \mu}\lbrace {\rm 2PC} \rbrace$ with AMPT calculations, together with the different contributions of muons from charm- and beauty-hadron decays, separately, and muons from charged pion and kaon decays in the p-going (left) and Pb-going (right) direction. The AMPT model generates a positive $v_2$ for all particle species. Its magnitude increases significantly in the low-$\pT$ region up to about 2--3.5~GeV/$c$ depending on the decay particle. At higher $\pT$, the $v_2$ signal decreases smoothly with increasing $\pT$ or saturates, except in the Pb-going direction where a slightly increase with $\pT$ is seen for muons from beauty-hadron decays. In line with hydrodynamic calculations~\cite{Zhao:2020wcd}, the AMPT model predicts a larger $v_2$ for muons from charged-pion and kaon decays than for muons from heavy-flavour hadron decays at low $\pT$ in high-multiplicity p--Pb collisions. As observed with the data, the calculated $v_2^{\rm \mu}$ values with the AMPT model are larger in the backward rapidity region compared to the forward rapidity region, which may be a consequence of rapidity-dependent flow-vector fluctuations~\cite{Khachatryan:2015oea,Aaboud:2017tql,ALICE:2017lyf,Acharya:2018pjd}. The AMPT predictions are in fair agreement with the measured inclusive muon $v_2^{\rm \mu}$, although the model tends to slightly overestimate the data in the backward rapidity region. It is important to note that finite $v_2^{\rm \mu \leftarrow b, c}$ values are indeed needed to reach such agreement. These AMPT comparisons suggest that the azimuthal anisotropies are mainly driven by the anisotropic parton escape mechanism where partons have a higher probability to escape along the shorter axis of the interaction zone, as discussed in Ref.~\cite{He:2015hfa}.

\begin{figure}[!h]
\begin{center}
\includegraphics[width=0.48\textwidth]{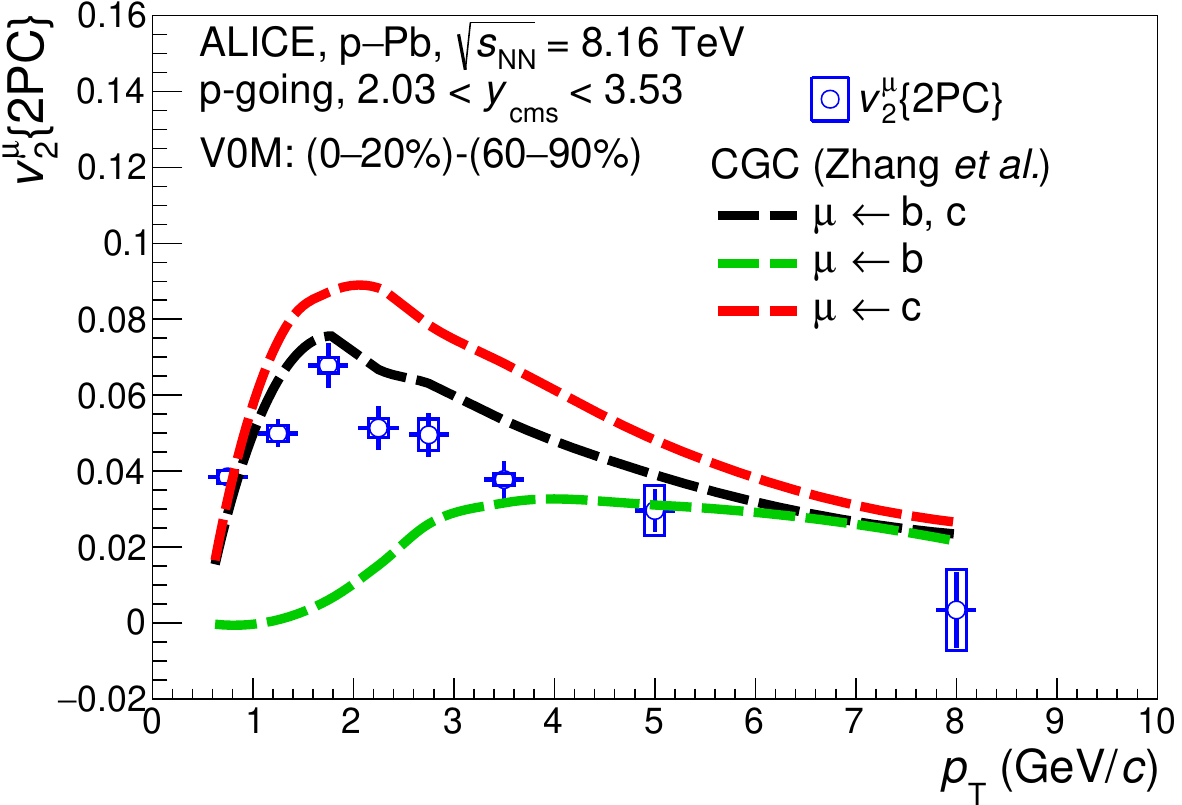}
\includegraphics[width=0.48\textwidth]{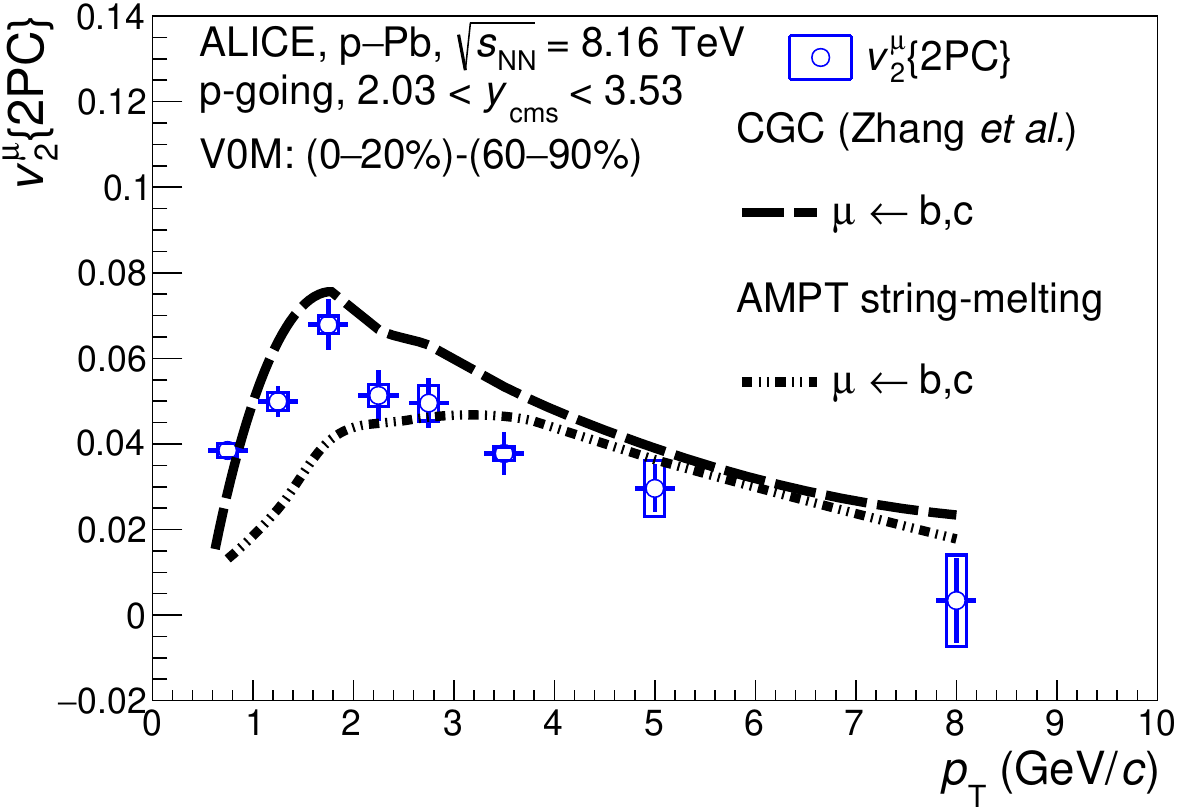}
\caption{Left: comparison of the $\pT$-differential $v_2^{\rm \mu} \lbrace {\rm 2PC} \rbrace$ of inclusive muons at forward rapidity in high-multiplicity p--Pb collisions at $\sNN =$~8.16 TeV with CGC-based calculations~\cite{Zhang:2019dth,Zhang:2020ayy}. The predictions are shown for muons from charm-hadron decays and beauty-hadron decays, separately, and for muons from the combination of the two sources. Right: comparison of the $\pT$-differential $v_2$ of muons from heavy-flavour hadron decays as obtained with CGC and AMPT calculations with the measured inclusive muon $v_2$. }
\label{Fig:cgc}
\end{center}
\end{figure}

Figure~\ref{Fig:cgc} (left) shows a comparison of the measured $\pT$-differential muon $v_2^{\rm \mu}\lbrace {\rm 2PC} \rbrace$ coefficient at forward rapidity with calculations based on the colour glass condensate (CGC) framework. The latter uses the dilute-dense formalism~\cite{Zhang:2019dth,Zhang:2020ayy} where interactions between partons from the proton projectile and dense gluons inside the target Pb nucleus at the early stage of the collision generate azimuthal anisotropies. The comparison is performed only for the p-going direction where the dilute-dense formalism is valid~\cite{Albacete:2010bs}. The predictions of the $v_2$ coefficient have been provided separately for $\rm D^0$ and B mesons from two-particle correlations.
The $v_2$ coefficient of muons from heavy-flavour hadron decays is further obtained implementing the same strategy as with the AMPT predictions.
One observes that in these CGC-based calculations, the correlations in the initial state generate a significant $v_2$ signal for muons from charm-hadron decays. Its magnitude increases significantly up to $\pT =$~2~GeV/$c$ where it reaches a maximum value of about 0.09, and it decreases smoothly with increasing $\pT$. The predicted $v_2$ signal of muons from beauty-hadron decays is less pronounced, the maximum being of about 0.03 at $\pT =$~3 GeV/$c$. In the highest $\pT$ region, the $v_2$ coefficient of muons from charm-hadron decays is similar to that of muons from beauty-hadron decays. The CGC-based calculations for muons from both charm- and beauty-hadron decays reproduce qualitatively the measured $v_2^{\rm \mu}$ coefficient of inclusive muons for $\pT > 2$~GeV/$c$. It is worth pointing out that the contribution of muons from pion and kaon decays which contributes significantly to the measured inclusive muon yield at low $\pT$ ($\pT < 2$~GeV/$c$) is not considered in the model predictions, preventing to draw any conclusion about the model comparisons in that kinematic region~\cite{Adam:2015bka}.
A similar agreement with these CGC calculations is found for prompt $\rm D^0$, prompt J/$\psi$, and $\rm D^0$ from beauty-hadron decays measured at midrapidity in p--Pb collisions at $\sNN =$~8.16 TeV with the CMS detector~\cite{Sirunyan:2020obi}. The qualitative agreement between data and model calculations may suggest possible contributions from initial-state effects to the measured positive $v_2$ in high-multiplicity p--Pb collisions. However, it should be mentioned that the measured inclusive muon $v_2^{\rm\mu}$ is obtained using charged particles for the calculation of the reference $v_2$ coefficient, which is in general interpreted as a final-state effect~\cite{ALICE:2017lyf}.

The CGC and AMPT predictions for muons from charm- and beauty-hadron decays are displayed together in the right panel of Fig.~\ref{Fig:cgc} for the p-going direction, and compared to the measured inclusive muon $v_2^{\rm\mu}$. The CGC-based calculations provide a larger $v_2$ signal for muons originating from heavy-flavour hadron decays at low $\pT$, up to about $\pT =$~3 GeV/$c$, compared to the AMPT calculations. Such behaviour indicates that the CGC calculations which do not incorporate muons from light-flavour hadron decays would have overestimated the data in this kinematic region. The two models provide compatible results and describe the data at high $\pT$, where the muons from charged-pion and kaon decays do not affect the $v_2^{\rm \mu}$ significantly.

\section{Conclusion}\label{sec:concl}

The second-order coefficient $v_2^{\rm \mu}$ of inclusive muons in high-multiplicity p--Pb collisions at $\sNN =$~8.16 TeV is measured with the ALICE detector at the LHC at forward ($2.03 < \ycms < 3.53$) and backward ($-4.46 < \ycms < -2.96$) rapidities. These new measurements extend previous inclusive muon ALICE results in p--Pb collisions at $\sNN =$~5.02 TeV to a significantly broader transverse momentum interval of $0.5 < \pT < 10$~GeV/$c$. The $v_2^{\rm \mu}$ coefficient is extracted using several event activity estimators with the well-known technique of two-particle correlations and, for the first time for open heavy-flavoured particles, with two-particle cumulants within the generic framework. Nonflow effects which cannot be neglected in the analyses involving correlations between two particles are subtracted using novel techniques.
A positive $v_2^{\rm \mu}$ signal is observed at forward and backward rapidities, with a significance larger than $4.7\sigma$ and $7.6\sigma$, respectively, in the region $2 < \pT < 6$~GeV/$c$ where muons from heavy-flavour hadron decays constitute the main source of muons. The results indicate that heavy quarks reveal a collective-like behaviour in high-multiplicity p--Pb collisions. The AMPT calculations which complement hydrodynamic models and address non-equilibrium dynamics provide a reasonable agreement with the data at both forward and backward rapidities over the whole $\pT$ interval. The measured $v_2^{\rm \mu}$ coefficient at forward rapidity (p-going direction) for $\pT > 2$~GeV/$c$ is in qualitative agreement with CGC calculations within the dilute-dense formalism. Possible contributions from initial-state effects are not fully excluded at high $\pT$.
These comprehensive results on $v_2^{\rm \mu}$ of inclusive muons spanning in a wide $\pT$ range, provide significant new insights for the understanding of the origin of the possible collective behaviour of heavy quarks in small systems such as p--Pb collisions and can be used to constrain the various approaches for modelling the azimuthal anisotropies in small collision systems.

\newenvironment{acknowledgement}{\relax}{\relax}
\begin{acknowledgement}
\section*{Acknowledgements}
The authors would like to thank Shu-Yi Wei for providing the CGC calculations and fruitful discussions.

The ALICE Collaboration would like to thank all its engineers and technicians for their invaluable contributions to the construction of the experiment and the CERN accelerator teams for the outstanding performance of the LHC complex.
The ALICE Collaboration gratefully acknowledges the resources and support provided by all Grid centres and the Worldwide LHC Computing Grid (WLCG) collaboration.
The ALICE Collaboration acknowledges the following funding agencies for their support in building and running the ALICE detector:
A. I. Alikhanyan National Science Laboratory (Yerevan Physics Institute) Foundation (ANSL), State Committee of Science and World Federation of Scientists (WFS), Armenia;
Austrian Academy of Sciences, Austrian Science Fund (FWF): [M 2467-N36] and Nationalstiftung f\"{u}r Forschung, Technologie und Entwicklung, Austria;
Ministry of Communications and High Technologies, National Nuclear Research Center, Azerbaijan;
Conselho Nacional de Desenvolvimento Cient\'{\i}fico e Tecnol\'{o}gico (CNPq), Financiadora de Estudos e Projetos (Finep), Funda\c{c}\~{a}o de Amparo \`{a} Pesquisa do Estado de S\~{a}o Paulo (FAPESP) and Universidade Federal do Rio Grande do Sul (UFRGS), Brazil;
Bulgarian Ministry of Education and Science, within the National Roadmap for Research Infrastructures 2020-2027 (object CERN), Bulgaria;
Ministry of Education of China (MOEC) , Ministry of Science \& Technology of China (MSTC) and National Natural Science Foundation of China (NSFC), China;
Ministry of Science and Education and Croatian Science Foundation, Croatia;
Centro de Aplicaciones Tecnol\'{o}gicas y Desarrollo Nuclear (CEADEN), Cubaenerg\'{\i}a, Cuba;
Ministry of Education, Youth and Sports of the Czech Republic, Czech Republic;
The Danish Council for Independent Research | Natural Sciences, the VILLUM FONDEN and Danish National Research Foundation (DNRF), Denmark;
Helsinki Institute of Physics (HIP), Finland;
Commissariat \`{a} l'Energie Atomique (CEA) and Institut National de Physique Nucl\'{e}aire et de Physique des Particules (IN2P3) and Centre National de la Recherche Scientifique (CNRS), France;
Bundesministerium f\"{u}r Bildung und Forschung (BMBF) and GSI Helmholtzzentrum f\"{u}r Schwerionenforschung GmbH, Germany;
General Secretariat for Research and Technology, Ministry of Education, Research and Religions, Greece;
National Research, Development and Innovation Office, Hungary;
Department of Atomic Energy Government of India (DAE), Department of Science and Technology, Government of India (DST), University Grants Commission, Government of India (UGC) and Council of Scientific and Industrial Research (CSIR), India;
National Research and Innovation Agency - BRIN, Indonesia;
Istituto Nazionale di Fisica Nucleare (INFN), Italy;
Japanese Ministry of Education, Culture, Sports, Science and Technology (MEXT) and Japan Society for the Promotion of Science (JSPS) KAKENHI, Japan;
Consejo Nacional de Ciencia (CONACYT) y Tecnolog\'{i}a, through Fondo de Cooperaci\'{o}n Internacional en Ciencia y Tecnolog\'{i}a (FONCICYT) and Direcci\'{o}n General de Asuntos del Personal Academico (DGAPA), Mexico;
Nederlandse Organisatie voor Wetenschappelijk Onderzoek (NWO), Netherlands;
The Research Council of Norway, Norway;
Commission on Science and Technology for Sustainable Development in the South (COMSATS), Pakistan;
Pontificia Universidad Cat\'{o}lica del Per\'{u}, Peru;
Ministry of Education and Science, National Science Centre and WUT ID-UB, Poland;
Korea Institute of Science and Technology Information and National Research Foundation of Korea (NRF), Republic of Korea;
Ministry of Education and Scientific Research, Institute of Atomic Physics, Ministry of Research and Innovation and Institute of Atomic Physics and University Politehnica of Bucharest, Romania;
Ministry of Education, Science, Research and Sport of the Slovak Republic, Slovakia;
National Research Foundation of South Africa, South Africa;
Swedish Research Council (VR) and Knut \& Alice Wallenberg Foundation (KAW), Sweden;
European Organization for Nuclear Research, Switzerland;
Suranaree University of Technology (SUT), National Science and Technology Development Agency (NSTDA), Thailand Science Research and Innovation (TSRI) and National Science, Research and Innovation Fund (NSRF), Thailand;
Turkish Energy, Nuclear and Mineral Research Agency (TENMAK), Turkey;
National Academy of  Sciences of Ukraine, Ukraine;
Science and Technology Facilities Council (STFC), United Kingdom;
National Science Foundation of the United States of America (NSF) and United States Department of Energy, Office of Nuclear Physics (DOE NP), United States of America.
In addition, individual groups or members have received support from:
Marie Sk\l{}odowska Curie, European Research Council, Strong 2020 - Horizon 2020 (grant nos. 950692, 824093, 896850), European Union;
Academy of Finland (Center of Excellence in Quark Matter) (grant nos. 346327, 346328), Finland;
Programa de Apoyos para la Superaci\'{o}n del Personal Acad\'{e}mico, UNAM, Mexico.
\end{acknowledgement}

\bibliographystyle{utphys} 
\bibliography{AliHFMv2pPb}

\providecommand{\href}[2]{#2}\begingroup\raggedright\begin{thebibliography}{100}

\bibitem{Muller:2006ee}
B.~Muller and J.~L. Nagle, ``{Results from the relativistic heavy ion
  collider}'',
  \href{http://dx.doi.org/10.1146/annurev.nucl.56.080805.140556}{{\em Ann. Rev.
  Nucl. Part. Sci.} {\bfseries 56} (2006) 93--135},
  \href{http://arxiv.org/abs/nucl-th/0602029}{{\ttfamily
  arXiv:nucl-th/0602029}}.

\bibitem{Bazavov:2011nk}
A.~Bazavov {\em et~al.}, ``{The chiral and deconfinement aspects of the QCD
  transition}'', \href{http://dx.doi.org/10.1103/PhysRevD.85.054503}{{\em Phys.
  Rev.} {\bfseries D85} (2012) 054503},
\href{http://arxiv.org/abs/1111.1710}{{\ttfamily arXiv:1111.1710 [hep-lat]}}.

\bibitem{Ollitrault:1992bk}
J.-Y. Ollitrault, ``{Anisotropy as a signature of transverse collective
  flow}'', \href{http://dx.doi.org/10.1103/PhysRevD.46.229}{{\em Phys. Rev. D}
  {\bfseries 46} (1992) 229--245}.

\bibitem{Voloshin:1994mz}
S.~Voloshin and Y.~Zhang, ``{Flow study in relativistic nuclear collisions by
  Fourier expansion of Azimuthal particle distributions}'',
  \href{http://dx.doi.org/10.1007/s002880050141}{{\em Z. Phys. C} {\bfseries
  70} (1996) 665--672}, \href{http://arxiv.org/abs/hep-ph/9407282}{{\ttfamily
  arXiv:hep-ph/9407282}}.

\bibitem{Averbeck:2013oga}
R.~Averbeck, ``{Heavy-flavor production in heavy-ion collisions and
  implications for the properties of hot QCD matter}'',
  \href{http://dx.doi.org/10.1016/j.ppnp.2013.01.001}{{\em Prog. Part. Nucl.
  Phys.} {\bfseries 70} (2013) 159--209},
  \href{http://arxiv.org/abs/1505.03828}{{\ttfamily arXiv:1505.03828
  [nucl-ex]}}.

\bibitem{Kolb:2003dz}
P.~F. Kolb and U.~W. Heinz, ``{Hydrodynamic description of ultrarelativistic
  heavy ion collisions}'',
  \href{http://arxiv.org/abs/nucl-th/0305084}{{\ttfamily
  arXiv:nucl-th/0305084}}.

\bibitem{Gyulassy:2000gk}
M.~Gyulassy, I.~Vitev, and X.~Wang, ``{High $p_{\rm T}$ azimuthal asymmetry in
  noncentral A+A at RHIC}'',
  \href{http://dx.doi.org/10.1103/PhysRevLett.86.2537}{{\em Phys. Rev. Lett.}
  {\bfseries 86} (2001) 2537--2540},
  \href{http://arxiv.org/abs/nucl-th/0012092}{{\ttfamily
  arXiv:nucl-th/0012092}}.

\bibitem{Shuryak:2001me}
E.~Shuryak, ``{The Azimuthal asymmetry at large $p_{\rm T}$ seems to be too
  large for a `jet quenching'}'',
  \href{http://dx.doi.org/10.1103/PhysRevC.66.027902}{{\em Phys. Rev. C}
  {\bfseries 66} (2002) 027902},
  \href{http://arxiv.org/abs/nucl-th/0112042}{{\ttfamily
  arXiv:nucl-th/0112042}}.

\bibitem{Greco:2003vf}
V.~Greco, C.~M. Ko, and R.~Rapp, ``{Quark coalescence for charmed mesons in
  ultrarelativistic heavy ion collisions}'',
  \href{http://dx.doi.org/10.1016/j.physletb.2004.06.064}{{\em Phys. Lett.}
  {\bfseries B595} (2004) 202--208},
\href{http://arxiv.org/abs/nucl-th/0312100}{{\ttfamily arXiv:nucl-th/0312100
  [nucl-th]}}.

\bibitem{Molnar:2004ph}
D.~Molnar, ``{Charm elliptic flow from quark coalescence dynamics}'',
  \href{http://dx.doi.org/10.1088/0954-3899/31/4/052}{{\em J. Phys. G}
  {\bfseries 31} (2005) S421--S428},
  \href{http://arxiv.org/abs/nucl-th/0410041}{{\ttfamily
  arXiv:nucl-th/0410041}}.

\bibitem{STAR:2017kkh}
{\bfseries STAR} Collaboration, L.~Adamczyk {\em et~al.}, ``{Measurement of
  $D^0$ Azimuthal Anisotropy at Midrapidity in Au+Au Collisions at
  $\sqrt{s_{NN}}$=200 GeV}'',
  \href{http://dx.doi.org/10.1103/PhysRevLett.118.212301}{{\em Phys. Rev.
  Lett.} {\bfseries 118} (2017) 212301},
  \href{http://arxiv.org/abs/1701.06060}{{\ttfamily arXiv:1701.06060
  [nucl-ex]}}.

\bibitem{ALICE:2021kfc}
{\bfseries ALICE} Collaboration, S.~Acharya {\em et~al.}, ``{Measurement of
  prompt $D_s^+$-meson production and azimuthal anisotropy in Pb--Pb collisions
  at $\sqrt {s_{\rm NN}} =$ 5.02 TeV}'',
  \href{http://dx.doi.org/10.1016/j.physletb.2022.136986}{{\em Phys. Lett. B}
  {\bfseries 827} (2022) 136986},
  \href{http://arxiv.org/abs/2110.10006}{{\ttfamily arXiv:2110.10006
  [nucl-ex]}}.

\bibitem{CMS:2020bnz}
{\bfseries CMS} Collaboration, A.~M. Sirunyan {\em et~al.}, ``{Measurement of
  prompt ${\mathrm{D^0}}$ and ${\mathrm{\overline{D}}{}^0}$ meson azimuthal
  anisotropy and search for strong electric fields in PbPb collisions at
  $\sqrt{s_\mathrm{NN}} =$ 5.02 TeV}'',
  \href{http://dx.doi.org/10.1016/j.physletb.2021.136253}{{\em Phys. Lett. B}
  {\bfseries 816} (2021) 136253},
  \href{http://arxiv.org/abs/2009.12628}{{\ttfamily arXiv:2009.12628
  [hep-ex]}}.

\bibitem{Aad:2020grf}
{\bfseries ATLAS} Collaboration, G.~Aad {\em et~al.}, ``{Measurement of
  azimuthal anisotropy of muons from charm and bottom hadrons in Pb+Pb
  collisions at $\sqrt {s_{\rm NN}} =$~5.02 TeV with the ATLAS detector}'',
  \href{http://dx.doi.org/10.1016/j.physletb.2020.135595}{{\em Phys. Lett. B}
  {\bfseries 807} (2020) 135595},
  \href{http://arxiv.org/abs/2003.03565}{{\ttfamily arXiv:2003.03565
  [nucl-ex]}}.

\bibitem{Acharya:2020qvg}
{\bfseries ALICE} Collaboration, S.~Acharya {\em et~al.}, ``{Elliptic Flow of
  Electrons from Beauty-Hadron Decays in Pb--Pb Collisions at $\sqrt {s_{\rm
  NN}} =$~5.02 TeV}'',
  \href{http://dx.doi.org/10.1103/PhysRevLett.126.162001}{{\em Phys. Rev.
  Lett.} {\bfseries 126} (2021) 162001},
  \href{http://arxiv.org/abs/2005.11130}{{\ttfamily arXiv:2005.11130
  [nucl-ex]}}.

\bibitem{ALICE:2020iug}
{\bfseries ALICE} Collaboration, S.~Acharya {\em et~al.},
  ``{Transverse-momentum and event-shape dependence of D-meson flow harmonics
  in Pb\textendash{}Pb collisions at $\sqrt {s_{NN}}$ = 5.02 TeV}'',
  \href{http://dx.doi.org/10.1016/j.physletb.2020.136054}{{\em Phys. Lett. B}
  {\bfseries 813} (2021) 136054},
  \href{http://arxiv.org/abs/2005.11131}{{\ttfamily arXiv:2005.11131
  [nucl-ex]}}.

\bibitem{Aaboud:2018bdg}
{\bfseries ATLAS} Collaboration, M.~Aaboud {\em et~al.}, ``{Measurement of the
  suppression and azimuthal anisotropy of muons from heavy-flavor decays in
  Pb+Pb collisions at $\sqrt{s_{\mathrm{NN}}} = 2.76$ TeV with the ATLAS
  detector}'', \href{http://dx.doi.org/10.1103/PhysRevC.98.044905}{{\em Phys.
  Rev. C} {\bfseries 98} (2018) 044905},
  \href{http://arxiv.org/abs/1805.05220}{{\ttfamily arXiv:1805.05220
  [nucl-ex]}}.

\bibitem{Acharya:2018bxo}
{\bfseries ALICE} Collaboration, S.~Acharya {\em et~al.}, ``{Event-shape
  engineering for the D-meson elliptic flow in mid-central Pb--Pb collisions at
  $\sqrt{s_{\rm NN}} =5.02$ TeV}'',
  \href{http://dx.doi.org/10.1007/JHEP02(2019)150}{{\em JHEP} {\bfseries 02}
  (2019) 150}, \href{http://arxiv.org/abs/1809.09371}{{\ttfamily
  arXiv:1809.09371 [nucl-ex]}}.

\bibitem{Sirunyan:2017plt}
{\bfseries CMS} Collaboration, A.~M. Sirunyan {\em et~al.}, ``{Measurement of
  prompt $D^0$ meson azimuthal anisotropy in PbPb collisions at $\sqrt{{s}_{\rm
  NN}} =$~5.02 TeV}'',
  \href{http://dx.doi.org/10.1103/PhysRevLett.120.202301}{{\em Phys. Rev.
  Lett.} {\bfseries 120} (2018) 202301},
  \href{http://arxiv.org/abs/1708.03497}{{\ttfamily arXiv:1708.03497
  [nucl-ex]}}.

\bibitem{ALICE:2017pbx}
{\bfseries ALICE} Collaboration, S.~Acharya {\em et~al.}, ``{$D$-meson
  azimuthal anisotropy in midcentral Pb-Pb collisions at $\sqrt{s_{\rm
  NN}}=5.02$ TeV}'',
  \href{http://dx.doi.org/10.1103/PhysRevLett.120.102301}{{\em Phys. Rev.
  Lett.} {\bfseries 120} (2018) 102301},
  \href{http://arxiv.org/abs/1707.01005}{{\ttfamily arXiv:1707.01005
  [nucl-ex]}}.

\bibitem{Adam:2016ssk}
{\bfseries ALICE} Collaboration, J.~Adam {\em et~al.}, ``{Elliptic flow of
  electrons from heavy-flavour hadron decays at mid-rapidity in Pb--Pb
  collisions at $ \sqrt{{\mathrm{s}}_{\mathrm{NN}}}=2.76 $ TeV}'',
  \href{http://dx.doi.org/10.1007/JHEP09(2016)028}{{\em JHEP} {\bfseries 09}
  (2016) 028}, \href{http://arxiv.org/abs/1606.00321}{{\ttfamily
  arXiv:1606.00321 [nucl-ex]}}.

\bibitem{CMS:2016mah}
{\bfseries CMS} Collaboration, V.~Khachatryan {\em et~al.}, ``{Suppression and
  azimuthal anisotropy of prompt and nonprompt ${\mathrm{J}}/\psi $ production
  in PbPb collisions at $\sqrt{{s_{_{\text {NN}}}}} =2.76$ $\,\mathrm{TeV}$}'',
  \href{http://dx.doi.org/10.1140/epjc/s10052-017-4781-1}{{\em Eur. Phys. J. C}
  {\bfseries 77} (2017) 252}, \href{http://arxiv.org/abs/1610.00613}{{\ttfamily
  arXiv:1610.00613 [nucl-ex]}}.

\bibitem{Adam:2015pga}
{\bfseries ALICE} Collaboration, J.~Adam {\em et~al.}, ``{Elliptic flow of
  muons from heavy-flavour hadron decays at forward rapidity in Pb--Pb
  collisions at $\sqrt{s_{\rm NN}}= 2.76$ TeV}'',
  \href{http://dx.doi.org/10.1016/j.physletb.2015.11.059}{{\em Phys. Lett. B}
  {\bfseries 753} (2016) 41--56},
  \href{http://arxiv.org/abs/1507.03134}{{\ttfamily arXiv:1507.03134
  [nucl-ex]}}.

\bibitem{Abelev:2014ipa}
{\bfseries ALICE} Collaboration, B.~B. Abelev {\em et~al.}, ``{Azimuthal
  anisotropy of D meson production in Pb--Pb collisions at $\sqrt{s_{\rm NN}} =
  2.76$ TeV}'', \href{http://dx.doi.org/10.1103/PhysRevC.90.034904}{{\em Phys.
  Rev. C} {\bfseries 90} (2014) 034904},
  \href{http://arxiv.org/abs/1405.2001}{{\ttfamily arXiv:1405.2001 [nucl-ex]}}.

\bibitem{ALICE:2021rxa}
{\bfseries ALICE} Collaboration, S.~Acharya {\em et~al.}, ``{Prompt D$^{0}$,
  D$^{+}$, and D$^{*+}$ production in Pb\textendash{}Pb collisions at $
  \sqrt{s_{\mathrm{NN}}} $ = 5.02 TeV}'',
  \href{http://dx.doi.org/10.1007/JHEP01(2022)174}{{\em JHEP} {\bfseries 01}
  (2022) 174}, \href{http://arxiv.org/abs/2110.09420}{{\ttfamily
  arXiv:2110.09420 [nucl-ex]}}.

\bibitem{ATLAS:2021xtw}
{\bfseries ATLAS} Collaboration, G.~Aad {\em et~al.}, ``{Measurement of the
  nuclear modification factor for muons from charm and bottom hadrons in Pb+Pb
  collisions at 5.02 TeV with the ATLAS detector}'',
  \href{http://dx.doi.org/10.1016/j.physletb.2022.137077}{{\em Phys. Lett. B}
  {\bfseries 829} (2022) 137077},
  \href{http://arxiv.org/abs/2109.00411}{{\ttfamily arXiv:2109.00411
  [nucl-ex]}}.

\bibitem{ALICE:2020sjb}
{\bfseries ALICE} Collaboration, S.~Acharya {\em et~al.}, ``{Production of
  muons from heavy-flavour hadron decays at high transverse momentum in Pb--Pb
  collisions at $\sqrt{s_{\rm NN}} =$~5.02 and 2.76 TeV}'',
  \href{http://dx.doi.org/10.1016/j.physletb.2021.136558}{{\em Phys. Lett. B}
  {\bfseries 820} (2021) 136558},
  \href{http://arxiv.org/abs/2011.05718}{{\ttfamily arXiv:2011.05718
  [nucl-ex]}}.

\bibitem{Acharya:2019mom}
{\bfseries ALICE} Collaboration, S.~Acharya {\em et~al.}, ``{Measurement of
  electrons from semileptonic heavy-flavour hadron decays at midrapidity in pp
  and Pb-Pb collisions at $\sqrt{s_{\rm{NN}}} =$~5.02 TeV}'',
  \href{http://dx.doi.org/10.1016/j.physletb.2020.135377}{{\em Phys. Lett. B}
  {\bfseries 804} (2020) 135377},
  \href{http://arxiv.org/abs/1910.09110}{{\ttfamily arXiv:1910.09110
  [nucl-ex]}}.

\bibitem{Acharya:2018upq}
{\bfseries ALICE} Collaboration, S.~Acharya {\em et~al.}, ``{Measurements of
  low-p$_{\rm T}$ electrons from semileptonic heavy-flavour hadron decays at
  mid-rapidity in pp and Pb--Pb collisions at $ \sqrt{s_{\mathrm{NN}}}=2.76 $
  TeV}'', \href{http://dx.doi.org/10.1007/JHEP10(2018)061}{{\em JHEP}
  {\bfseries 10} (2018) 061},
\href{http://arxiv.org/abs/1805.04379}{{\ttfamily arXiv:1805.04379 [nucl-ex]}}.

\bibitem{Acharya:2018hre}
{\bfseries ALICE} Collaboration, S.~Acharya {\em et~al.}, ``{Measurement of
  D$^{0}$, D$^{+}$, D$^{*+}$ and D$_{s}^{+}$ production in Pb-Pb collisions at
  $ \sqrt{{\mathrm{s}}_{\mathrm{NN}}}=5.02 $ TeV}'',
  \href{http://dx.doi.org/10.1007/JHEP10(2018)174}{{\em JHEP} {\bfseries 10}
  (2018) 174}, \href{http://arxiv.org/abs/1804.09083}{{\ttfamily
  arXiv:1804.09083 [nucl-ex]}}.

\bibitem{Sirunyan:2018ktu}
{\bfseries CMS} Collaboration, A.~M. Sirunyan {\em et~al.}, ``{Studies of
  Beauty Suppression via Nonprompt $\rm D^0$ Mesons in PbPb Collisions at $Q^2
  = 4$ $\rm GeV^2$}'',
  \href{http://dx.doi.org/10.1103/PhysRevLett.123.022001}{{\em Phys. Rev.
  Lett.} {\bfseries 123} (2019) 022001},
\href{http://arxiv.org/abs/1810.11102}{{\ttfamily arXiv:1810.11102 [hep-ex]}}.

\bibitem{CMS:2018eso}
{\bfseries CMS} Collaboration, A.~M. Sirunyan {\em et~al.}, ``{Measurement of
  B$^0_\mathrm{s}$ meson production in pp and PbPb collisions at
  $\sqrt{s_\mathrm{NN}} =$ 5.02 TeV}'',
  \href{http://dx.doi.org/10.1016/j.physletb.2019.07.014}{{\em Phys. Lett. B}
  {\bfseries 796} (2019) 168--190},
  \href{http://arxiv.org/abs/1810.03022}{{\ttfamily arXiv:1810.03022
  [hep-ex]}}.

\bibitem{ATLAS:2018hqe}
{\bfseries ATLAS} Collaboration, M.~Aaboud {\em et~al.}, ``{Prompt and
  non-prompt $J/\psi $ and $\psi (2\mathrm {S})$ suppression at high transverse
  momentum in $5.02~\mathrm {TeV}$ Pb+Pb collisions with the ATLAS
  experiment}'', \href{http://dx.doi.org/10.1140/epjc/s10052-018-6219-9}{{\em
  Eur. Phys. J. C} {\bfseries 78} (2018) 762},
  \href{http://arxiv.org/abs/1805.04077}{{\ttfamily arXiv:1805.04077
  [nucl-ex]}}.

\bibitem{Sirunyan:2017xss}
{\bfseries CMS} Collaboration, A.~M. Sirunyan {\em et~al.}, ``{Nuclear
  modification factor of D$^0$ mesons in PbPb collisions at
  $\sqrt{s_\mathrm{NN}} = 5.02$ TeV}'',
  \href{http://dx.doi.org/10.1016/j.physletb.2018.05.074}{{\em Phys. Lett.}
  {\bfseries B782} (2018) 474--496},
\href{http://arxiv.org/abs/1708.04962}{{\ttfamily arXiv:1708.04962 [nucl-ex]}}.

\bibitem{Sirunyan:2017oug}
{\bfseries CMS} Collaboration, A.~M. Sirunyan {\em et~al.}, ``{Measurement of
  the $\rm {B}^{\pm}$ Meson Nuclear Modification Factor in PbPb Collisions at
  $\sqrt{{s}_{\rm NN}}=5.02\text{ }\text{ }\mathrm{TeV}$}'',
  \href{http://dx.doi.org/10.1103/PhysRevLett.119.152301}{{\em Phys. Rev.
  Lett.} {\bfseries 119} (2017) 152301},
\href{http://arxiv.org/abs/1705.04727}{{\ttfamily arXiv:1705.04727 [hep-ex]}}.

\bibitem{Sirunyan:2017isk}
{\bfseries CMS} Collaboration, A.~M. Sirunyan {\em et~al.}, ``{Measurement of
  prompt and nonprompt charmonium suppression in $\text {PbPb}$ collisions at
  5.02 $\,\text {Te}\text {V}$}'',
  \href{http://dx.doi.org/10.1140/epjc/s10052-018-5950-6}{{\em Eur. Phys. J.}
  {\bfseries C78} (2018) 509},
\href{http://arxiv.org/abs/1712.08959}{{\ttfamily arXiv:1712.08959 [nucl-ex]}}.

\bibitem{Adam:2016khe}
{\bfseries ALICE} Collaboration, J.~Adam {\em et~al.}, ``{Measurement of the
  production of high-$p_{\rm T}$ electrons from heavy-flavour hadron decays in
  Pb--Pb collisions at $\mathbf{\sqrt{\it s_{\rm{NN}}}}$ = 2.76 TeV}'',
  \href{http://dx.doi.org/10.1016/j.physletb.2017.05.060}{{\em Phys. Lett.}
  {\bfseries B771} (2017) 467--481},
\href{http://arxiv.org/abs/1609.07104}{{\ttfamily arXiv:1609.07104 [nucl-ex]}}.

\bibitem{Eskola:2009uj}
K.~J. Eskola, H.~Paukkunen, and C.~A. Salgado, ``{EPS09: A New Generation of
  NLO and LO Nuclear Parton Distribution Functions}'',
  \href{http://dx.doi.org/10.1088/1126-6708/2009/04/065}{{\em JHEP} {\bfseries
  04} (2009) 065},
\href{http://arxiv.org/abs/0902.4154}{{\ttfamily arXiv:0902.4154 [hep-ph]}}.

\bibitem{Kopeliovich:2002yh}
B.~Z. Kopeliovich, J.~Nemchik, A.~Schafer, and A.~V. Tarasov, ``{Cronin effect
  in hadron production off nuclei}'',
  \href{http://dx.doi.org/10.1103/PhysRevLett.88.232303}{{\em Phys. Rev. Lett.}
  {\bfseries 88} (2002) 232303},
\href{http://arxiv.org/abs/hep-ph/0201010}{{\ttfamily arXiv:hep-ph/0201010
  [hep-ph]}}.

\bibitem{Kang:2014hha}
Z.-B. Kang, I.~Vitev, E.~Wang, H.~Xing, and C.~Zhang, ``{Multiple scattering
  effects on heavy meson production in p+A collisions at backward rapidity}'',
  \href{http://dx.doi.org/10.1016/j.physletb.2014.11.024}{{\em Phys. Lett. B}
  {\bfseries 740} (2015) 23--29},
  \href{http://arxiv.org/abs/1409.2494}{{\ttfamily arXiv:1409.2494 [hep-ph]}}.

\bibitem{CMS:2012qk}
{\bfseries CMS} Collaboration, S.~Chatrchyan {\em et~al.}, ``{Observation of
  Long-Range Near-Side Angular Correlations in Proton-Lead Collisions at the
  LHC}'', \href{http://dx.doi.org/10.1016/j.physletb.2012.11.025}{{\em Phys.
  Lett. B} {\bfseries 718} (2013) 795--814},
  \href{http://arxiv.org/abs/1210.5482}{{\ttfamily arXiv:1210.5482 [nucl-ex]}}.

\bibitem{Abelev:2012ola}
{\bfseries ALICE} Collaboration, B.~Abelev {\em et~al.}, ``{Long-range angular
  correlations on the near and away side in pPb collisions at $\sqrt{s_{\rm
  NN}}=5.02$ TeV}'',
  \href{http://dx.doi.org/10.1016/j.physletb.2013.01.012}{{\em Phys. Lett. B}
  {\bfseries 719} (2013) 29--41},
  \href{http://arxiv.org/abs/1212.2001}{{\ttfamily arXiv:1212.2001 [nucl-ex]}}.

\bibitem{Aad:2012gla}
{\bfseries ATLAS} Collaboration, G.~Aad {\em et~al.}, ``{Observation of
  Associated Near-Side and Away-Side Long-Range Correlations in $\sqrt{s_{\rm
  NN}}$ = 5.02 TeV Proton-Lead Collisions with the ATLAS Detector}'',
  \href{http://dx.doi.org/10.1103/PhysRevLett.110.182302}{{\em Phys. Rev.
  Lett.} {\bfseries 110} (2013) 182302},
  \href{http://arxiv.org/abs/1212.5198}{{\ttfamily arXiv:1212.5198 [hep-ex]}}.

\bibitem{ABELEV:2013wsa}
{\bfseries ALICE} Collaboration, B.~B. Abelev {\em et~al.}, ``{Long-range
  angular correlations of $\rm \pi$, K and p in p--Pb collisions at
  $\sqrt{s_{\rm NN}}$ = 5.02 TeV}'',
  \href{http://dx.doi.org/10.1016/j.physletb.2013.08.024}{{\em Phys. Lett. B}
  {\bfseries 726} (2013) 164--177},
  \href{http://arxiv.org/abs/1307.3237}{{\ttfamily arXiv:1307.3237 [nucl-ex]}}.

\bibitem{Aaboud:2016yar}
{\bfseries ATLAS} Collaboration, M.~Aaboud {\em et~al.}, ``{Measurements of
  long-range azimuthal anisotropies and associated Fourier coefficients for pp
  collisions at $\sqrt{s}=5.02$ and $13$ TeV and p+Pb collisions at
  $\sqrt{s_{\mathrm{NN}}}=5.02$ TeV with the ATLAS detector}'',
  \href{http://dx.doi.org/10.1103/PhysRevC.96.024908}{{\em Phys. Rev. C}
  {\bfseries 96} (2017) 024908},
  \href{http://arxiv.org/abs/1609.06213}{{\ttfamily arXiv:1609.06213
  [nucl-ex]}}.

\bibitem{Adam:2015sza}
{\bfseries ALICE} Collaboration, J.~Adam {\em et~al.}, ``{Transverse momentum
  dependence of D-meson production in Pb--Pb collisions at $
  \sqrt{{\mathrm{s}}_{\mathrm{NN}}}=$ 2.76 TeV}'',
  \href{http://dx.doi.org/10.1007/JHEP03(2016)081}{{\em JHEP} {\bfseries 03}
  (2016) 081},
\href{http://arxiv.org/abs/1509.06888}{{\ttfamily arXiv:1509.06888 [nucl-ex]}}.

\bibitem{Aaij:2015qcq}
{\bfseries LHCb} Collaboration, R.~Aaij {\em et~al.}, ``{Measurements of
  long-range near-side angular correlations in $\sqrt{s_{\text{\rm NN}}} = 5$
  TeV proton-lead collisions in the forward region}'',
  \href{http://dx.doi.org/10.1016/j.physletb.2016.09.064}{{\em Phys. Lett. B}
  {\bfseries 762} (2016) 473--483},
  \href{http://arxiv.org/abs/1512.00439}{{\ttfamily arXiv:1512.00439
  [nucl-ex]}}.

\bibitem{Adam:2015bka}
{\bfseries ALICE} Collaboration, J.~Adam {\em et~al.}, ``{Forward-central
  two-particle correlations in p--Pb collisions at $\sqrt{s_{\rm NN}}$ = 5.02
  TeV}'', \href{http://dx.doi.org/10.1016/j.physletb.2015.12.010}{{\em Phys.
  Lett. B} {\bfseries 753} (2016) 126--139},
  \href{http://arxiv.org/abs/1506.08032}{{\ttfamily arXiv:1506.08032
  [nucl-ex]}}.

\bibitem{Acharya:2017tfn}
{\bfseries ALICE} Collaboration, S.~Acharya {\em et~al.}, ``{Search for
  collectivity with azimuthal J/$\psi$-hadron correlations in high multiplicity
  p--Pb collisions at $\sqrt{s_{\rm NN}}$ = 5.02 and 8.16 TeV}'',
  \href{http://dx.doi.org/10.1016/j.physletb.2018.02.039}{{\em Phys. Lett. B}
  {\bfseries 780} (2018) 7--20},
  \href{http://arxiv.org/abs/1709.06807}{{\ttfamily arXiv:1709.06807
  [nucl-ex]}}.

\bibitem{Sirunyan:2018kiz}
{\bfseries CMS} Collaboration, A.~M. Sirunyan {\em et~al.}, ``{Observation of
  prompt J/$\psi$ meson elliptic flow in high-multiplicity pPb collisions at
  $\sqrt{s_\mathrm{NN}} =$ 8.16 TeV}'',
  \href{http://dx.doi.org/10.1016/j.physletb.2019.02.018}{{\em Phys. Lett. B}
  {\bfseries 791} (2019) 172--194},
  \href{http://arxiv.org/abs/1810.01473}{{\ttfamily arXiv:1810.01473
  [hep-ex]}}.

\bibitem{ALICE:2016clc}
{\bfseries ALICE} Collaboration, J.~Adam {\em et~al.}, ``{Measurement of
  azimuthal correlations of D mesons and charged particles in pp collisions at
  $\sqrt{s}=7$ TeV and p--Pb collisions at $\sqrt{s_{\rm NN}}=5.02$ TeV}'',
  \href{http://dx.doi.org/10.1140/epjc/s10052-017-4779-8}{{\em Eur. Phys. J. C}
  {\bfseries 77} (2017) 245}, \href{http://arxiv.org/abs/1605.06963}{{\ttfamily
  arXiv:1605.06963 [nucl-ex]}}.

\bibitem{Acharya:2019icl}
{\bfseries ALICE} Collaboration, S.~Acharya {\em et~al.}, ``{Azimuthal
  correlations of prompt D mesons with charged particles in pp and
  p\textendash{}Pb collisions at $\sqrt{s_{\rm NN}}$ = 5.02 TeV}'',
  \href{http://dx.doi.org/10.1140/epjc/s10052-020-8118-0}{{\em Eur. Phys. J. C}
  {\bfseries 80} (2020) 979}, \href{http://arxiv.org/abs/1910.14403}{{\ttfamily
  arXiv:1910.14403 [nucl-ex]}}.

\bibitem{Sirunyan:2018toe}
{\bfseries CMS} Collaboration, A.~M. Sirunyan {\em et~al.}, ``{Elliptic flow of
  charm and strange hadrons in high-multiplicity pPb collisions at
  $\sqrt{s_{_\mathrm{NN}}} =$ 8.16 TeV}'',
  \href{http://dx.doi.org/10.1103/PhysRevLett.121.082301}{{\em Phys. Rev.
  Lett.} {\bfseries 121} (2018) 082301},
  \href{http://arxiv.org/abs/1804.09767}{{\ttfamily arXiv:1804.09767
  [hep-ex]}}.

\bibitem{Sirunyan:2020obi}
{\bfseries CMS} Collaboration, A.~M. Sirunyan {\em et~al.}, ``{Studies of charm
  and beauty hadron long-range correlations in pp and pPb collisions at LHC
  energies}'', \href{http://dx.doi.org/10.1016/j.physletb.2020.136036}{{\em
  Phys. Lett. B} {\bfseries 813} (2021) 136036},
  \href{http://arxiv.org/abs/2009.07065}{{\ttfamily arXiv:2009.07065
  [hep-ex]}}.

\bibitem{Acharya:2018dxy}
{\bfseries ALICE} Collaboration, S.~Acharya {\em et~al.}, ``{Azimuthal
  Anisotropy of Heavy-Flavor Decay Electrons in $p$-Pb Collisions at
  $\sqrt{s_{\rm NN}}$ = 5.02 TeV}'',
  \href{http://dx.doi.org/10.1103/PhysRevLett.122.072301}{{\em Phys. Rev.
  Lett.} {\bfseries 122} (2019) 072301},
  \href{http://arxiv.org/abs/1805.04367}{{\ttfamily arXiv:1805.04367
  [nucl-ex]}}.

\bibitem{ATLAS:2019xqc}
{\bfseries ATLAS} Collaboration, G.~Aad {\em et~al.}, ``{Measurement of
  azimuthal anisotropy of muons from charm and bottom hadrons in $pp$
  collisions at $\sqrt{s}=13$ TeV with the ATLAS detector}'',
  \href{http://dx.doi.org/10.1103/PhysRevLett.124.082301}{{\em Phys. Rev.
  Lett.} {\bfseries 124} (2020) 082301},
  \href{http://arxiv.org/abs/1909.01650}{{\ttfamily arXiv:1909.01650
  [nucl-ex]}}.

\bibitem{Adare:2013piz}
{\bfseries PHENIX} Collaboration, A.~Adare {\em et~al.}, ``{Quadrupole
  Anisotropy in Dihadron Azimuthal Correlations in Central $d$$+$Au Collisions
  at $\sqrt{s_{_{\rm NN}}}$ = 200 GeV}'',
  \href{http://dx.doi.org/10.1103/PhysRevLett.111.212301}{{\em Phys. Rev.
  Lett.} {\bfseries 111} (2013) 212301},
  \href{http://arxiv.org/abs/1303.1794}{{\ttfamily arXiv:1303.1794 [nucl-ex]}}.

\bibitem{Adare:2014keg}
{\bfseries PHENIX} Collaboration, A.~Adare {\em et~al.}, ``{Measurement of
  long-range angular correlation and quadrupole anisotropy of pions and
  (anti)protons in central $d$$+$Au collisions at $\sqrt{s_{_{NN}}}$ = 200
  GeV}'', \href{http://dx.doi.org/10.1103/PhysRevLett.114.192301}{{\em Phys.
  Rev. Lett.} {\bfseries 114} (2015) 192301},
  \href{http://arxiv.org/abs/1404.7461}{{\ttfamily arXiv:1404.7461 [nucl-ex]}}.

\bibitem{Adamczyk:2015xjc}
{\bfseries STAR} Collaboration, L.~Adamczyk {\em et~al.}, ``{Long-range
  pseudorapidity dihadron correlations in $d$+Au collisions at $\sqrt{s_{\rm
  NN}}=200$ GeV}'',
  \href{http://dx.doi.org/10.1016/j.physletb.2015.05.075}{{\em Phys. Lett. B}
  {\bfseries 747} (2015) 265--271},
  \href{http://arxiv.org/abs/1502.07652}{{\ttfamily arXiv:1502.07652
  [nucl-ex]}}.

\bibitem{Adare:2015ctn}
{\bfseries PHENIX} Collaboration, A.~Adare {\em et~al.}, ``{Measurements of
  elliptic and triangular flow in high-multiplicity $^{3}$He$+$Au collisions at
  $\sqrt{s_{_{\rm NN}}} = 200$ GeV}'',
  \href{http://dx.doi.org/10.1103/PhysRevLett.115.142301}{{\em Phys. Rev.
  Lett.} {\bfseries 115} (2015) 142301},
  \href{http://arxiv.org/abs/1507.06273}{{\ttfamily arXiv:1507.06273
  [nucl-ex]}}.

\bibitem{PHENIX:2018hho}
{\bfseries PHENIX} Collaboration, A.~Adare {\em et~al.}, ``{Pseudorapidity
  Dependence of Particle Production and Elliptic Flow in Asymmetric Nuclear
  Collisions of $p+$Al, $p+$Au, $d+$Au, and $^{3}$He$+$Au at
  $\sqrt{s_{_{NN}}}=200$ GeV}'',
  \href{http://dx.doi.org/10.1103/PhysRevLett.121.222301}{{\em Phys. Rev.
  Lett.} {\bfseries 121} (2018) 222301},
  \href{http://arxiv.org/abs/1807.11928}{{\ttfamily arXiv:1807.11928
  [nucl-ex]}}.

\bibitem{Aad:2014lta}
{\bfseries ATLAS} Collaboration, G.~Aad {\em et~al.}, ``{Measurement of
  long-range pseudorapidity correlations and azimuthal harmonics in
  $\sqrt{s_{NN}}=5.02$ TeV proton-lead collisions with the ATLAS detector}'',
  \href{http://dx.doi.org/10.1103/PhysRevC.90.044906}{{\em Phys. Rev. C}
  {\bfseries 90} (2014) 044906},
  \href{http://arxiv.org/abs/1409.1792}{{\ttfamily arXiv:1409.1792 [hep-ex]}}.

\bibitem{Aad:2019ajj}
{\bfseries ATLAS} Collaboration, G.~Aad {\em et~al.}, ``{Transverse momentum
  and process dependent azimuthal anisotropies in $\sqrt{s_{\mathrm{NN}}}=8.16$
  TeV p+Pb collisions with the ATLAS detector}'',
  \href{http://dx.doi.org/10.1140/epjc/s10052-020-7624-4}{{\em Eur. Phys. J. C}
  {\bfseries 80} (2020) 73}, \href{http://arxiv.org/abs/1910.13978}{{\ttfamily
  arXiv:1910.13978 [nucl-ex]}}.

\bibitem{ATLAS:2018ngv}
{\bfseries ATLAS} Collaboration, M.~Aaboud {\em et~al.}, ``{Correlated
  long-range mixed-harmonic fluctuations measured in $pp$, $p$+Pb and
  low-multiplicity Pb+Pb collisions with the ATLAS detector}'',
  \href{http://dx.doi.org/10.1016/j.physletb.2018.11.065}{{\em Phys. Lett. B}
  {\bfseries 789} (2019) 444--471},
  \href{http://arxiv.org/abs/1807.02012}{{\ttfamily arXiv:1807.02012
  [nucl-ex]}}.

\bibitem{Khachatryan:2015waa}
{\bfseries CMS} Collaboration, V.~Khachatryan {\em et~al.}, ``{Evidence for
  Collective Multiparticle Correlations in pPb Collisions}'',
  \href{http://dx.doi.org/10.1103/PhysRevLett.115.012301}{{\em Phys. Rev.
  Lett.} {\bfseries 115} (2015) 012301},
  \href{http://arxiv.org/abs/1502.05382}{{\ttfamily arXiv:1502.05382
  [nucl-ex]}}.

\bibitem{Aaboud:2017acw}
{\bfseries ATLAS} Collaboration, M.~Aaboud {\em et~al.}, ``{Measurement of
  multi-particle azimuthal correlations in $pp$, $p+$Pb and low-multiplicity
  Pb$+$Pb collisions with the ATLAS detector}'',
  \href{http://dx.doi.org/10.1140/epjc/s10052-017-4988-1}{{\em Eur. Phys. J. C}
  {\bfseries 77} (2017) 428}, \href{http://arxiv.org/abs/1705.04176}{{\ttfamily
  arXiv:1705.04176 [hep-ex]}}.

\bibitem{Aaboud:2017blb}
{\bfseries ATLAS} Collaboration, M.~Aaboud {\em et~al.}, ``{Measurement of
  long-range multiparticle azimuthal correlations with the subevent cumulant
  method in pp and p + Pb collisions with the ATLAS detector at the CERN Large
  Hadron Collider}'', \href{http://dx.doi.org/10.1103/PhysRevC.97.024904}{{\em
  Phys. Rev. C} {\bfseries 97} (2018) 024904},
  \href{http://arxiv.org/abs/1708.03559}{{\ttfamily arXiv:1708.03559
  [hep-ex]}}.

\bibitem{Sirunyan:2019pbr}
{\bfseries CMS} Collaboration, A.~M. Sirunyan {\em et~al.}, ``{Multiparticle
  correlation studies in pPb collisions at $\sqrt{s_\mathrm{NN}} =$ 8.16
  TeV}'', \href{http://dx.doi.org/10.1103/PhysRevC.101.014912}{{\em Phys. Rev.
  C} {\bfseries 101} (2020) 014912},
  \href{http://arxiv.org/abs/1904.11519}{{\ttfamily arXiv:1904.11519
  [hep-ex]}}.

\bibitem{Acharya:2019vdf}
{\bfseries ALICE} Collaboration, S.~Acharya {\em et~al.}, ``{Investigations of
  Anisotropic Flow Using Multiparticle Azimuthal Correlations in pp, p-Pb,
  Xe-Xe, and Pb-Pb Collisions at the LHC}'',
  \href{http://dx.doi.org/10.1103/PhysRevLett.123.142301}{{\em Phys. Rev.
  Lett.} {\bfseries 123} (2019) 142301},
  \href{http://arxiv.org/abs/1903.01790}{{\ttfamily arXiv:1903.01790
  [nucl-ex]}}.

\bibitem{Abelev:2014mda}
{\bfseries ALICE} Collaboration, B.~B. Abelev {\em et~al.}, ``{Multiparticle
  azimuthal correlations in p--Pb and Pb--Pb collisions at the CERN Large
  Hadron Collider}'', \href{http://dx.doi.org/10.1103/PhysRevC.90.054901}{{\em
  Phys. Rev. C} {\bfseries 90} (2014) 054901},
  \href{http://arxiv.org/abs/1406.2474}{{\ttfamily arXiv:1406.2474 [nucl-ex]}}.

\bibitem{Adam:2014qja}
{\bfseries ALICE} Collaboration, J.~Adam {\em et~al.}, ``{Centrality dependence
  of particle production in p-Pb collisions at $\sqrt{s_{\rm NN} }$ = 5.02
  TeV}'', \href{http://dx.doi.org/10.1103/PhysRevC.91.064905}{{\em Phys. Rev.
  C} {\bfseries 91} (2015) 064905},
  \href{http://arxiv.org/abs/1412.6828}{{\ttfamily arXiv:1412.6828 [nucl-ex]}}.

\bibitem{Aad:2016zif}
{\bfseries ATLAS} Collaboration, G.~Aad {\em et~al.}, ``{Transverse momentum,
  rapidity, and centrality dependence of inclusive charged-particle production
  in $\sqrt{s_{\rm NN}}=5.02$ TeV $p$ + Pb collisions measured by the ATLAS
  experiment}'', \href{http://dx.doi.org/10.1016/j.physletb.2016.10.053}{{\em
  Phys. Lett. B} {\bfseries 763} (2016) 313--336},
  \href{http://arxiv.org/abs/1605.06436}{{\ttfamily arXiv:1605.06436
  [hep-ex]}}.

\bibitem{Zhang:2013oca}
X.~Zhang and J.~Liao, ``{Jet Quenching and Its Azimuthal Anisotropy in AA and
  possibly High Multiplicity pA and dA Collisions}'',
  \href{http://arxiv.org/abs/1311.5463}{{\ttfamily arXiv:1311.5463 [nucl-th]}}.

\bibitem{Dusling:2015gta}
K.~Dusling, W.~Li, and B.~Schenke, ``{Novel collective phenomena in high-energy
  proton--proton and proton--nucleus collisions}'',
  \href{http://dx.doi.org/10.1142/S0218301316300022}{{\em Int. J. Mod. Phys. E}
  {\bfseries 25} (2016) 1630002},
  \href{http://arxiv.org/abs/1509.07939}{{\ttfamily arXiv:1509.07939
  [nucl-ex]}}.

\bibitem{Bozek:2011if}
P.~Bozek, ``{Collective flow in p-Pb and d-Pb collisions at TeV energies}'',
  \href{http://dx.doi.org/10.1103/PhysRevC.85.014911}{{\em Phys. Rev. C}
  {\bfseries 85} (2012) 014911},
  \href{http://arxiv.org/abs/1112.0915}{{\ttfamily arXiv:1112.0915 [hep-ph]}}.

\bibitem{Bozek:2012gr}
P.~Bozek and W.~Broniowski, ``{Correlations from hydrodynamic flow in p--Pb
  collisions}'', \href{http://dx.doi.org/10.1016/j.physletb.2012.12.051}{{\em
  Phys. Lett. B} {\bfseries 718} (2013) 1557--1561},
  \href{http://arxiv.org/abs/1211.0845}{{\ttfamily arXiv:1211.0845 [nucl-th]}}.

\bibitem{Nagle:2018nvi}
J.~L. Nagle and W.~A. Zajc, ``{Small System Collectivity in Relativistic
  Hadronic and Nuclear Collisions}'',
  \href{http://dx.doi.org/10.1146/annurev-nucl-101916-123209}{{\em Ann. Rev.
  Nucl. Part. Sci.} {\bfseries 68} (2018) 211--235},
  \href{http://arxiv.org/abs/1801.03477}{{\ttfamily arXiv:1801.03477
  [nucl-ex]}}.

\bibitem{Dusling:2012cg}
K.~Dusling and R.~Venugopalan, ``{Evidence for BFKL and saturation dynamics
  from dihadron spectra at the LHC}'',
  \href{http://dx.doi.org/10.1103/PhysRevD.87.051502}{{\em Phys. Rev. D}
  {\bfseries 87} (2013) 051502},
  \href{http://arxiv.org/abs/1210.3890}{{\ttfamily arXiv:1210.3890 [hep-ph]}}.

\bibitem{Dusling:2013oia}
K.~Dusling and R.~Venugopalan, ``{Comparison of the color glass condensate to
  dihadron correlations in proton-proton and proton-nucleus collisions}'',
  \href{http://dx.doi.org/10.1103/PhysRevD.87.094034}{{\em Phys. Rev. D}
  {\bfseries 87} (2013) 094034},
  \href{http://arxiv.org/abs/1302.7018}{{\ttfamily arXiv:1302.7018 [hep-ph]}}.

\bibitem{He:2015hfa}
L.~He, T.~Edmonds, Z.-W. Lin, F.~Liu, D.~Molnar, and F.~Wang, ``{Anisotropic
  parton escape is the dominant source of azimuthal anisotropy in transport
  models}'', \href{http://dx.doi.org/10.1016/j.physletb.2015.12.051}{{\em Phys.
  Lett. B} {\bfseries 753} (2016) 506--510},
  \href{http://arxiv.org/abs/1502.05572}{{\ttfamily arXiv:1502.05572
  [nucl-th]}}.

\bibitem{Lin:2004en}
Z.-W. Lin, C.~M. Ko, B.-A. Li, B.~Zhang, and S.~Pal, ``{A Multi-phase transport
  model for relativistic heavy ion collisions}'',
  \href{http://dx.doi.org/10.1103/PhysRevC.72.064901}{{\em Phys. Rev. C}
  {\bfseries 72} (2005) 064901},
  \href{http://arxiv.org/abs/nucl-th/0411110}{{\ttfamily
  arXiv:nucl-th/0411110}}.

\bibitem{Li:2018leh}
H.~Li, Z.-W. Lin, and F.~Wang, ``{Charm quarks are more hydrodynamic than light
  quarks in final-state elliptic flow}'',
  \href{http://dx.doi.org/10.1103/PhysRevC.99.044911}{{\em Phys. Rev. C}
  {\bfseries 99} (2019) 044911},
  \href{http://arxiv.org/abs/1804.02681}{{\ttfamily arXiv:1804.02681
  [hep-ph]}}.

\bibitem{Lin:2021mdn}
Z.-W. Lin and L.~Zheng, ``{Further developments of a multi-phase transport
  model for relativistic nuclear collisions}'',
  \href{http://dx.doi.org/10.1007/s41365-021-00944-5}{{\em Nucl. Sci. Tech.}
  {\bfseries 32} (2021) 113}, \href{http://arxiv.org/abs/2110.02989}{{\ttfamily
  arXiv:2110.02989 [nucl-th]}}.

\bibitem{Zhang:1997ej}
B.~Zhang, ``{ZPC 1.0.1: A Parton cascade for ultrarelativistic heavy ion
  collisions}'', \href{http://dx.doi.org/10.1016/S0010-4655(98)00010-1}{{\em
  Comput. Phys. Commun.} {\bfseries 109} (1998) 193--206},
  \href{http://arxiv.org/abs/nucl-th/9709009}{{\ttfamily
  arXiv:nucl-th/9709009}}.

\bibitem{Li:1995pra}
B.-A. Li and C.~M. Ko, ``{Formation of superdense hadronic matter in
  high-energy heavy ion collisions}'',
  \href{http://dx.doi.org/10.1103/PhysRevC.52.2037}{{\em Phys. Rev. C}
  {\bfseries 52} (1995) 2037--2063},
  \href{http://arxiv.org/abs/nucl-th/9505016}{{\ttfamily
  arXiv:nucl-th/9505016}}.

\bibitem{Bilandzic:2013kga}
A.~Bilandzic, C.~H. Christensen, K.~Gulbrandsen, A.~Hansen, and Y.~Zhou,
  ``{Generic framework for anisotropic flow analyses with multiparticle
  azimuthal correlations}'',
  \href{http://dx.doi.org/10.1103/PhysRevC.89.064904}{{\em Phys. Rev. C}
  {\bfseries 89} (2014) 064904},
  \href{http://arxiv.org/abs/1312.3572}{{\ttfamily arXiv:1312.3572 [nucl-ex]}}.

\bibitem{Aamodt:2008zz}
{\bfseries ALICE} Collaboration, K.~Aamodt {\em et~al.}, ``{The ALICE
  experiment at the CERN LHC}'',
\href{http://dx.doi.org/10.1088/1748-0221/3/08/S08002}{{\em JINST} {\bfseries
  3} (2008) S08002}.

\bibitem{Abelev:2014ffa}
{\bfseries ALICE} Collaboration, B.~B. Abelev {\em et~al.}, ``{Performance of
  the ALICE Experiment at the CERN LHC}'',
  \href{http://dx.doi.org/10.1142/S0217751X14300440}{{\em Int. J. Mod. Phys.}
  {\bfseries A29} (2014) 1430044},
\href{http://arxiv.org/abs/1402.4476}{{\ttfamily arXiv:1402.4476 [nucl-ex]}}.

\bibitem{Bossu:2012jt}
{\bfseries ALICE} Collaboration, F.~Bossu, M.~Gagliardi, and M.~Marchisone,
  ``{Performance of the RPC-based ALICE muon trigger system at the LHC}'',
  \href{http://dx.doi.org/10.1088/1748-0221/7/12/T12002}{{\em JINST} {\bfseries
  7} (2012) T12002}, \href{http://arxiv.org/abs/1211.1948}{{\ttfamily
  arXiv:1211.1948 [physics.ins-det]}}.

\bibitem{Acharya:2018egz}
{\bfseries ALICE} Collaboration, S.~Acharya {\em et~al.}, ``{Charged-particle
  pseudorapidity density at mid-rapidity in p--Pb collisions at
  $\sqrt{s_{\rm{NN}}}$ = 8.16 TeV}'',
  \href{http://dx.doi.org/10.1140/epjc/s10052-019-6801-9}{{\em Eur. Phys. J. C}
  {\bfseries 79} (2019) 307}, \href{http://arxiv.org/abs/1812.01312}{{\ttfamily
  arXiv:1812.01312 [nucl-ex]}}.

\bibitem{ALICE:2018wma}
{\bfseries ALICE} Collaboration, S.~Acharya {\em et~al.}, ``{Charged-particle
  pseudorapidity density at mid-rapidity in p-Pb collisions at
  $\sqrt{s_{\rm{NN}}}$ = 8.16 TeV}'',
  \href{http://dx.doi.org/10.1140/epjc/s10052-019-6801-9}{{\em Eur. Phys. J. C}
  {\bfseries 79} (2019) 307}, \href{http://arxiv.org/abs/1812.01312}{{\ttfamily
  arXiv:1812.01312 [nucl-ex]}}.

\bibitem{pubPbPb}
{\bfseries ALICE} Collaboration, S.~Acharya {\em et~al.}, ``{Production of
  muons from heavy-flavour hadron decays at high transverse momentum in Pb-Pb
  collisions at $\sqrt{s_{\rm NN}}=5.02$ and 2.76 TeV}'',
  \href{http://dx.doi.org/10.1016/j.physletb.2021.136558}{{\em Phys. Lett. B}
  {\bfseries 820} (2021) 136558},
  \href{http://arxiv.org/abs/2011.05718}{{\ttfamily arXiv:2011.05718
  [nucl-ex]}}.

\bibitem{Acharya:2019mky}
{\bfseries ALICE} Collaboration, S.~Acharya {\em et~al.}, ``{Production of
  muons from heavy-flavour hadron decays in pp collisions at $ \sqrt{s} $ =
  5.02 TeV}'', \href{http://dx.doi.org/10.1007/JHEP09(2019)008}{{\em JHEP}
  {\bfseries 09} (2019) 008},
\href{http://arxiv.org/abs/1905.07207}{{\ttfamily arXiv:1905.07207 [nucl-ex]}}.

\bibitem{Abelev:2014mva}
{\bfseries ALICE} Collaboration, B.~B. Abelev {\em et~al.}, ``{Multiplicity
  dependence of jet-like two-particle correlation structures in p--Pb
  collisions at $\sqrt{s_{NN}}$=5.02 TeV}'',
  \href{http://dx.doi.org/10.1016/j.physletb.2014.11.028}{{\em Phys. Lett. B}
  {\bfseries 741} (2015) 38--50},
  \href{http://arxiv.org/abs/1406.5463}{{\ttfamily arXiv:1406.5463 [nucl-ex]}}.

\bibitem{Roesler:2000he}
S.~Roesler, R.~Engel, and J.~Ranft,
  \href{http://dx.doi.org/10.1007/978-3-642-18211-2_166}{``{The Monte Carlo
  event generator DPMJET-III}'',} in {\em {International Conference on Advanced
  Monte Carlo for Radiation Physics, Particle Transport Simulation and
  Applications (MC 2000)}}, pp.~1033--1038.
\newblock 12, 2000.
\newblock \href{http://arxiv.org/abs/hep-ph/0012252}{{\ttfamily
  arXiv:hep-ph/0012252}}.

\bibitem{GEANT4:2002zbu}
{\bfseries GEANT4} Collaboration, S.~Agostinelli {\em et~al.}, ``{GEANT4--a
  simulation toolkit}'',
  \href{http://dx.doi.org/10.1016/S0168-9002(03)01368-8}{{\em Nucl. Instrum.
  Meth. A} {\bfseries 506} (2003) 250--303}.

\bibitem{Asai:2015xno}
{\bfseries GEANT4} Collaboration, M.~Asai, A.~Dotti, M.~Verderi, and D.~H.
  Wright, ``{Recent developments in Geant4}'',
  \href{http://dx.doi.org/10.1016/j.anucene.2014.08.021}{{\em Annals Nucl.
  Energy} {\bfseries 82} (2015) 19--28}.

\bibitem{Masera:2009zz}
M.~Masera, G.~Ortona, M.~G. Poghosyan, and F.~Prino, ``{Anisotropic transverse
  flow introduction in Monte Carlo generators for heavy ion collisions}'',
  \href{http://dx.doi.org/10.1103/PhysRevC.79.064909}{{\em Phys. Rev. C}
  {\bfseries 79} (2009) 064909}.

\bibitem{Xu:2011fi}
J.~Xu and C.~M. Ko, ``{Pb-Pb collisions at $\sqrt{s_{NN}}=2.76$ TeV in a
  multiphase transport model}'',
  \href{http://dx.doi.org/10.1103/PhysRevC.83.034904}{{\em Phys. Rev. C}
  {\bfseries 83} (2011) 034904},
  \href{http://arxiv.org/abs/1101.2231}{{\ttfamily arXiv:1101.2231 [nucl-th]}}.

\bibitem{Efron:98913}
B.~Efron, {\em {The Jackknife, the bootstrap and other resampling plans}}.
\newblock CBMS-NSF Regional Conference Series in Applied Mathematics. SIAM,
  Philadelphia, PA, 1982.
\newblock \url{https://cds.cern.ch/record/98913}.
\newblock Lectures given at Bowling Green State Univ., June 1980.

\bibitem{Khachatryan:2015oea}
{\bfseries CMS} Collaboration, V.~Khachatryan {\em et~al.}, ``{Evidence for
  transverse momentum and pseudorapidity dependent event plane fluctuations in
  PbPb and pPb collisions}'',
  \href{http://dx.doi.org/10.1103/PhysRevC.92.034911}{{\em Phys. Rev. C}
  {\bfseries 92} (2015) 034911},
  \href{http://arxiv.org/abs/1503.01692}{{\ttfamily arXiv:1503.01692
  [nucl-ex]}}.

\bibitem{Aaboud:2017tql}
{\bfseries ATLAS} Collaboration, M.~Aaboud {\em et~al.}, ``{Measurement of
  longitudinal flow decorrelations in Pb+Pb collisions at $\sqrt{s_{\rm NN}} =
  2.76$ and 5.02 TeV with the ATLAS detector}'',
  \href{http://dx.doi.org/10.1140/epjc/s10052-018-5605-7}{{\em Eur. Phys. J. C}
  {\bfseries 78} (2018) 142}, \href{http://arxiv.org/abs/1709.02301}{{\ttfamily
  arXiv:1709.02301 [nucl-ex]}}.

\bibitem{ALICE:2017lyf}
{\bfseries ALICE} Collaboration, S.~Acharya {\em et~al.}, ``{Searches for
  transverse momentum dependent flow vector fluctuations in Pb--Pb and p--Pb
  collisions at the LHC}'',
  \href{http://dx.doi.org/10.1007/JHEP09(2017)032}{{\em JHEP} {\bfseries 09}
  (2017) 032}, \href{http://arxiv.org/abs/1707.05690}{{\ttfamily
  arXiv:1707.05690 [nucl-ex]}}.

\bibitem{Acharya:2018pjd}
{\bfseries ALICE} Collaboration, S.~Acharya {\em et~al.}, ``{Study of J/$\psi$
  azimuthal anisotropy at forward rapidity in Pb--Pb collisions at $
  \sqrt{s_{\mathrm{NN}}}=5.02 $ TeV}'',
  \href{http://dx.doi.org/10.1007/JHEP02(2019)012}{{\em JHEP} {\bfseries 02}
  (2019) 012}, \href{http://arxiv.org/abs/1811.12727}{{\ttfamily
  arXiv:1811.12727 [nucl-ex]}}.

\bibitem{Cacciari:1998it}
M.~Cacciari, M.~Greco, and P.~Nason, ``{The $p_{\rm T}$ spectrum in heavy
  flavor hadroproduction}'',
  \href{http://dx.doi.org/10.1088/1126-6708/1998/05/007}{{\em JHEP} {\bfseries
  05} (1998) 007},
\href{http://arxiv.org/abs/hep-ph/9803400}{{\ttfamily arXiv:hep-ph/9803400
  [hep-ph]}}.

\bibitem{Cacciari:2012ny}
M.~Cacciari, S.~Frixione, N.~Houdeau, M.~L. Mangano, P.~Nason, and G.~Ridolfi,
  ``{Theoretical predictions for charm and bottom production at the LHC}'',
  \href{http://dx.doi.org/10.1007/JHEP10(2012)137}{{\em JHEP} {\bfseries 10}
  (2012) 137},
\href{http://arxiv.org/abs/1205.6344}{{\ttfamily arXiv:1205.6344 [hep-ph]}}.

\bibitem{ALICE-PUBLIC-2018-011}
{\bfseries ALICE} Collaboration, ``Centrality determination in heavy ion
  collisions.'' {ALICE-PUBLIC-2018-011}, Aug., 2018.
\newblock \url{http://cds.cern.ch/record/2636623}.

\bibitem{ALICE:2013snk}
{\bfseries ALICE} Collaboration, B.~B. Abelev {\em et~al.}, ``{Long-range
  angular correlations of $\rm \pi$, K and p in p-Pb collisions at
  $\sqrt{s_{\rm NN}}$ = 5.02 TeV}'',
  \href{http://dx.doi.org/10.1016/j.physletb.2013.08.024}{{\em Phys. Lett. B}
  {\bfseries 726} (2013) 164--177},
  \href{http://arxiv.org/abs/1307.3237}{{\ttfamily arXiv:1307.3237 [nucl-ex]}}.

\bibitem{ALICE:2022wpn}
{\bfseries ALICE} Collaboration, ``{The ALICE experiment -- A journey through
  QCD}'', \href{http://arxiv.org/abs/2211.04384}{{\ttfamily arXiv:2211.04384
  [nucl-ex]}}.

\bibitem{Sjostrand:2014zea}
T.~Sj{\"o}strand, S.~Ask, J.~R. Christiansen, R.~Corke, N.~Desai, P.~Ilten,
  S.~Mrenna, S.~Prestel, C.~O. Rasmussen, and P.~Z. Skands, ``{An Introduction
  to PYTHIA 8.2}'', \href{http://dx.doi.org/10.1016/j.cpc.2015.01.024}{{\em
  Comput. Phys. Commun.} {\bfseries 191} (2015) 159--177},
\href{http://arxiv.org/abs/1410.3012}{{\ttfamily arXiv:1410.3012 [hep-ph]}}.

\bibitem{Zhao:2020wcd}
W.~Zhao, C.~M. Ko, Y.-X. Liu, G.-Y. Qin, and H.~Song, ``{Probing the Partonic
  Degrees of Freedom in High-Multiplicity p--Pb collisions at $\sqrt {s_{\rm
  NN}}$ = 5.02 TeV}'',
  \href{http://dx.doi.org/10.1103/PhysRevLett.125.072301}{{\em Phys. Rev.
  Lett.} {\bfseries 125} (2020) 072301},
  \href{http://arxiv.org/abs/1911.00826}{{\ttfamily arXiv:1911.00826
  [nucl-th]}}.

\bibitem{Zhang:2019dth}
C.~Zhang, C.~Marquet, G.-Y. Qin, S.-Y. Wei, and B.-W. Xiao, ``{Elliptic Flow of
  Heavy Quarkonia in $pA$ Collisions}'',
  \href{http://dx.doi.org/10.1103/PhysRevLett.122.172302}{{\em Phys. Rev.
  Lett.} {\bfseries 122} (2019) 172302},
  \href{http://arxiv.org/abs/1901.10320}{{\ttfamily arXiv:1901.10320
  [hep-ph]}}.

\bibitem{Zhang:2020ayy}
C.~Zhang, C.~Marquet, G.-Y. Qin, Y.~Shi, L.~Wang, S.-Y. Wei, and B.-W. Xiao,
  ``{Collectivity of heavy mesons in proton-nucleus collisions}'',
  \href{http://dx.doi.org/10.1103/PhysRevD.102.034010}{{\em Phys. Rev. D}
  {\bfseries 102} (2020) 034010},
  \href{http://arxiv.org/abs/2002.09878}{{\ttfamily arXiv:2002.09878
  [hep-ph]}}.

\bibitem{Albacete:2010bs}
J.~L. Albacete and C.~Marquet, ``{Single Inclusive Hadron Production at RHIC
  and the LHC from the Color Glass Condensate}'',
  \href{http://dx.doi.org/10.1016/j.physletb.2010.02.073}{{\em Phys. Lett. B}
  {\bfseries 687} (2010) 174--179},
  \href{http://arxiv.org/abs/1001.1378}{{\ttfamily arXiv:1001.1378 [hep-ph]}}.

\end{thebibliography}\endgroup

\newpage
\appendix

\section{The ALICE Collaboration}
\label{app:collab}
\begin{flushleft} 
\small

S.~Acharya\,\orcidlink{0000-0002-9213-5329}\,$^{\rm 124}$, 
D.~Adamov\'{a}\,\orcidlink{0000-0002-0504-7428}\,$^{\rm 85}$, 
A.~Adler$^{\rm 69}$, 
G.~Aglieri Rinella\,\orcidlink{0000-0002-9611-3696}\,$^{\rm 32}$, 
M.~Agnello\,\orcidlink{0000-0002-0760-5075}\,$^{\rm 29}$, 
N.~Agrawal\,\orcidlink{0000-0003-0348-9836}\,$^{\rm 50}$, 
Z.~Ahammed\,\orcidlink{0000-0001-5241-7412}\,$^{\rm 132}$, 
S.~Ahmad\,\orcidlink{0000-0003-0497-5705}\,$^{\rm 15}$, 
S.U.~Ahn\,\orcidlink{0000-0001-8847-489X}\,$^{\rm 70}$, 
I.~Ahuja\,\orcidlink{0000-0002-4417-1392}\,$^{\rm 37}$
% ...

\section*{Affiliation Notes}

$^{\rm I}$ Deceased\\
$^{\rm II}$ Also at: Max-Planck-Institut f\"{u}r Physik, Munich, Germany\\
$^{\rm III}$ Also at: Italian National Agency for New Technologies, Energy and Sustainable Economic Development (ENEA), Bologna, Italy\\
$^{\rm IV}$ Also at: Dipartimento DET del Politecnico di Torino, Turin, Italy\\
$^{\rm V}$ Also at: Department of Applied Physics, Aligarh Muslim University, Aligarh, India\\
$^{\rm VI}$ Also at: Institute of Theoretical Physics, University of Wroclaw, Poland\\
$^{\rm VII}$ Also at: An institution covered by a cooperation agreement with CERN\\

\section*{Collaboration Institutes}

$^{1}$ A.I. Alikhanyan National Science Laboratory (Yerevan Physics Institute) Foundation, Yerevan, Armenia\\
$^{2}$ AGH University of Krakow, Cracow, Poland\\
$^{3}$ Bogolyubov Institute for Theoretical Physics, National Academy of Sciences of Ukraine, Kiev, Ukraine\\
$^{4}$ Bose Institute, Department of Physics  and Centre for Astroparticle Physics and Space Science (CAPSS), Kolkata, India\\
$^{5}$ California Polytechnic State University, San Luis Obispo, California, United States\\
$^{6}$ Central China Normal University, Wuhan, China\\
$^{7}$ Centro de Aplicaciones Tecnol\'{o}gicas y Desarrollo Nuclear (CEADEN), Havana, Cuba\\
$^{8}$ Centro de Investigaci\'{o}n y de Estudios Avanzados (CINVESTAV), Mexico City and M\'{e}rida, Mexico\\
$^{9}$ Chicago State University, Chicago, Illinois, United States\\
$^{10}$ China Institute of Atomic Energy, Beijing, China\\
$^{11}$ Chungbuk National University, Cheongju, Republic of Korea\\
$^{12}$ Comenius University Bratislava, Faculty of Mathematics, Physics and Informatics, Bratislava, Slovak Republic\\
$^{13}$ COMSATS University Islamabad, Islamabad, Pakistan\\
$^{14}$ Creighton University, Omaha, Nebraska, United States\\
$^{15}$ Department of Physics, Aligarh Muslim University, Aligarh, India\\
$^{16}$ Department of Physics, Pusan National University, Pusan, Republic of Korea\\
$^{17}$ Department of Physics, Sejong University, Seoul, Republic of Korea\\
$^{18}$ Department of Physics, University of California, Berkeley, California, United States\\
$^{19}$ Department of Physics, University of Oslo, Oslo, Norway\\
$^{20}$ Department of Physics and Technology, University of Bergen, Bergen, Norway\\
$^{21}$ Dipartimento di Fisica, Universit\`{a} di Pavia, Pavia, Italy\\
$^{22}$ Dipartimento di Fisica dell'Universit\`{a} and Sezione INFN, Cagliari, Italy\\
$^{23}$ Dipartimento di Fisica dell'Universit\`{a} and Sezione INFN, Trieste, Italy\\
$^{24}$ Dipartimento di Fisica dell'Universit\`{a} and Sezione INFN, Turin, Italy\\
$^{25}$ Dipartimento di Fisica e Astronomia dell'Universit\`{a} and Sezione INFN, Bologna, Italy\\
$^{26}$ Dipartimento di Fisica e Astronomia dell'Universit\`{a} and Sezione INFN, Catania, Italy\\
$^{27}$ Dipartimento di Fisica e Astronomia dell'Universit\`{a} and Sezione INFN, Padova, Italy\\
$^{28}$ Dipartimento di Fisica `E.R.~Caianiello' dell'Universit\`{a} and Gruppo Collegato INFN, Salerno, Italy\\
$^{29}$ Dipartimento DISAT del Politecnico and Sezione INFN, Turin, Italy\\
$^{30}$ Dipartimento di Scienze MIFT, Universit\`{a} di Messina, Messina, Italy\\
$^{31}$ Dipartimento Interateneo di Fisica `M.~Merlin' and Sezione INFN, Bari, Italy\\
$^{32}$ European Organization for Nuclear Research (CERN), Geneva, Switzerland\\
$^{33}$ Faculty of Electrical Engineering, Mechanical Engineering and Naval Architecture, University of Split, Split, Croatia\\
$^{34}$ Faculty of Engineering and Science, Western Norway University of Applied Sciences, Bergen, Norway\\
$^{35}$ Faculty of Nuclear Sciences and Physical Engineering, Czech Technical University in Prague, Prague, Czech Republic\\
$^{36}$ Faculty of Physics, Sofia University, Sofia, Bulgaria\\
$^{37}$ Faculty of Science, P.J.~\v{S}af\'{a}rik University, Ko\v{s}ice, Slovak Republic\\
$^{38}$ Frankfurt Institute for Advanced Studies, Johann Wolfgang Goethe-Universit\"{a}t Frankfurt, Frankfurt, Germany\\
$^{39}$ Fudan University, Shanghai, China\\
$^{40}$ Gangneung-Wonju National University, Gangneung, Republic of Korea\\
$^{41}$ Gauhati University, Department of Physics, Guwahati, India\\
$^{42}$ Helmholtz-Institut f\"{u}r Strahlen- und Kernphysik, Rheinische Friedrich-Wilhelms-Universit\"{a}t Bonn, Bonn, Germany\\
$^{43}$ Helsinki Institute of Physics (HIP), Helsinki, Finland\\
$^{44}$ High Energy Physics Group,  Universidad Aut\'{o}noma de Puebla, Puebla, Mexico\\
$^{45}$ Horia Hulubei National Institute of Physics and Nuclear Engineering, Bucharest, Romania\\
$^{46}$ Indian Institute of Technology Bombay (IIT), Mumbai, India\\
$^{47}$ Indian Institute of Technology Indore, Indore, India\\
$^{48}$ INFN, Laboratori Nazionali di Frascati, Frascati, Italy\\
$^{49}$ INFN, Sezione di Bari, Bari, Italy\\
$^{50}$ INFN, Sezione di Bologna, Bologna, Italy\\
$^{51}$ INFN, Sezione di Cagliari, Cagliari, Italy\\
$^{52}$ INFN, Sezione di Catania, Catania, Italy\\
$^{53}$ INFN, Sezione di Padova, Padova, Italy\\
$^{54}$ INFN, Sezione di Pavia, Pavia, Italy\\
$^{55}$ INFN, Sezione di Torino, Turin, Italy\\
$^{56}$ INFN, Sezione di Trieste, Trieste, Italy\\
$^{57}$ Inha University, Incheon, Republic of Korea\\
$^{58}$ Institute for Gravitational and Subatomic Physics (GRASP), Utrecht University/Nikhef, Utrecht, Netherlands\\
$^{59}$ Institute of Experimental Physics, Slovak Academy of Sciences, Ko\v{s}ice, Slovak Republic\\
$^{60}$ Institute of Physics, Homi Bhabha National Institute, Bhubaneswar, India\\
$^{61}$ Institute of Physics of the Czech Academy of Sciences, Prague, Czech Republic\\
$^{62}$ Institute of Space Science (ISS), Bucharest, Romania\\
$^{63}$ Institut f\"{u}r Kernphysik, Johann Wolfgang Goethe-Universit\"{a}t Frankfurt, Frankfurt, Germany\\
$^{64}$ Instituto de Ciencias Nucleares, Universidad Nacional Aut\'{o}noma de M\'{e}xico, Mexico City, Mexico\\
$^{65}$ Instituto de F\'{i}sica, Universidade Federal do Rio Grande do Sul (UFRGS), Porto Alegre, Brazil\\
$^{66}$ Instituto de F\'{\i}sica, Universidad Nacional Aut\'{o}noma de M\'{e}xico, Mexico City, Mexico\\
$^{67}$ iThemba LABS, National Research Foundation, Somerset West, South Africa\\
$^{68}$ Jeonbuk National University, Jeonju, Republic of Korea\\
$^{69}$ Johann-Wolfgang-Goethe Universit\"{a}t Frankfurt Institut f\"{u}r Informatik, Fachbereich Informatik und Mathematik, Frankfurt, Germany\\
$^{70}$ Korea Institute of Science and Technology Information, Daejeon, Republic of Korea\\
$^{71}$ KTO Karatay University, Konya, Turkey\\
$^{72}$ Laboratoire de Physique Subatomique et de Cosmologie, Universit\'{e} Grenoble-Alpes, CNRS-IN2P3, Grenoble, France\\
$^{73}$ Lawrence Berkeley National Laboratory, Berkeley, California, United States\\
$^{74}$ Lund University Department of Physics, Division of Particle Physics, Lund, Sweden\\
$^{75}$ Nagasaki Institute of Applied Science, Nagasaki, Japan\\
$^{76}$ Nara Women{'}s University (NWU), Nara, Japan\\
$^{77}$ National and Kapodistrian University of Athens, School of Science, Department of Physics , Athens, Greece\\
$^{78}$ National Centre for Nuclear Research, Warsaw, Poland\\
$^{79}$ National Institute of Science Education and Research, Homi Bhabha National Institute, Jatni, India\\
$^{80}$ National Nuclear Research Center, Baku, Azerbaijan\\
$^{81}$ National Research and Innovation Agency - BRIN, Jakarta, Indonesia\\
$^{82}$ Niels Bohr Institute, University of Copenhagen, Copenhagen, Denmark\\
$^{83}$ Nikhef, National institute for subatomic physics, Amsterdam, Netherlands\\
$^{84}$ Nuclear Physics Group, STFC Daresbury Laboratory, Daresbury, United Kingdom\\
$^{85}$ Nuclear Physics Institute of the Czech Academy of Sciences, Husinec-\v{R}e\v{z}, Czech Republic\\
$^{86}$ Oak Ridge National Laboratory, Oak Ridge, Tennessee, United States\\
$^{87}$ Ohio State University, Columbus, Ohio, United States\\
$^{88}$ Physics department, Faculty of science, University of Zagreb, Zagreb, Croatia\\
$^{89}$ Physics Department, Panjab University, Chandigarh, India\\
$^{90}$ Physics Department, University of Jammu, Jammu, India\\
$^{91}$ Physics Program and International Institute for Sustainability with Knotted Chiral Meta Matter (SKCM2), Hiroshima University, Hiroshima, Japan\\
$^{92}$ Physikalisches Institut, Eberhard-Karls-Universit\"{a}t T\"{u}bingen, T\"{u}bingen, Germany\\
$^{93}$ Physikalisches Institut, Ruprecht-Karls-Universit\"{a}t Heidelberg, Heidelberg, Germany\\
$^{94}$ Physik Department, Technische Universit\"{a}t M\"{u}nchen, Munich, Germany\\
$^{95}$ Politecnico di Bari and Sezione INFN, Bari, Italy\\
$^{96}$ Research Division and ExtreMe Matter Institute EMMI, GSI Helmholtzzentrum f\"ur Schwerionenforschung GmbH, Darmstadt, Germany\\
$^{97}$ Saga University, Saga, Japan\\
$^{98}$ Saha Institute of Nuclear Physics, Homi Bhabha National Institute, Kolkata, India\\
$^{99}$ School of Physics and Astronomy, University of Birmingham, Birmingham, United Kingdom\\
$^{100}$ Secci\'{o}n F\'{\i}sica, Departamento de Ciencias, Pontificia Universidad Cat\'{o}lica del Per\'{u}, Lima, Peru\\
$^{101}$ Stefan Meyer Institut f\"{u}r Subatomare Physik (SMI), Vienna, Austria\\
$^{102}$ SUBATECH, IMT Atlantique, Nantes Universit\'{e}, CNRS-IN2P3, Nantes, France\\
$^{103}$ Sungkyunkwan University, Suwon City, Republic of Korea\\
$^{104}$ Suranaree University of Technology, Nakhon Ratchasima, Thailand\\
$^{105}$ Technical University of Ko\v{s}ice, Ko\v{s}ice, Slovak Republic\\
$^{106}$ The Henryk Niewodniczanski Institute of Nuclear Physics, Polish Academy of Sciences, Cracow, Poland\\
$^{107}$ The University of Texas at Austin, Austin, Texas, United States\\
$^{108}$ Universidad Aut\'{o}noma de Sinaloa, Culiac\'{a}n, Mexico\\
$^{109}$ Universidade de S\~{a}o Paulo (USP), S\~{a}o Paulo, Brazil\\
$^{110}$ Universidade Estadual de Campinas (UNICAMP), Campinas, Brazil\\
$^{111}$ Universidade Federal do ABC, Santo Andre, Brazil\\
$^{112}$ University of Cape Town, Cape Town, South Africa\\
$^{113}$ University of Houston, Houston, Texas, United States\\
$^{114}$ University of Jyv\"{a}skyl\"{a}, Jyv\"{a}skyl\"{a}, Finland\\
$^{115}$ University of Kansas, Lawrence, Kansas, United States\\
$^{116}$ University of Liverpool, Liverpool, United Kingdom\\
$^{117}$ University of Science and Technology of China, Hefei, China\\
$^{118}$ University of South-Eastern Norway, Kongsberg, Norway\\
$^{119}$ University of Tennessee, Knoxville, Tennessee, United States\\
$^{120}$ University of the Witwatersrand, Johannesburg, South Africa\\
$^{121}$ University of Tokyo, Tokyo, Japan\\
$^{122}$ University of Tsukuba, Tsukuba, Japan\\
$^{123}$ University Politehnica of Bucharest, Bucharest, Romania\\
$^{124}$ Universit\'{e} Clermont Auvergne, CNRS/IN2P3, LPC, Clermont-Ferrand, France\\
$^{125}$ Universit\'{e} de Lyon, CNRS/IN2P3, Institut de Physique des 2 Infinis de Lyon, Lyon, France\\
$^{126}$ Universit\'{e} de Strasbourg, CNRS, IPHC UMR 7178, F-67000 Strasbourg, France, Strasbourg, France\\
$^{127}$ Universit\'{e} Paris-Saclay, Centre d'Etudes de Saclay (CEA), IRFU, D\'{e}partment de Physique Nucl\'{e}aire (DPhN), Saclay, France\\
$^{128}$ Universit\'{e}  Paris-Saclay, CNRS/IN2P3, IJCLab, Orsay, France\\
$^{129}$ Universit\`{a} degli Studi di Foggia, Foggia, Italy\\
$^{130}$ Universit\`{a} del Piemonte Orientale, Vercelli, Italy\\
$^{131}$ Universit\`{a} di Brescia, Brescia, Italy\\
$^{132}$ Variable Energy Cyclotron Centre, Homi Bhabha National Institute, Kolkata, India\\
$^{133}$ Warsaw University of Technology, Warsaw, Poland\\
$^{134}$ Wayne State University, Detroit, Michigan, United States\\
$^{135}$ Westf\"{a}lische Wilhelms-Universit\"{a}t M\"{u}nster, Institut f\"{u}r Kernphysik, M\"{u}nster, Germany\\
$^{136}$ Wigner Research Centre for Physics, Budapest, Hungary\\
$^{137}$ Yale University, New Haven, Connecticut, United States\\
$^{138}$ Yonsei University, Seoul, Republic of Korea\\
$^{139}$  Zentrum  f\"{u}r Technologie und Transfer (ZTT), Worms, Germany\\
$^{140}$ Affiliated with an institute covered by a cooperation agreement with CERN\\
$^{141}$ Affiliated with an international laboratory covered by a cooperation agreement with CERN.\\

\end{flushleft} 

\end{document}